\documentclass[sigconf]{acmart}

\newcommand{\code}[1]{\texttt{\detokenize{#1}}}
\usepackage{arydshln}
\usepackage[utf8]{inputenc}
\usepackage[russian, greek, english]{babel}
\usepackage[T1]{fontenc}
\usepackage[autostyle, english = american]{csquotes}
\MakeOuterQuote{"}

\usepackage{makecell}
\usepackage{booktabs}
\usepackage{graphicx}
\usepackage{multirow}
\usepackage[toc]{appendix}
\usepackage{longtable}
\usepackage{eurosym}

\usepackage{pifont}
\usepackage{multicol}
\usepackage{pgfplots}
\usepackage{array}
\usepackage{colortbl}

\usepackage{tikz}

\usepackage{caption}
\usepackage{subcaption}
\usepackage{hyperref}
\AtBeginDocument{%
  }

\copyrightyear{2025}
\acmYear{2025}
\setcopyright{cc}
\setcctype{by-nc-sa}
\acmConference[CHI '25]{CHI Conference on Human Factors in Computing Systems}{April 26-May 1, 2025}{Yokohama, Japan}
\acmBooktitle{CHI Conference on Human Factors in Computing Systems (CHI '25), April 26-May 1, 2025, Yokohama, Japan}\acmDOI{10.1145/3706598.3713648}
\acmISBN{979-8-4007-1394-1/2025/04}

\begin{document}

% \title{\code{[anonymised].eu}: Robust Analysis of Consent Interfaces on the Web}
% \title{How to Scrape a Cookie Banner: A Cross-Country Analysis of GDPR Consent and a Flexible Method For Doing It}
\title{A Cross-Country Analysis of GDPR Cookie Banners and Flexible Methods for Scraping Them}
% \title{\code{consent-observatory.eu}: Robust Analysis of Consent Interfaces on the Web}

\author{Midas Nouwens}
\affiliation{%
  \institution{Digital Design and Information Studies}
  \city{Aarhus University}
  \country{Denmark}
}
\email{midasnouwens@cc.au.dk}

\author{Janus Bager Kristensen}
\affiliation{%
  \institution{Center for Advanced Visualisation and Interaction}
  \city{Aarhus University}
  \country{Denmark}
}
\email{jbk@cavi.dk}

\author{Kristjan Maalt}
\affiliation{%
  \institution{Business Development and Technology}
  \city{Aarhus University}
  \country{Denmark}}
\email{krma@btech.au.dk}

\author{Rolf Bagge}
\affiliation{%
 \institution{Center for Advanced Visualisation and Interaction}
 \city{Aarhus University}
 \country{Denmark}
 }
\email{rolf@cavi.dk}

\renewcommand{\shortauthors}{Nouwens et al.}

\begin{abstract}
    Online tracking remains problematic, with compliance and ethical issues persisting despite regulatory efforts. 
    Consent interfaces, the visible manifestation of this industry, have seen significant attention over the years. 
    We present robust automated methods to study the presence, design, and third-party suppliers of consent interfaces at scale and the web service \href{https://consent-observatory.eu/}{consent-observatory.eu} to do it with.
    We examine the top 10,000 websites across 31 countries under the ePrivacy Directive and GDPR (n=254.148). 
    Our findings show that 67\% of websites use consent interfaces, but only 15\% are minimally compliant, mostly because they lack a reject option.
    Consent management platforms (CMPs) are powerful intermediaries in this space: 67\% of interfaces are provided by CMPs, and three organisations hold 37\% of the market.
    There is little evidence that regulators’ guidance and fines have impacted compliance rates, but 18\% of compliance variance is explained by CMPs.
    Researchers should take an infrastructural perspective on online tracking and study the factual control of intermediaries to identify effective leverage points.
\end{abstract}

% national differences
% popups mostly on top
% CMPs mostly used by top websites




\maketitle

\section{Introduction}
European digital regulation continues to struggle with the online tracking industry, despite widespread concerns that much of what is happening violates individuals' fundamental rights to both data protection and privacy, and threatens states' electoral integrity~\cite{breyer2024} and national security~\cite{ICCL}.
The way individuals are tracked online, their data flowing to hundreds of actors, remains largely invisible to an average Web user~\cite{zhang2024they}. The visible surface is typically a consent interface --- colloquially `cookie banners' or `pop-ups'.\footnote{Research refers to these variously as consent banners, consent pop-ups, consent notices, consent management platforms, cookie banners, cookie pop-ups, cookie walls, tracking walls, notice \& consent or notice \& choice. Here we use `consent interfaces' as a more neutral term that does not say anything about its design (in a way that banner, pop-up, or wall do) or the technology used (which `cookie' does, given that such interfaces supposedly control other mechanisms). Arguably the word consent is also not entirely representative of these interfaces, however, as sites also claim to use them to inform or establish alternative legal bases such as `legitimate interest', however legally questionable~\cite{vealeAdtechRealTimeBidding2022}.} These interfaces claim to help individuals understand what data is collected, what for, and who it is shared with. In the process, the companies selling these interfaces make the questionable claim to websites that installing the interface will ensure their compliance with laws that, when considered more closely, would appear to prohibit these contemporary tracking practices~\cite{vealeAdtechRealTimeBidding2022,vealeImpossibleAsksCan2022}.
 
The visibility of these interfaces means they are often a topic of public commentary in the news and by politicians.\footnote{See, for example, the extensive reporting by Natasha Lomas for TechCrunch: \url{https://techcrunch.com/tag/cookie-consent/}.} 
Wide-ranging laws like the ePrivacy Directive (ePD) and the General Data Protection Regulation (GDPR) are so strongly associated with them (and suffer reputational damage because of it) that their much broader scope and ambitions are often forgotten. % (so much so that the ePD was dubbed the ``EU Cookie Law'')
Even regulatory authorities can be accused of prioritising interface design over considering the legality of the actual tracking that goes on behind them~\cite{APcookies}. 

Academic researchers have mostly participated in this power struggle by collecting empirical evidence about various aspects of these consent interfaces, for example about the user's perception of the pop-ups~\cite{Kulyk_Hilt_Gerber_Volkamer_2018, Jayakumar_2021}, their impact on consent behaviour~\cite{bielova2024effect, Nouwens_Liccardi_Veale_Karger_Kagal_2020}, legal interpretations of their compliance~\cite{Matte_Bielova_Santos_2020, Zuiderveen_Borgesius_2017, santos2021consent,vealeAdtechRealTimeBidding2022}, and the prevalence of different interface designs~\cite{Krisam_Dietmann_Volkamer_Kulyk_2021, Kirkman_Vaniea_Woods_2023, Bollinger_Kubicek_Cotrini_Basin_2022}. 
These papers have been used by EU data protection authorities (DPAs) to prioritise limited enforcement capacity and justify interpretation~\cite{cnilfacebook}.

% TODO: Rewrite this after the research gap is clear
% Methodologically, studies analysing the prevalence and designs of consent interfaces have focused on interfaces of a few third-party Consent Management Platforms (CMPs), studied a limited number of countries, and often used hard-coded detection methods. 

Existing studies, although useful, have been methodologically limited in their ability to observe the entire ecosystem. 
Many countries within relevant jurisdictions have never been studied, meaning comparative knowledge about effective enforcement strategies is lacking. 
Previous research has also shown that the off-the-shelf interfaces, `consent management platforms' (CMPs), which are the ones primarily hardcoded for study by the literature, are mostly used by medium-popularity websites~\cite{Hils_Woods_Bohme_2020}, and so we know less about popular sites (which might implement their own) or less popular sites (with little resource or incentive). 
% Furthermore, hard-coded detection methods are time-consuming to build and maintain, so research is biased towards a handful of popular interface vendors, even though a long tail of other formats exist.
% TODO: bring in the point about finding leverage points?
% TODO: make explicit that this paper focuses on Country and Popularity (maybe CMP?)?
Only by taking an expansive view of these interfaces are we able to reckon with its complex, heavily-intermediated nature, and identify the most promising leverage points for intervention.

This paper makes three contributions:
\begin{itemize}
    \item \textit{Methodological}: open-source methods that can accurately detect the presence of a consent interface, the provider of the interface, the presence of user options, the visual prominence of those options, and the status of granular consent toggles;
    \item \textit{Empirical}: an analysis of consent interfaces on the top 10,000 most popular websites across 31 countries under the ePrivacy Directive and GDPR;
    \item \textit{Technological}: \href{https://consent-observatory.eu/}{consent-observatory.eu}, an online scraping service that makes studying consent interfaces easier by removing technical barriers and more accessible to a broader audience, such as journalists and regulators.
\end{itemize}

\section{Background \& Related Work}

Consent interfaces have been studied by many disciplinary communities.
HCI studies have focussed on user perception of such interfaces~\cite{Kulyk_Hilt_Gerber_Volkamer_2018} or their impact on behaviours~\cite{Habib_Li_Young_Cranor_2022}. 
Computer Science studies have often undertaken large scale interface measurements~\cite{Bollinger_Kubicek_Cotrini_Basin_2022}, including the impact of these interfaces on actual tracking~\cite{Demir_Urban_Pohlmann_Wressnegger_2024}.
Legal studies analyse what various laws say about the interfaces~\cite{Kosta_2013}, and to what extent interfaces could or do comply~\cite{Fouad_2020,vealeAdtechRealTimeBidding2022,vealeImpossibleAsksCan2022}. Lessons for each of these fields can still be learned from synthesising the above findings. 
Below we draw out salient aspects from each that inform the study in this paper.
% TODO: this next sentence could be in the intro
%There is now a sizeable enough body of work that multiple methodological approaches have been tried, and it is possible to compare their results, combine the best approaches, and provide more comprehensive analyses.
% TODO: perhaps add refs to the subsections here
%After briefly reviewing the relevant legislation, we will discuss 1) which websites and countries have been studied, 2) what aspects of the interfaces have been analysed, and 3) what methods have been used. 

\subsection{What is the legal context?}
In Europe, there are two key pieces of legislation that shaped the emergence and design of consent interfaces.

The 2002 ePrivacy Directive regulates confidentiality of communications in what was then a rapidly digitising world~\cite{eprivadirective2002}. A significant change occurred in 2009~\cite{eprivadirective2009}, where the storage or reading of information on an individual's ``terminal device'' (e.g. their phone, or browser) moved from an opt-out provision (which already created some banners) to an opt-in provision, particularly in response to Sony Music/BMG's covert installation of rootkits on people's devices when they inserted CDs as an anti-piracy measure~\cite{Kosta_2013}. Like a cookie, this counted as storing and reading information from a terminal device, and like a cookie, needs consent. Yet while the law is generic across many different technological areas (including malware, fingerprinting, HTML storage, or even connected device technologies like RFID~\cite{EDPBepriv}), it became known as the ``Cookie Law''. 
Some national implementations exempt consent if a website reads or writes data that is ``strictly necessary'' for providing the service or assessing its quality and effectiveness, provided it does not affect the user's private life.

The 2016 General Data Protection Regulation (GDPR) regulates the processing of personal data~\cite{gdpr}. It became enforceable in 2018 --- around the same time that the consent interface market began professionalising into `consent management platforms', and websites started to install more interfaces~\cite{Hils_Woods_Bohme_2020}. However, the law did not make a material change to the ePrivacy Directive in the context of online tracking, which already required consent for \textit{any} data access. Although ePrivacy law relies on the GDPR's (and its predecessor's) definition of consent by referring to it, and the statutory text of this definition did get stronger in 2018 (such as emphasising that a pre-checked box, or continuing to use a website could not be consent~\cite{article29workingpartyGuidelinesConsentRegulation2018}), the European Court of Justice later clarified that the old definition was to be interpreted in just as strong a way, and so little had really changed in what websites were supposed to do~\cite{planet49ag}. What likely explains the compliance rush is a panic around fines and consequences, which \emph{did} change 
% --- although fines for ePrivacy violations alone remain small as the reform to that law never passed --- 
and firms were concerned that regulatory action would be harsh, when in practice, limited enforcement occurred around online tracking in the years directly following the GDPR going into effect. Furthermore, uncertainty does still exist in relation to diverging national guidance~\cite{bielova2024two}. This includes whether closing a consent banner should constitute a rejection (the French DPA says yes~\cite{CNIL}, the Swedish DPA no~\cite{IMY}) or how accept and reject options should be presented visually (the Danish DPA states green is an impermissible nudge~\cite{Datatilsynet}, but few other DPAs express views on colours).
The spread of these interfaces, and the question whether they truly empower users, has also reignited discussions about the legal theory of consent in privacy and data protection laws~\cite{solove2024murky, jones2024character}.

While the vast majority of work focuses on European law{~\cite{birrell2024sok}, privacy laws exist in over 160 countries, with many copied from the GDPR or very similar in nature~\cite{greenleafGlobalTablesData2023,greenleafGlobalDataPrivacy2023a}. Studies on consent interfaces are now emerging with comparable regulations such as the California Consumer Privacy Act (CCPA) in the United States~\cite{Zhang_Meng_Zhou_Ren_2024, Mazumdar_Timko_Rahman_2023, O’Connor_Nurwono_Siebel_Birrell_2021}.

\subsection{What part of the Web is studied?}
% TODO: if time, make a table where you actually quantify this for each
Academic research on consent interfaces do not all study the same part of the Web.
The Web is not a singular entity~\cite{o2018four} and users have different experiences due to the device and browser they use, their position in the material infrastructure, geo-restrictions, personalisation, regulatory jurisdiction, etc.
There are three variables that impact whose point of view is represented in these studies: \textit{which online services are looked at}, \textit{how many domains are included}, and the \textit{vantage point from which the data is collected}.

Many studies use popularity-based `top lists' of websites -- created by services such as Tranco, Alexa, Majestic, or Google -- rather than a manually curated set of domains.\footnote{Exceptions include supplementing lists with handpicked domains~\cite{Sheil_Malone_2022,Kampanos_Shahandashti_2021}, a UN database~\cite{Alharbi_Albesher_Wahsheh_2023}, and incorporating lists from magazines/news outlets~\cite{Soe_Nordberg_Guribye_Slavkovik_2020}.}
Often, these studies simply use the list of globally most popular sites~\cite{Bollinger_Kubicek_Cotrini_Basin_2022, Habib_Li_Young_Cranor_2022, Sanchez-Rola_Dell’Amico_Kotzias_Balzarotti_Bilge_Vervier_Santos_2019, Kyi_2023, Zhang_Meng_Zhou_Ren_2024, Morel_Santos_Fredholm_Thunberg_2023}, whereas others create subsets from these top lists.
For example, by filtering the list using the country code top-level domain (ccTLD, such as .be for Belgium \cite{Bouhoula_Kubicek_Zac_Cotrini_Basin_2024}), the language of the site (mostly English~\cite{Gundelach_Herrmann_2023, Khandelwal_Nayak_Harkous_Fawaz_2023}), the website categorisation (e.g. shopping, business, entertainment, health~\cite{Sanchez-Rola_Dell’Amico_Kotzias_Balzarotti_Bilge_Vervier_Santos_2019, Jha_Trevisan_Vassio_Mellia_2022, Wesselkamp_Fouad_Santos_Boussad_Bielova_Legout_2021}), different popularity segments such as combining the top 500 with a samples from the larger top million~\cite{Kirkman_Vaniea_Woods_2023}, or some combination of these \cite{Matte_Bielova_Santos_2020}.
These lists all come with their own pros and cons --- they are based on global traffic, biased towards international sites and transnational platforms, suppressing nationally popular domains including government, newspaper, or e-commerce~\cite{Ruth_Fass_Azose_Pearson_Thomas_Sadowski_Durumeric_2022}.
While a country's web can be studied by filtering lists using the ccTLD, many nationally specific sites also use .com or .org, foreign sites use ccTLD's for novelty effect (e.g., youtu.be, linktr.ee, remove.bg), while some national subunits make heavy use of gTLDs (e.g. Catalonia's .cat, Scotland's .scot).
It is empirically unclear what the impact of these flaws is --- whether the .fr subset of the global list is a reasonable approximation of the top websites visited by people in France. Only a handful of studies create lists based on national traffic~\cite{Bouhoula_Kubicek_Zac_Cotrini_Basin_2024, Rasaii_Gosain_Gasser_2023}, likely because until the Chrome User Experience Report in 2021, no available list included the location of the visitor.

The number of domains included seems to matter too, as previous studies show consent interfaces and their designs are not distributed at random~\cite{Hils_Woods_Bohme_2020}. The top 50 most popular sites appear to implement their own designs, while the middle 100 to 10,000 seem more likely to use off-the-shelf CMPs.
%The number of domains that are analysed is another factor where studies differ from each other in a way that influences their results.
%has shown that consent interfaces and their specific designs are not distributed evenly across popularity ranking: the top 50 most popular sites implement their own designs and are often more compliant, whereas the middle section of the top 100 until 10,000 are more likely to use third-party Consent Management Platforms.
%The size of the list and the ranking of the sites in that list, then, impacts the proportion of consent interfaces that are found or the frequency of certain interface elements (such as buttons or checkboxes).
This makes the studies hard to compare, as they vary from small lists of just 100~\cite{Jha_Trevisan_Vassio_Mellia_2022} or 500~\cite{Singh_Upadhyaya_Seth_Hu_Sastry_Mondal_2022, Krisam_Dietmann_Volkamer_Kulyk_2021} domains; medium lists between 1,000~\cite{Matte_Bielova_Santos_2020} and 50,000~\cite{Kampanos_Shahandashti_2021}, and some in the millions~\cite{Bollinger_Kubicek_Cotrini_Basin_2022, Morel_Santos_Fredholm_Thunberg_2023, Hils_Woods_Bohme_2020}.
Naturally, if consent interfaces are not equally as prevalent across all popularity segments, studying different segments will result in different distributions.

The vantage point of the study --- the IP address from which a website is accessed --- also matters, although perhaps less so. Many papers omit it, although some consider differences across countries~\cite{Rasaii_Gosain_Gasser_2023, rasaii2023exploring}. The most comprehensive investigation of the impact of vantage points looked across many EU countries as well as some outside it~\cite{Eijk_Asghari_Winter_Narayanan_2019}. The broad conclusion from these papers is that little difference is found within the EU, some difference is found when crawling from outside the EU, but websites seem to adjust their compliance tactics more with their national base (as proxied by ccTLD) rather than visitor vantage point.

In summary, our knowledge of consent interfaces is generally based on analyses of the most popular websites according to global traffic, visited from a vantage point within the EU. As lists in particular vary so much across studies, and as consent interface design and presence are both affected strongly by the domains chosen~\cite{Hils_Woods_Bohme_2020,Eijk_Asghari_Winter_Narayanan_2019}, studies are extremely hard to rigorously compare. 

%There is a lot of variation in the subsets of those lists that are analysed, based on, for example, popularity (random samples, or just a small amount within the top 500) or the top-level domain (mostly ccTLDs) of the site. 
%This variation makes it hard to compare their results, because studies such as \citet{Hils_Woods_Bohme_2020} and \citet{Eijk_Asghari_Winter_Narayanan_2019} show that a domain's popularity and ccTLDs are two variables that greatly impact the presence and design characteristics of consent interfaces. 
% In this study, we look at national traffic to that country's ccTLDs across granular 

\subsection{What design elements are analysed?}
All studies in our review analyse whether there is a consent interface on the website or app, but differ in the other design elements that they consider, often motivated by legal requirements and how those translate into a user interface.
For example, studies have looked at whether the interface is a paywall~\cite{Rasaii_Gosain_Gasser_2023} and if so, the prices asked~\cite{Morel_Santos_Lintao_Human_2022}, the legal bases they attempt to establish (e.g., consent or legitimate interest)~\cite{Kyi_2023}, the position of the interface on the screen~\cite{Mehrnezhad_2020}, whether it uses a banner or barrier style design~\cite{Nouwens_Liccardi_Veale_Karger_Kagal_2020}, the presence of different kinds of buttons (e.g., accept, reject, settings)~\cite{Krisam_Dietmann_Volkamer_Kulyk_2021}, the visual prominence of those buttons~\cite{Kirkman_Vaniea_Woods_2023, Mehrnezhad_2020}, the text labels of those buttons~\cite{Sheil_Malone_2022}, the purposes that are mentioned for which data is collected~\cite{Bouhoula_Kubicek_Zac_Cotrini_Basin_2024}, the legal quality of those purpose descriptions~\cite{Santos_Rossi_Sanchez_2021}, whether any purpose checkboxes are pre-checked~\cite{Sheil_Malone_2022, Matte_Bielova_Santos_2020}, whether the website uses implied consent~\cite{Matte_Bielova_Santos_2020, Nouwens_Liccardi_Veale_Karger_Kagal_2020}, or whether the interface is designed by a consent management platform and if so, which one~\cite{rasaii2023exploring}.

In response to regulatory efforts to bring online tracking into compliance, different parts of these interfaces become more or less salient. For example, as guidance was released and legal cases were fought about the requirements for valid consent (e.g., clarifying the illegality of implied consent and pre-checked purposes), legitimate interest appeared as a new legal basis and the pay-or-OK approach became more prevalent (although primarily only in Germany~\cite{Morel_Santos_Fredholm_Thunberg_2023}). 
Academic studies follow these developments and have helped to either provide evidence of new practices for regulators to react to or evaluated the compliance levels after regulatory activity.

\begin{table*}[h]
\caption{The prevalence of consent interfaces and their design elements as found by previous research, in chronological order. Button prevalence is only reported if they were present on the initial screen of the interface. Note that there are many methodological differences that impact these numbers and that some of these percentages are our reconstructions based on limited information from the paper and not directly reported. These percentages should be understood as approximations rather than precise measurements. }
% \small
\resizebox{\textwidth}{!}{%
\setlength{\extrarowheight}{0.5pt}
\begin{tabular}{@{}llllllllll@{}}
\toprule
\makecell[l]{\\\textbf{Source}} & \makecell[l]{\\\textbf{Year}} & \makecell[l]{\textbf{Analysis}\\\textbf{method}} & \makecell[l]{\textbf{Sample}\\\textbf{size}} & \makecell[l]{\textbf{Interface}\\\textbf{prevalence}} & \makecell[l]{\textbf{CMP}} & \makecell[l]{\textbf{Accept}\\\textbf{option}} & \makecell[l]{\textbf{Reject}\\\textbf{option}} & \makecell[l]{\textbf{Visually} \\\textbf{unequal}} & \makecell[l]{\textbf{Pre-checked}\\ \textbf{purposes}} 
\\ \midrule
\ \ \citet{Leenes_Kosta_2015} & \citeyear{Leenes_Kosta_2015} & Manual & 100 & 50\% & & & & & \\ \cdashline{1-10}[0.5pt/2pt]
%NL
\ \ \citet{Degeling_Utz_Lentzsch_Hosseini_Schaub_Holz_2019} & \citeyear{Degeling_Utz_Lentzsch_Hosseini_Schaub_Holz_2019} & Manual & 6.357 & 62\% & 15\%  & & & & \\ \cdashline{1-10}[0.5pt/2pt]
%EU
\ \ \citet{Eijk_Asghari_Winter_Narayanan_2019} & \citeyear{Eijk_Asghari_Winter_Narayanan_2019} & Automated&1.500 & 40\% & & & & & \\ \cdashline{1-10}[0.5pt/2pt]
%EU, US, CA
\ \ \citet{Sanchez-Rola_Dell’Amico_Kotzias_Balzarotti_Bilge_Vervier_Santos_2019} & \citeyear{Sanchez-Rola_Dell’Amico_Kotzias_Balzarotti_Bilge_Vervier_Santos_2019} & Manual&2.000 & & 3\% & & & & \\ \cdashline{1-10}[0.5pt/2pt]
%Global
\ \ \citet{Utz_Degeling_Fahl_Schaub_Holz_2019} & \citeyear{Utz_Degeling_Fahl_Schaub_Holz_2019} & Manual&1.000 & & & 71\% & 3\% & & \\ \cdashline{1-10}[0.5pt/2pt]
%?
\ \ \citet{Hils_Woods_Bohme_2020} & \citeyear{Hils_Woods_Bohme_2020} & Automated&4.222.704 & & 9\% & & & & \\ \cdashline{1-10}[0.5pt/2pt]
%?
\ \ \citet{Matte_Bielova_Santos_2020} & \citeyear{Matte_Bielova_Santos_2020} & Automated/Manual&22.949/560 & & 6\% & & & & 47\% \\ \cdashline{1-10}[0.5pt/2pt]
%EU + global
\ \ \citet{Nouwens_Liccardi_Veale_Karger_Kagal_2020}  & \citeyear{Nouwens_Liccardi_Veale_Karger_Kagal_2020} & Automated&10.000  & & 7\% & & 13\% & & 56\% \\ \cdashline{1-10}[0.5pt/2pt]
%UK
\ \ \citet{Mehrnezhad_2020} & \citeyear{Mehrnezhad_2020} & Manual&116 & 91\% & & 80\% & 6\%  & 39\% & \\ \cdashline{1-10}[0.5pt/2pt]
%?
\ \ \citet{Bornschein_Schmidt_Maier_2020} & \citeyear{Bornschein_Schmidt_Maier_2020} & Manual&360 & 64\% & & 45\% & 4\%  & & \\ \cdashline{1-10}[0.5pt/2pt]
%?
\ \ \citet{Soe_Nordberg_Guribye_Slavkovik_2020} & \citeyear{Soe_Nordberg_Guribye_Slavkovik_2020} & Manual&300 & & & & 5\%  & & \\ \cdashline{1-10}[0.5pt/2pt]
%NO
\ \ \citet{Kampanos_Shahandashti_2021} & \citeyear{Kampanos_Shahandashti_2021} & Automated&17.737 & 45\% & & 89\% & 9\%  & & \\ \cdashline{1-10}[0.5pt/2pt]
%UK + GR
\ \ \citet{Krisam_Dietmann_Volkamer_Kulyk_2021} & \citeyear{Krisam_Dietmann_Volkamer_Kulyk_2021} & Manual&389 & & & 89\% & 7\%  & & \\ \cdashline{1-10}[0.5pt/2pt]
%DE
\ \ \citet{Wesselkamp_Fouad_Santos_Boussad_Bielova_Legout_2021} & \citeyear{Wesselkamp_Fouad_Santos_Boussad_Bielova_Legout_2021} & Manual&385 & 60\% & & & & & \\ \cdashline{1-10}[0.5pt/2pt]
% Germany, Austria, France, Belgium, and Ireland.
\ \ \citet{Habib_Li_Young_Cranor_2022} & \citeyear{Habib_Li_Young_Cranor_2022} & Automated/Manual&10.000/191 & & & & & 78\% & 26\% \\ \cdashline{1-10}[0.5pt/2pt]
%global
\ \ \citet{Jha_Trevisan_Vassio_Mellia_2022} & \citeyear{Jha_Trevisan_Vassio_Mellia_2022} & Automated&12.277 & 63\% & & & & & \\ \cdashline{1-10}[0.5pt/2pt]
%France, Germany, Italy, Spain and US
\ \ \citet{Sheil_Malone_2022} & \citeyear{Sheil_Malone_2022} & Automated/Manual&3.735 & 55\% & & 76\% & 16\% & & 8\% \\ \cdashline{1-10}[0.5pt/2pt]
%?
\ \ \citet{rasaii2023exploring} & \citeyear{rasaii2023exploring} & Automated&10.000 & 47\% & 13\%  & 87\% & & 26\% & \\ \cdashline{1-10}[0.5pt/2pt]
%Global
\ \ \citet{Kirkman_Vaniea_Woods_2023} & \citeyear{Kirkman_Vaniea_Woods_2023} & Automated&10.992 & 24\% & & & 34\% & & \\ \cdashline{1-10}[0.5pt/2pt]
\ \ \citet{Alharbi_Albesher_Wahsheh_2023} & \citeyear{Alharbi_Albesher_Wahsheh_2023} & Manual&243 & 41\% & & 95\% & 61\% & 20\% & \\ \cdashline{1-10}[0.5pt/2pt]
\ \ \citet{Khandelwal_Nayak_Harkous_Fawaz_2023} & \citeyear{Khandelwal_Nayak_Harkous_Fawaz_2023} & Automated&85.470 & 53\% & & & 22\% & & \\ \cdashline{1-10}[0.5pt/2pt]
%?
\ \ \citet{Gundelach_Herrmann_2023}  & \citeyear{Gundelach_Herrmann_2023} & Manual&1.000  & 47\% & & & 52\% & 46\% & \\ \cdashline{1-10}[0.5pt/2pt]
%?
\ \ \citet{Bouhoula_Kubicek_Zac_Cotrini_Basin_2024} & \citeyear{Bouhoula_Kubicek_Zac_Cotrini_Basin_2024} & Automated&85.443 & 45\% & & & 43\% & & \\ \cdashline{1-10}[0.5pt/2pt]
%?
\ \ \citet{Zhang_Meng_Zhou_Ren_2024} & \citeyear{Zhang_Meng_Zhou_Ren_2024} & Automated/Manual&82.624/239 & & 8\% & & & & 71\% \\ \midrule
%global
Average  & & & & 53\% & 9\% & 79\% & 21\% & 42\% & 42\% \\ 
\bottomrule
\end{tabular}%
}
\Description[Overview of previous literature]{A table summarising previous work on consent interfaces between 2015 and 2024.}
\label{litresultsoverview}
\end{table*}

\subsection{What detection methods are used?}\label{methods}
A variety of methods have been explored to improve the accuracy and coverage of the analysis of consent interfaces.
Many studies rely on a manual analysis of the interfaces (see Table \ref{litresultsoverview}).
% \changed{~\cite{Habib_Li_Young_Cranor_2022, Sanchez-Rola_Dell’Amico_Kotzias_Balzarotti_Bilge_Vervier_Santos_2019, Alharbi_Albesher_Wahsheh_2023, Degeling_Utz_Lentzsch_Hosseini_Schaub_Holz_2019, Mehrnezhad_2020, O’Connor_Nurwono_Siebel_Birrell_2021, Soe_Nordberg_Guribye_Slavkovik_2020, Wesselkamp_Fouad_Santos_Boussad_Bielova_Legout_2021, Data_Protection_Commission_2020}}. 
This makes them both accurate and flexible with regards to new and interesting interface elements, but labour intensive and thus limited in the number of sites evaluated.
Automated detection methods are easier to deploy at scale, but have struggled to deal with the diverse and often devious ways consent interfaces are programmed.
We summarise the different detection methods for various interface elements below and the accuracy of these methods (as reported by the authors) where possible.

To automatically detect a \textit{consent interface}, a few methods can be used. Various studies \cite{Eijk_Asghari_Winter_Narayanan_2019,Kampanos_Shahandashti_2021,Bouhoula_Kubicek_Zac_Cotrini_Basin_2024,Kirkman_Vaniea_Woods_2023} use CSS selectors from browser extensions such as ``I don't care about cookies''~\cite{idontcareaboutcookies} and ``EasyList Cookie List''~\cite{easylist}, sometimes supplemented by the authors~\cite{Kampanos_Shahandashti_2021,Bouhoula_Kubicek_Zac_Cotrini_Basin_2024}. Some studies use the visibility of the element, including the \code{z-index} of an element (the higher the index, the higher in the `stack' it is, like a pop-up)~\cite{Bouhoula_Kubicek_Zac_Cotrini_Basin_2024,Khandelwal_Nayak_Harkous_Fawaz_2023,rasaii2023exploring}, its coordinates~\cite{Gundelach_Herrmann_2023} or its `screenshot-ability'~\cite{Kirkman_Vaniea_Woods_2023}. Many studies analyse the text on the page, ranging from string matches for `cookie(s)'~\cite{Kampanos_Shahandashti_2021,Gundelach_Herrmann_2023} or a broader corpus~\cite{Kirkman_Vaniea_Woods_2023,rasaii2023exploring}, to fine-tuned language models~\cite{Khandelwal_Nayak_Harkous_Fawaz_2023,Bouhoula_Kubicek_Zac_Cotrini_Basin_2024}. \citet{Gundelach_Herrmann_2023} compare four different methods: CSS selector lists, searching for the word `cookie' combined with heuristics, a language model combined with heuristics, and searching for the word `cookie' combined with computer vision. 
They conclude that word-matching, checking whether the element's coordinates are inside the viewport, and then picking the element with the most amount of text has the best results, and report an accuracy of 96.5\% (F1=0.96) (based on a manual inspection of 1,000 sites). %F1 = 0,96, accuracy = 96,5%
\citet{rasaii2023exploring} select possible candidates by analysing the \code{z-index} and \code{position} of an element and then searching inside the element for a match against a multi-language word corpus. 
They report an accuracy of 99\% (F1=0.99) --- the highest in the literature --- based on a manual inspection of 1,000 sites. %99% accuracy, F1 = 0.9914

To detect \textit{consent management platforms} (CMPs), some researchers similarly use CSS selectors~\cite{Nouwens_Liccardi_Veale_Karger_Kagal_2020}, while others monitor network traffic during page load to check for domains that are known to be hosted by specific CMP providers~\cite{Habib_Li_Young_Cranor_2022, Hils_Woods_Bohme_2020, Kirkman_Vaniea_Woods_2023}.
Some studies take advantage of the fact that many CMPs are part of the Transparency \& Consent Framework (TCF) --- a data-sharing network created by the Interactive Advertising Bureau \cite{IABTCF} --- which requires all members to expose specific APIs that return a CMP identifier~\cite{Zhang_Meng_Zhou_Ren_2024, rasaii2023exploring, Matte_Bielova_Santos_2020}.

To detect \textit{user options}, such as links or buttons, most studies search the page for words matching a multi-language corpus of common terms (often machine translated from English rather than specifically created for a particular language). 
\citet{rasaii2023exploring} use this approach with a corpus of 172 words and accurately find accept buttons 97\% of the time, and reject buttons 87.4\% of the time. % accept accuracy 97% reject accuracy 87.4%
Some studies include heuristics to first detect interactive elements~\cite{Klein_Musch_Barber_Kopmann_Johns_2022,Bouhoula_Kubicek_Zac_Cotrini_Basin_2024,Khandelwal_Nayak_Harkous_Fawaz_2023,Gundelach_Herrmann_2023}.
\citet{Bouhoula_Kubicek_Zac_Cotrini_Basin_2024} query for all elements with the property \code{role="button"} (checking that it has a non-negative \code{tabindex} value, and a non-null \code{onclick} attribute).
They classify those elements into accept, reject, close, save or settings options using a language model, and report an overall (rather than per-item) accuracy of 95.1\% for all options (F1=0.91). %Accuracy = 95.1%
%\citet{Klein_Musch_Barber_Kopmann_Johns_2022} first checked whether the element had a \code{.click()} method and if the text was less than 200 characters long, matched this against their corpus of 43 words, and then applied some scoring mechanism in case too many items were found. They reported an accuracy score of 79.5\% for the accept button.
\citet{Khandelwal_Nayak_Harkous_Fawaz_2023} find all user-focusable elements by simulating tab key-presses, filter out all elements that redirect the user to a different page, and then classify these elements into accept, reject, save, or settings options through a combination of clicking on them and observing the effect it has on the page, and a machine learning model.
Their overall accuracy score (including more than just user option detection) is 93.7\%.
%\citet{Gundelach_Herrmann_2023} first detected interactive elements by hovering over an item and checking whether it changed the mouse cursor, and then compard screenshots before and after clicking on this item. They reported an accuracy of 38\%.

Only two studies try to automatically analyse the \textit{visual prominence} of user options. 
\citet{Gundelach_Herrmann_2023} extract the colour of each pixel of a detected element and compute the dominant colour (76\% accurate), while \citet{Kirkman_Vaniea_Woods_2023} convert a screenshot to black and white and look at the contrast between the detected element and the background (no accuracy reported).

The \textit{status of checkboxes} or toggles to control processing purposes is to our knowledge only analysed through automated means by \citet{Nouwens_Liccardi_Veale_Karger_Kagal_2020}, using known CSS selectors of specific CMPs.

In summary, many different methods have been tried, only some of which report their accuracy in a granular way.
It seems that generally speaking, a combination of a word corpus with heuristics outperforms both simple hard-coded CSS selectors and more complex machine learning approaches.

\subsection{What are the results?}
The results of consent interface studies vary wildly for all the reasons above, on top of the changes we might expect over time (see Table \ref{litresultsoverview}).
If we take the average of these results --- with all the serious caveats of doing so based on the methodological differences discussed above --- we can estimate that 53\% of sites have a consent interface, and a marginal 9\% of sites use CMPs. 
Of these interfaces, 79\% have an accept options, while only 21\% have a reject option. 
If these interfaces do have both options, 42\% of the time they are not visually equal but instead have one option that is more prominent than the other.
Granular controls, such as processing purposes or individual vendors, are pre-checked in 42\% of the interfaces.

\section{Method}
\subsection{Data collection}
\subsubsection{Country selection}
The scope of the study is each country that transposed the ePD and adopted the GDPR, the two laws from the European Union that triggered the online tracking industry to develop consent interfaces. 
This includes all countries in the European Economic Area (EEA) plus the United Kingdom.\footnote{The included countries are: Austria, Belgium, Bulgaria, Croatia, Cyprus, Czech Republic, Denmark, Estonia, Finland, France, Germany, Greece, Hungary, Iceland, Ireland, Italy, Latvia, Liechtenstein, Lithuania, Luxembourg, Malta, the Netherlands, Norway, Poland, Portugal, Romania, Slovakia, Slovenia, Spain, Sweden, and the United Kingdom (which has not substantively altered the laws since leaving the EU).}
For each of these countries, we crawled the top 10,000 locally most popular websites. 

\subsubsection{Website selection}
% An average user's web usage includes a transnationally similar set of activities (such as major international search engines, social networks, and adult content) and a more nationally specific subset of sites, such as those which are regionally, culturally or linguistically specific~\cite{Ruth_Fass_Azose_Pearson_Thomas_Sadowski_Durumeric_2022}. 
We aim to study differences in consent interfaces between countries, as this is a crucial level for any intervention that wants to change the status quo: regulatory guidance, enforcement responsibilities and priorities are all predominantly national in scope. 
% National practices already diverge significantly, such as the widely reported emergence of controversial `consent-or-pay' banners in certain European jurisdictions over others~\cite{Rasaii_Gosain_Gasser_2023, Morel_Santos_Fredholm_Thunberg_2023}.
% Given such variance appears to exist, it is furthermore methodologically important not to ignore it, and instead to explore and attempt to explain it.
Previous analyses of consent interfaces which incorporate a national element typically use the country code top-level domain (ccTLD) as a proxy for the average browsing experience of a person within that country (e.g., .nl for the Netherlands, or .pl for Poland)~\cite{Eijk_Asghari_Winter_Narayanan_2019, Sheil_Malone_2022, Matte_Bielova_Santos_2020, Kampanos_Shahandashti_2021}, even though this misses out on popular non-national sites (such as .com or .org domains). 
In part, this is because there is no free service which generates ranked lists of the most popular websites per location, which would enable an estimation of a user's real experience across both the national and transnational elements of their browsing.\footnote{Only the Chrome User Experience list (CrUX) has lists based on location-level traffic, but their popularity ranking uses very coarse bins (1k, 5k, 10k) and previous work has indicated that popularity ranking is an important factor that affects consent interface design~\cite{Hils_Woods_Bohme_2020}, so this level of granularity might obscure important dynamics.} 
% TODO: although Kirkman and Rasaii already show some evidence this is no longer true
Other studies use a list of globally popular websites and visit them from the vantage point of a specific location~\cite{rasaii2023exploring, Hils_Woods_Bohme_2020, Zhang_Meng_Zhou_Ren_2024}, but this approach misses out on important local sites and erases the experiences of users in small jurisdictions~\cite{Ruth_Fass_Azose_Pearson_Thomas_Sadowski_Durumeric_2022}, and in any case it appears websites rarely adjust their consent interfaces based on user location~\cite{Eijk_Asghari_Winter_Narayanan_2019}.

We also adopt a ccTLD approach in this study, but 
% unlike most prior studies 
want to clarify the consequences of this clearly for our analysis: the results do not represent the average consent experience of a user in the country, but instead captures how national web cultures and enforcement actions correlate with the design of consent interfaces. 
% We cannot say anything about how large this component is, and it is likely to differ per person --- internationalised or disapora communities may use few local sites. However, crucially, it is the most important component to inform national decision-making, as the large multinationals, at least under the GDPR, can only be enforced against in the jurisdictions in which they have their `main establishment', typically Ireland or Luxembourg~\cite{gentileDeficientDesignTransnational2022}. 
By omitting or limiting the transnational element, we heighten the statistical focus on national differences where there is competence to make change. 

\subsubsection{List generation}\label{listgeneration}
The top lists for each country were generated using Tranco~\cite{LePochat2019} by aggregating all domains mentioned in the CrUX, Majestic, Cloudflare Radar, and Cisco Umbrella lists, averaging the domain's popularity in each list from 13-07-2024 to 11-08-2024, and using harmonic progression to generate a single popularity score per domain.
We used the CrUX list, the only list that has location-specific traffic information, to filter out any national domains that were not actually visited by people in that country (e.g., some globally popular sites use a ccTLD for novelty effect, such as youtu.be or twitch.tv). 
The final domain lists per country are linked to in Appendix \ref{toplists}.

\subsubsection{Scraping process}
We collected the data between 12-08-2024 and 04-09-2024. 
The websites were scraped from a vantage point in Denmark.
We limited our concurrency to four sites, to reduce the chances of being detected as a bot.
Each site was scraped in its own virtual container.
We successfully collected data from 94\% of the target domains; 4.66\% of sites were unreachable, and 1.34\% of domains blocked the scraper (see Table \ref{scrapecounts}).

\begin{table*}[]
    \centering
    \begin{minipage}[b]{0.4265\textwidth}
    \captionsetup{width=0.95\textwidth}
    \caption{Scraper success rates for collecting data about a website. `Target' is the number of attempted URLs. `Successful' means data was retrieved. `Unreachable' means a failed server connection. `Blocked' means access was denied.}
        \resizebox{\textwidth}{!}{%
    \setlength{\extrarowheight}{1pt}
    \begin{tabular}{@{}lllllllllllllllll@{}}
        \cmidrule(lr){1-8} \cmidrule(lr){9-16}
        \multicolumn{8}{c}{\textbf{Scraper success rates}}\\
        \cmidrule(lr){1-8} \cmidrule(lr){9-16} 
        & & \multicolumn{2}{l}{\emph{Successful}} &  \multicolumn{2}{l}{\emph{Unreachable}}  & \multicolumn{2}{l}{\emph{Blocked}} \\
        \cmidrule(lr){3-4} \cmidrule(lr){5-6} \cmidrule(lr){7-8}
        & \emph{Target} & \textit{n} & \textit{\%} & \textit{n} & \textit{\%} & \textit{n} & \textit{\%} \ \\
        \cmidrule(lr){1-8}
        \ \ \textbf{Austria} & 10000 & 9467 & 95 & 430 & 4 & 103 & 1 \\ \cdashline{1-8}[0.5pt/2pt]
        \ \ \textbf{Belgium} & 10000 & 9502 & 95 & 360 & 4 & 138 & 1 \\ \cdashline{1-8}[0.5pt/2pt]
        \ \ \textbf{Bulgaria} & 10000 & 9526 & 95 & 474 & 5 & 0 & 0 \\ \cdashline{1-8}[0.5pt/2pt]
        \ \ \textbf{Croatia} & 10000 & 9471 & 95 & 529 & 5 & 0 & 0 \\ \cdashline{1-8}[0.5pt/2pt]
        \ \ \textbf{Cyprus} & 1889 & 1724 & 91 & 130 & 7 & 35 & 2 \\ \cdashline{1-8}[0.5pt/2pt]
        \ \ \textbf{Czech Republic} & 10000 & 9583 & 96 & 241 & 2 & 176 & 2 \\ \cdashline{1-8}[0.5pt/2pt]
        \ \ \textbf{Denmark} & 10000 & 9533 & 95 & 372 & 4 & 95 & 1 \\ \cdashline{1-8}[0.5pt/2pt]
        \ \ \textbf{Estonia} & 10000 & 9665 & 97 & 192 & 2 & 143 & 1 \\ \cdashline{1-8}[0.5pt/2pt]
        \ \ \textbf{Finland} & 10000 & 9510 & 95 & 358 & 4 & 132 & 1 \\ \cdashline{1-8}[0.5pt/2pt]
        \ \ \textbf{France} & 10000 & 9057 & 91 & 671 & 7 & 272 & 3 \\ \cdashline{1-8}[0.5pt/2pt]
        \ \ \textbf{Germany} & 10000 & 9386 & 94 & 434 & 4 & 180 & 2 \\ \cdashline{1-8}[0.5pt/2pt]
        \ \ \textbf{Greece} & 10000 & 9413 & 94 & 349 & 3 & 238 & 2 \\ \cdashline{1-8}[0.5pt/2pt]
        \ \ \textbf{Hungary} & 10000 & 9484 & 95 & 516 & 5 & 0 & 0 \\ \cdashline{1-8}[0.5pt/2pt]
        \ \ \textbf{Iceland} & 4692 & 4535 & 97 & 95 & 2 & 62 & 1 \\ \cdashline{1-8}[0.5pt/2pt]
        \ \ \textbf{Ireland} & 10000 & 9463 & 95 & 377 & 4 & 160 & 2 \\ \cdashline{1-8}[0.5pt/2pt]
        \ \ \textbf{Italy} & 10000 & 8666 & 87 & 1082 & 11 & 252 & 3 \\ \cdashline{1-8}[0.5pt/2pt]
        \ \ \textbf{Latvia} & 10000 & 9487 & 95 & 513 & 5 & 0 & 0 \\ \cdashline{1-8}[0.5pt/2pt]
        \ \ \textbf{Liechtenstein} & 124 & 122 & 98 & 2 & 2 & 0 & 0 \\ \cdashline{1-8}[0.5pt/2pt]
        \ \ \textbf{Lithuania} & 10000 & 9425 & 94 & 245 & 2 & 330 & 3 \\ \cdashline{1-8}[0.5pt/2pt]
        \ \ \textbf{Luxembourg} & 2374 & 2243 & 94 & 87 & 4 & 44 & 2 \\ \cdashline{1-8}[0.5pt/2pt]
        \ \ \textbf{Malta} & 1290 & 1222 & 95 & 37 & 3 & 31 & 2 \\ \cdashline{1-8}[0.5pt/2pt]
        \ \ \textbf{Netherlands} & 10000 & 9529 & 95 & 303 & 3 & 168 & 2 \\ \cdashline{1-8}[0.5pt/2pt]
        \ \ \textbf{Norway} & 10000 & 9653 & 97 & 266 & 3 & 81 & 1 \\ \cdashline{1-8}[0.5pt/2pt]
        \ \ \textbf{Poland} & 10000 & 9587 & 96 & 256 & 3 & 157 & 2 \\ \cdashline{1-8}[0.5pt/2pt]
        \ \ \textbf{Portugal} & 10000 & 9029 & 90 & 770 & 8 & 201 & 2 \\ \cdashline{1-8}[0.5pt/2pt]
        \ \ \textbf{Romania} & 10000 & 9392 & 94 & 608 & 6 & 0 & 0 \\ \cdashline{1-8}[0.5pt/2pt]
        \ \ \textbf{Slovakia} & 10000 & 9559 & 96 & 441 & 4 & 0 & 0 \\ \cdashline{1-8}[0.5pt/2pt]
        \ \ \textbf{Slovenia} & 10000 & 9299 & 93 & 701 & 7 & 0 & 0 \\ \cdashline{1-8}[0.5pt/2pt]
        \ \ \textbf{Spain} & 10000 & 8901 & 89 & 876 & 9 & 223 & 2 \\ \cdashline{1-8}[0.5pt/2pt]
        \ \ \textbf{Sweden} & 10000 & 9552 & 96 & 360 & 4 & 88 & 1 \\ \cdashline{1-8}[0.5pt/2pt]
        \ \ \textbf{United Kingdom} & 10000 & 9163 & 92 & 533 & 5 & 304 & 3 \\  \cmidrule(lr){1-8}
        \ \ \textbf{Total} & 270369 & 254148 & 94 & 12608 & 5 & 3613 & 1 \\ \cmidrule(lr){1-8}
        \end{tabular}
        }\label{scrapecounts}
        \Description[Data about scrape returns]{Describes for each country whether the scraper was able to collect data or whether it ran into problems. 94 percent was successful, 5 percent unreachable, and 1 percent blocked.}
    \end{minipage}
    \hspace{-12pt}
    % \begingroup
    \setlength{\tabcolsep}{2pt}
    % \begin{minipage}[b]{0.5845\textwidth}
    \begin{minipage}[b]{0.5849\textwidth}
    \captionsetup{width=0.95\textwidth}
    \caption{Accuracy scores of each automated detection method, based on 100 sites per country. `FP' and `FN' refer to False Negative and False Positive results; values are rounded percentages. Missing values means that the randomised evaluation sample lacked data points for this element, which means no accuracy score could be calculated.}
            \resizebox{\textwidth}{!}{%
            \setlength{\extrarowheight}{1pt}
            \begin{tabular}{@{}llllllllllllllllllllllll@{}}
                \cmidrule(lr){1-24} 
        \multicolumn{24}{c}{\textbf{Detection method accuracy score}}\\
        \cmidrule(lr){1-24}
        \multicolumn{3}{l}{\emph{Interface}} &  \multicolumn{3}{l}{\emph{Accept}}  & \multicolumn{3}{l}{\emph{Reject}} & \multicolumn{3}{l}{\emph{Settings}} &  \multicolumn{3}{l}{\emph{Save}}  & \multicolumn{3}{l}{\emph{Pay}} & \multicolumn{3}{l}{\emph{Checkbox}} &  \multicolumn{3}{l}{\emph{Pre-checked}} \\
        \cmidrule(lr){1-3} \cmidrule(lr){4-6} \cmidrule(lr){7-9} \cmidrule(lr){10-12} \cmidrule(lr){13-15} \cmidrule(lr){16-18} \cmidrule(lr){19-21} \cmidrule(lr){22-24} 
        \ \ \emph{F1} & \textit{FP} & \textit{FN} &  \emph{F1} & \textit{FP} & \textit{FN} &\emph{F1} & \textit{FP} & \textit{FN} &\emph{F1} & \textit{FP} & \textit{FN} &\emph{F1} & \textit{FP} & \textit{FN} &\emph{F1} & \textit{FP} & \textit{FN} &\emph{F1} & \textit{FP} & \textit{FN} &\emph{F1} & \textit{FP} & \textit{FN}  \\
        \cmidrule(lr){1-24}
\ \ 0.99 & 0 & 1 & 0.99 & 0 & 1 & 0.99 & 1 & 0 & 0.99 & 1 & 0 & 0.98 & 0 & 1 & - & - & - & 0.71 & 1 & 9 & 0.00 & 0 & 2 \\ \cdashline{1-24}[0.5pt/2pt]
\ \ 1.00 & 0 & 0 & 0.97 & 0 & 4 & 0.97 & 0 & 2 & 0.95 & 0 & 4 & 1.00 & 0 & 0 & - & - & - & 1.00 & 0 & 0 & 1.00 & 0 & 0 \\ \cdashline{1-24}[0.5pt/2pt]
\ \ 0.95 & 4 & 2 & 0.96 & 2 & 2 & 0.90 & 2 & 0 & 0.83 & 4 & 4 & 1.00 & 0 & 0 & - & - & - & 1.00 & 0 & 0 & - & - & - \\ \cdashline{1-24}[0.5pt/2pt]
\ \ 0.97 & 4 & 0 & 0.96 & 0 & 4 & 1.00 & 0 & 0 & 1.00 & 0 & 0 & 1.00 & 0 & 0 & - & - & - & 0.80 & 0 & 2 & 1.00 & 0 & 0 \\ \cdashline{1-24}[0.5pt/2pt]
\ \ 0.94 & 2 & 4 & 0.93 & 0 & 6 & 0.96 & 0 & 2 & 0.97 & 0 & 2 & - & - & - & - & - & - & 1.00 & 0 & 0 & - & - & - \\ \cdashline{1-24}[0.5pt/2pt]
\ \ 0.97 & 0 & 4 & 0.97 & 0 & 4 & 0.98 & 0 & 2 & 0.98 & 0 & 2 & 1.00 & 0 & 0 & - & - & - & 0.78 & 4 & 0 & - & - & - \\ \cdashline{1-24}[0.5pt/2pt]
\ \ 1.00 & 0 & 0 & 1.00 & 0 & 0 & 1.00 & 0 & 0 & 1.00 & 0 & 0 & 1.00 & 0 & 0 & - & - & - & 0.89 & 0 & 9 & 1.00 & 0 & 0 \\ \cdashline{1-24}[0.5pt/2pt]
\ \ 1.00 & 0 & 0 & 1.00 & 0 & 0 & 1.00 & 0 & 0 & 1.00 & 0 & 0 & 1.00 & 0 & 0 & - & - & - & 0.00 & 0 & 2 & 0.00 & 0 & 2 \\ \cdashline{1-24}[0.5pt/2pt]
\ \ 0.98 & 0 & 2 & 0.98 & 0 & 2 & 0.96 & 0 & 2 & 1.00 & 0 & 0 & 1.00 & 0 & 0 & - & - & - & 0.55 & 2 & 3 & - & - & - \\ \cdashline{1-24}[0.5pt/2pt]
\ \ 0.99 & 0 & 2 & 0.96 & 0 & 6 & 0.96 & 0 & 4 & 0.96 & 0 & 4 & 1.00 & 0 & 0 & - & - & - & 1.00 & 0 & 0 & - & - & - \\ \cdashline{1-24}[0.5pt/2pt]
\ \ 0.99 & 0 & 2 & 0.96 & 0 & 6 & 0.98 & 0 & 2 & 0.98 & 0 & 2 & 0.94 & 0 & 2 & 0.67 & 0 & 2 & 0.91 & 2 & 2 & 0.50 & 2 & 2 \\ \cdashline{1-24}[0.5pt/2pt]
\ \ 1.00 & 0 & 0 & 1.00 & 0 & 0 & 1.00 & 0 & 0 & 1.00 & 0 & 0 & 1.00 & 0 & 0 & - & - & - & 0.89 & 2 & 0 & - & - & - \\ \cdashline{1-24}[0.5pt/2pt]
\ \ 0.92 & 6 & 6 & 0.96 & 0 & 6 & 0.90 & 0 & 6 & 0.96 & 0 & 4 & 0.00 & 0 & 2 & - & - & - & 0.75 & 2 & 2 & - & - & - \\ \cdashline{1-24}[0.5pt/2pt]
\ \ 1.00 & 0 & 0 & 1.00 & 0 & 0 & 1.00 & 0 & 0 & 1.00 & 0 & 0 & - & - & - & - & - & - & - & - & - & - & - & - \\ \cdashline{1-24}[0.5pt/2pt]
\ \ 0.97 & 4 & 0 & 1.00 & 0 & 0 & 1.00 & 0 & 0 & 1.00 & 0 & 0 & 1.00 & 0 & 0 & - & - & - & 0.90 & 2 & 0 & - & - & - \\ \cdashline{1-24}[0.5pt/2pt]
\ \ 1.00 & 0 & 0 & 1.00 & 0 & 0 & 1.00 & 0 & 0 & 1.00 & 0 & 0 & - & - & - & - & - & - & 1.00 & 0 & 0 & - & - & - \\ \cdashline{1-24}[0.5pt/2pt]
\ \ 0.98 & 2 & 0 & 1.00 & 0 & 0 & 1.00 & 0 & 0 & 0.97 & 2 & 0 & - & - & - & - & - & - & 0.67 & 2 & 0 & - & - & - \\ \cdashline{1-24}[0.5pt/2pt]
\ \ 1.00 & 0 & 0 & 1.00 & 0 & 0 & 1.00 & 0 & 0 & 1.00 & 0 & 0 & 1.00 & 0 & 0 & - & - & - & 1.00 & 0 & 0 & - & - & - \\ \cdashline{1-24}[0.5pt/2pt]
\ \ 1.00 & 0 & 0 & 1.00 & 0 & 0 & 1.00 & 0 & 0 & 1.00 & 0 & 0 & - & - & - & - & - & - & - & - & - & - & - & - \\ \cdashline{1-24}[0.5pt/2pt]
\ \ 0.99 & 0 & 2 & 0.99 & 0 & 2 & 0.98 & 0 & 2 & 0.98 & 0 & 2 & 0.88 & 2 & 0 & - & - & - & 0.88 & 2 & 0 & 0.00 & 2 & 0 \\ \cdashline{1-24}[0.5pt/2pt]
\ \ 1.00 & 0 & 0 & 1.00 & 0 & 0 & 1.00 & 0 & 0 & 1.00 & 0 & 0 & - & - & - & - & - & - & 1.00 & 0 & 0 & - & - & - \\ \cdashline{1-24}[0.5pt/2pt]
\ \ 1.00 & 0 & 0 & 1.00 & 0 & 0 & 1.00 & 0 & 0 & 1.00 & 0 & 0 & 1.00 & 0 & 0 & - & - & - & 0.70 & 4 & 2 & 0.67 & 0 & 2 \\ \cdashline{1-24}[0.5pt/2pt]
\ \ 1.00 & 0 & 0 & 0.98 & 2 & 0 & 1.00 & 0 & 0 & 1.00 & 0 & 0 & 1.00 & 0 & 0 & - & - & - & 1.00 & 0 & 0 & - & - & - \\ \cdashline{1-24}[0.5pt/2pt]
\ \ 1.00 & 0 & 0 & 1.00 & 0 & 0 & 1.00 & 0 & 0 & 0.98 & 2 & 0 & 0.86 & 2 & 0 & - & - & - & 0.94 & 0 & 2 & 1.00 & 0 & 0 \\ \cdashline{1-24}[0.5pt/2pt]
\ \ 0.98 & 2 & 0 & 1.00 & 0 & 0 & 1.00 & 0 & 0 & 1.00 & 0 & 0 & 1.00 & 0 & 0 & - & - & - & 0.50 & 2 & 2 & 0.67 & 0 & 2 \\ \cdashline{1-24}[0.5pt/2pt]
\ \ 0.95 & 2 & 5 & 0.96 & 0 & 5 & 0.95 & 0 & 3 & 0.94 & 0 & 5 & - & - & - & - & - & - & 0.67 & 3 & 2 & 0.00 & 2 & 2 \\ \cdashline{1-24}[0.5pt/2pt]
\ \ 0.94 & 8 & 0 & 0.98 & 2 & 0 & 0.98 & 2 & 0 & 0.98 & 2 & 0 & 1.00 & 0 & 0 & - & - & - & 1.00 & 0 & 0 & - & - & - \\ \cdashline{1-24}[0.5pt/2pt]
\ \ 0.94 & 4 & 5 & 0.98 & 0 & 2 & 0.96 & 2 & 0 & 1.00 & 0 & 0 & - & - & - & - & - & - & 0.50 & 2 & 2 & 0.67 & 0 & 2 \\ \cdashline{1-24}[0.5pt/2pt]
\ \ 0.98 & 2 & 2 & 0.99 & 0 & 2 & 1.00 & 0 & 0 & 1.00 & 0 & 0 & 1.00 & 0 & 0 & - & - & - & 0.88 & 2 & 0 & 0.83 & 2 & 0 \\ \cdashline{1-24}[0.5pt/2pt]
\ \ 1.00 & 0 & 0 & 1.00 & 0 & 0 & 0.94 & 4 & 0 & 0.97 & 2 & 0 & 1.00 & 0 & 0 & - & - & - & 0.92 & 0 & 4 & 1.00 & 0 & 0 \\ \cdashline{1-24}[0.5pt/2pt]
\ \ 0.98 & 2 & 2 & 0.98 & 2 & 2 & 0.97 & 2 & 0 & 1.00 & 0 & 0 & - & - & - & - & - & - & 1.00 & 0 & 0 & - & - & - \\  \cmidrule(lr){1-24}
\ \ 0.99 & 1 & 1 & 0.98 & 0 & 2 & 0.99 & 0 & 1 & 0.99 & 0 & 1 & 1.00 & 0 & 0 & - & - & - & 0.82 & 1 & 2 & 0.67 & 0 & 1 \\  \cmidrule(lr){1-24}
        \end{tabular}
        }\label{accuracyTab}
        \Description[Accuracy of detection methods]{Describes the F1, false negative, and false positive scores of different detection methods for each country.}
    \end{minipage}
\end{table*}

\subsection{Scraper setup}
% Technical scraper tool description
% Scraper settings used
% URL list used and how generated
% Process of data collection
\subsubsection{Domain preparation}
We collected the data using a scraper built with \emph{puppeteer-extra} (v. 3.1.15), an open source framework that extends Google's Puppeteer library (v. 5.5.0) and makes it possible to add plugin functionality. 
We used the \emph{puppeteer-extra-plugin-stealth} (v. 2.6.5), which uses a number of different mechanisms to avoid being detected as a bot.
We complemented this with our own scripts to detect whether a website was showing a CAPTCHA challenge, a Cloudflare `Are You Human' challenge, or an HTTP 403 Forbidden error.
% If any of these were true, we either waited until we were forwarded to the actual website or skipped the domain. %TODO: does it skip it then? or what? My notes say "wait until it isn't any more"
% kill after 2 minutes
Once the scraper navigated to the target domain, we redefined any closed ShadowDOM as open (since some websites place their interface code inside this DOM, making it unreadable to code) and then waited for the \code{DOMContentLoaded} event to be fired to make sure the page was done loading.
After this signal, the scraper waited for an additional 10 seconds and simulated some mouse movement (since we observed some websites do not load their consent interface until after a delay or until user behaviour is detected). If the domain was not successfully navigated to, loaded, or scraped after 2 minutes, it was skipped. 

\subsubsection{Data extraction}
Once the preparatory steps were completed, we extracted the actual information relevant for this study. 
This included the presence of a consent interface, the identification of the third-party CMP, which options were available on the initial screen of the interface for the user to click on (e.g., accept, reject), how these options were visually styled, and whether there were any purpose-level control options and what their status was. 
The methods to detect these components are described below, and the open-source code is available at: \href{https://github.com/cavi-au/consent-observatory.eu}{github.com/cavi-au/consent-observatory.eu}
% TODO: CHECK IF URL IS GONNA BE OK WITH JANUS?
We only considered the first page of the consent interface because 1) the law requires accepting and rejecting to be equally as easy, which regulators have made clear means they should be on the same page~\cite{vealeAdtechRealTimeBidding2022}; and because 2) previous research has shown that people rarely interact with the second page (users went to the second page only 6.9\% of the time~\cite{Nouwens_Liccardi_Veale_Karger_Kagal_2020}).

\subsection{Interface detection}
The process to detect the presence of a consent interface extends the approach developed by \citet{rasaii2023exploring}.
% It uses the positioning of the element on the page and whether it contains a word from a list of keywords as the heuristic.
First, the scraper checks whether there is any element on the page that has a \code{z-index} larger than 10 and a \code{position} equal to \code{fixed}.
It looks for these elements in the DOM, the ShadowDOM, and iframes, which returns a list of potential consent interfaces.
The scraper then checks whether the items in this list contain a word from a corpus of commonly found words in consent interfaces, either in the element's text content or attribute values.
The corpus includes 363 phrases from all the official languages of the countries in the scope of this study (see Appendix \ref{popupcorpus}).\footnote{The full list is: Bulgarian, Croatian, Czech, Danish, Dutch, English, Estonian, Finnish, French, German, Greek, Hungarian, Icelandic, Irish, Italian, Latvian, Lithuanian, Luxembourgish, Maltese, Norwegian, Polish, Portuguese, Romanian, Sami (northern), Slovak, Slovene, Spanish, Swedish, Turkish}
This corpus was developed iteratively by manually looking at consent interfaces on the Tranco top 500 most popular ccTLD sites for each country and other hand-picked sites if certain national languages were not represented in that top list (e.g, Sami, Irish).
The scraper also includes a negative corpus of word patterns (e.g., \code{\\d+ years or older}) and an analysis of the visibility of the detected elements (see Section \ref{visualprominence}), which can be used for filtering in post-processing.

\subsection{CMP detection}
We detect whether a consent interface is a consent management platform (CMP): a third-party software service that provides the interface for a website owner.
To do this, we used the TCF API approach~\cite{Matte_Bielova_Santos_2020, rasaii2023exploring, zhang2024they} and called the \code{__tcfapi} and \code{__cmp} APIs, which return an object with (among other things) the \code{cmpId}.
We cross referenced this id with IAB's public list~\cite{IABTCFCMPID} to get the CMP provider name.
We supplemented this approach with a custom list of CSS selectors to detect CMPs that do not implement the TCF standard.
We first extracted all \code{id} and \code{class} values from the HTML of the detected interface.
We then manually matched these selectors with their corresponding CMP provider, and used regular expression pattern matching to find these same selectors in the HTML of other detected interfaces (see Appendix \ref{selectors} for the full list).

\subsection{User options detection}
Developing the detection method for user options in the consent interface followed a two step process. 
% TODO: say which method this was inspired by
\subsubsection{Word corpus generation}
We developed a heuristic to detect any elements within an interface that a user can interact with.
This heuristic looks for any elements with the tag \code{button} or \code{a}, elements which have a class, input type, or role that indicate it might be a button (e.g., \code{.btn} or \code{input[role="submit"]}), or elements which have an \code{eventListener} attached to them.
We scraped the top 500 most popular ccTLD websites of each country for elements that met this heuristic, extracted their labels, and then manually categorised all those labels as an `accept', `reject', `settings', `save', or `pay' option. 
As much as possible, this classification was done by, or with the help of, people who spoke the language,\footnote{This was the case in English, Danish, Norwegian, Swedish, Dutch, French, Greek, Italian, Spanish, Portuguese, Estonian, German, Serbian, Croatian, and Bulgarian.} or otherwise by visiting a website that contained that word and using context clues and automated translation services to determine which category it belonged to.
We then normalised the list of labels using compatibility decomposition (NFKD) (which, for example, decomposes \'e into e and `) and regular expressions to remove various characters (e.g., control characters, spacing marks, punctuation, modifier symbols, separators).
This resulted in a corpus of 4081 labels across all languages (1288 for accept, 1009 for reject, 1531 for settings, 188 for save, and 65 for pay).

The classification of ambiguous words is an inherently interpretive exercise. 
Some words are used for elements of interest in one interface, but are irrelevant in another. For example, "learn more" is used by some interfaces to go to settings, but in others it takes the user to a privacy policy.
Some words are used for two different categories of user options at the same time (e.g., "accept necessary cookies").
% There is also no guarantee that the user options actually function correctly.
In addition to more obvious labels, we counted variations of "close" and "dismiss" as accept options, "accept necessary" as reject, "learn more" and "see details" as settings, and "login" as pay.
In cases of doubt, we chose to be more inclusive and risk false positives to ensure our analysis represents the best case scenario for the presence of user options.

\subsubsection{Data extraction}
To detect user options in a consent interface, we check the attribute values and text content of all child elements inside a detected consent interface (under 256 characters). 
We normalise the text and compare it to the corpus, looking for the best match within a Levenshtein distance of 1.
We also check for event listeners (on the element or an ancestor) and perform a visual analysis (see Section \ref{visualprominence}), which can be used for post-processing to filter out non-visible or non-interactable (e.g., a header with the word "Consent") elements.
When multiple user options were detected that fell into the same category (for example when there is a settings button but there is also a settings link in the text), we chose to use the most visible element for our analysis.

% TODO: the figure numbers don't make sense
%TC:ignore
\begin{figure*}[]
    \centering
    % First row of images
    \begin{minipage}[t]{0.23\linewidth}
        \centering
        \fbox{\includegraphics[width=\linewidth, trim=0cm 0cm 0cm -0.05cm, clip]{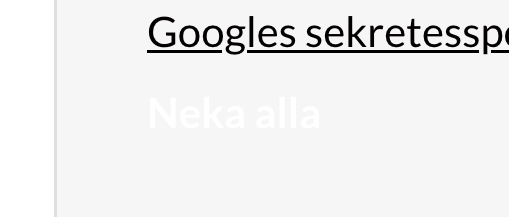}}
        \fbox{\includegraphics[width=\linewidth, trim=0cm 0cm 0cm 0.39cm, clip]{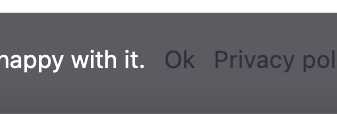}}
        \subcaption{\textbf{Nearly invisible}. The top image says "Neka alla" ("Reject all") and the bottom says "Ok". Both images have very low contrast between the text and the background, and no other features that make it stand out.}
        % \label
    \end{minipage}
    \hspace{0.01\linewidth} % Space between columns
    \begin{minipage}[t]{0.23\linewidth}
        \centering
        \fbox{\includegraphics[width=\linewidth, trim=0cm 0.5cm 0cm 0.04cm, clip]{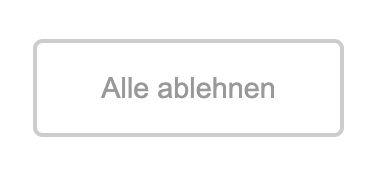}}
        \fbox{\includegraphics[width=\linewidth]{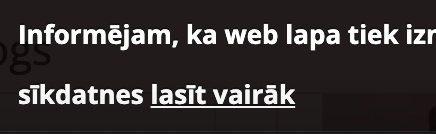}}
        \subcaption{\textbf{Subtle}. The top element has no contrast between the button background and parent background and low contrast between the background and the text and border. The bottom element is merely an underlined link.}
    \end{minipage}
    \hspace{0.01\linewidth} % Space between columns
    \begin{minipage}[t]{0.23\linewidth}
        \centering
        \fbox{\includegraphics[width=\linewidth,trim=0cm 0cm 0cm 0.2cm, clip]{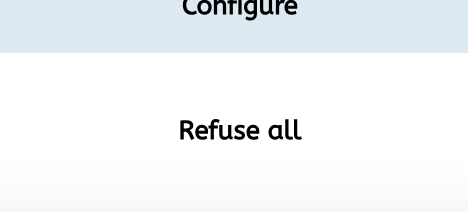}}
        \fbox{\includegraphics[width=\linewidth,trim=0cm 0cm 0cm 0.55cm, clip]{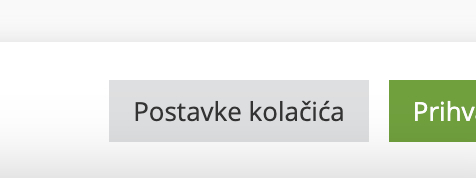}}
        \subcaption{\textbf{Visible}. The top element has high contrast between the text and the background, but no other colour or border. The bottom element is a button with a background, but with medium contrast.}
    \end{minipage}
    \hspace{0.01\linewidth} % Space between columns
    \begin{minipage}[t]{0.23\linewidth}
        \centering
        \fbox{\includegraphics[width=\linewidth, trim=0cm 0.4cm 0cm 0.25cm, clip]{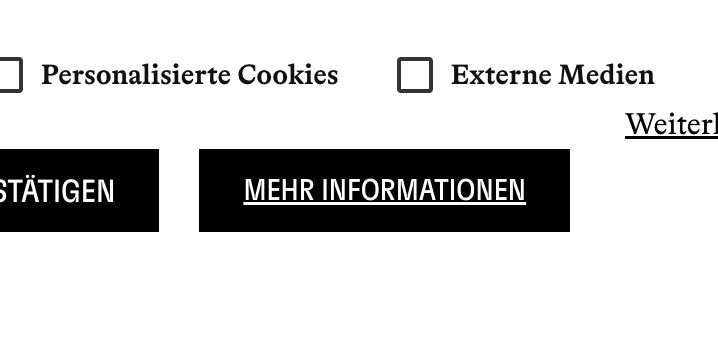}}
        \fbox{\includegraphics[width=\linewidth, trim=0cm 1.05cm 0cm 0.5cm, clip]{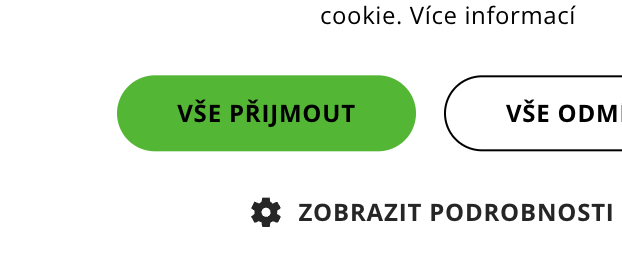}}
        \subcaption{\textbf{Prominent}. The top element has a high contrast between the parent background, button background, and the text, and it is underlined. The bottom element has lower contrast, but a vibrant colour.}
    \end{minipage}
    
    % Final caption
    \caption{Examples of elements with different visibility scores.}
    \label{fig:visibility}
    \Description[Screenshots of user options with different visibility]{Shows user options that are nearly invisible, subtle, visible, or prominent.}
\end{figure*}
%TC:endignore

% [TODO: should visual equivalence be normalised to between 0 and 1?]
% [add examples of images and their scores]

\subsection{Visual prominence analysis}\label{visualprominence}
We developed a method to calculate how visually prominent and clickable an element is on the page.
The analysis of visibility is based on the element's size, colour, background colour, border width, border colour, border radius, and text decoration. 
% [TODO: need more help from Janus to describe how exactly this is done. Currently unclear from this text exactly what the number values are.. e.g., what value does white and red get, and how do you calculate their contrast?]
% TODO: actually write out the formula
% TODO: mention that cultural meaning of colours is not taken into account (i.e., red and green, which also few regulators have said something about)
% TODO: have more image examples of things that are considered equal, rather than the images of the individual scores? that gives the wrong idea
It uses these properties to calculate the colour saturation, contrast between the element and its background, the element's border and its background, the element's text and its background, and the text's decoration. 
These values can be either negative or positive (or 0 if no value could be extracted) and are weighted based on the extent to which we believe the specific design element contributes to an element's visual prominence. 
The saturation has the biggest impact (factor 1.5), followed by the contrast between the element and its background (1), border contrast and text contrast (0.5), and border saturation contrast and text decoration (0.2)
The final score is the sum of these individual scores (see Figure \ref{fig:visibility} for examples). 
% The contrast between the element's colour and the background contribute to the score at a factor of 1, the difference in saturation between the element and the background at factor 1.5, the border contrast and text contrast at factor 0.5, and the border saturation contrast and text decoration at 0.2. 

While the scores themselves may not be meaningful alone (e.g., a score of 0.35 has no inherent meaning), they allow for consistent comparisons between elements.
Elements are considered equally as visually prominent if the difference between their score is less than 0.5. 
We decided on this threshold by finding the bimodal separation point in the distribution of differences and manually inspecting sets of elements that fell on either side of this cut-off point to see if it was a fair and meaningful distinction.

It is important to recognise that the visibility of an element on a screen is subjective and is co-constituted between the user and the environment. 
For instance, green and red are visually different to computers and some people, but not to those with red-green colour blindness. Factors like screen brightness and size also affect visibility depending on the user. Despite this subjectivity, quantification is necessary for large-scale, automated comparisons, and the resulting scores were deemed reasonable upon manual inspection.

% To account for the subjectivity of these scores, we categorize them into broad ranges rather than relying on precise numerical values, which incorporates a margin of error to better accommodate varying perceptions of visibility.
% The four resulting categories are (nearly invisible, subtle, visible, and prominent

% [manually evaluated. accuracy rate per country = table]

\subsection{Purpose controls detection}
We detect whether there are any purpose-level controls on the first page of the interface. These are toggles, checkboxes, or buttons that allow a user to give more granular consent than just accepting or rejecting all data processing. For example, ``Create profiles for personalised advertising'' or ``Measure advertising performance''~\cite{IABTCFPurposes}. 
This heuristic looks for all items inside the detected consent interface that satisfy the CSS selector \code{input[type="checkbox"]} or \code{[role="checkbox"]}. It returns its current status (checked or not) and whether it is disabled (often the ``mandatory purposes'' have a checkbox that cannot be interacted with).

\subsection{Detection accuracy}\label{accuracy}
We manually evaluated 3.100 random website screenshots (100 for each country) to establish a ground truth to compare our detection methods against (see Table \ref{accuracyTab}).

On average, the detection method for the consent interface achieves an F1 score of 0.99, similar to \citet{rasaii2023exploring}, with Hungary scoring the lowest at 0.92.
% The difference in accuracy score is likely because their analysis used a global top list (which they report is 77\% in English), whereas we use country specific top lists, and thus encountered more linguistic diversity: the corpus from \citet{rasaii2023exploring} covered twelve languages, whereas we support thirty-one languages.
% Our accuracy score for English is similarly 99\%.
We obtain an F1 score of 0.98 for accept options, 0.99 for reject options, and 0.99 for settings options, surpassing previous work. 
The F1 score for save options is 1.00, but because it is based on a much smaller sample, it should be interpreted cautiously. 
We were unable to calculate a score for pay options, as these were mostly absent from the sample of screenshots. 
The detection of purpose controls, such as checkboxes, results in a relatively lower F1 score of 0.82 (with the lowest for Finland at 0.55), while the detection of checkbox status yields an F1 score of 0.67, primarily due to false negatives.
% Visual equivalence? how is that evaluated?

% If a user option has a label that is not in our word corpus and is not within a Levenshtein distance of 1, it will result in a false negative.
% Any discrepancy, whether false negative or false positive, was marked as an incorrect answer.

\section{Results}

% TODO FUTURE:
% Check the numbers of the popups. They don't quite add up across column and row totals.
% Increase row spacing
% Dotted line above colour cells
% Numbers not C but left aligned
% Make rows centred vertically? They are closer to the top of the row now

\subsection{How prevalent are consent interfaces?}
On average, 67\% of websites use a consent interface (see Table \ref{tab:PopupStats}).
% TODO: add stats about how many have more than 1 popup? Although I should be pretty confident about those not being false positives then

\subsubsection{Consent interfaces across countries}
There are substantial differences in the prevalence of consent interfaces across countries (see column \textit{Interfaces} in Table \ref{tab:PopupStats}).
For instance, Germany has the highest proportion of consent interfaces across their top 10.000 with 78\%, while Estonian websites have the lowest share with only 41\%.\footnote{Iceland and Liechtenstein have lower shares, but are left out of consideration because of the small sample size.} 
There is no clear way to separate countries into `consent-heavy' or `consent-light' groups: the prevalence of consent interfaces gradually decreases between the two extremes, without a clear dividing line.

\subsubsection{Consent interfaces across popularity ranks}
Consent interfaces are quite evenly spread across popularity ranks (see column \textit{Popularity Rank} in Table \ref{tab:PopupStats}) -- the difference between the top 1,000 and the bottom 1,000 is just 12 percentage points -- although consent interfaces are somewhat more prevalent in the top 3,000 most popular websites.

There are national differences in the way consent interfaces are spread across popularity ranks. 
In some countries, the proportion stays quite even no matter how popular the sites are, such as in Poland or Slovakia, where the difference between the highest and lowest proportion is minimal (6 and 7 points respectively).
Other countries, such as Ireland, Estonia, or Latvia, exhibit a `top-heavy' pattern where the most popular sites have higher shares of consent interfaces, which then quickly drops off.
A third category of countries have distributions with peaks and valleys, where popularity rises or dips across the ranks. 
In Germany, for example, the middle popularity segment has the highest share of interfaces, whereas in the United Kingdom this segment actually has the lowest share, and the Netherlands and Portugal instead show a ripple pattern.
% There does not seem to be any correlation between the overall prevalence of consent interface and the distribution of those interfaces across popularity ranks. 
% A higher share of websites with pop-ups (such as Germany and Spain) does not mean that all websites in that country are equally likely to have them (r=0.49), or that the most and least popular sites have very similar rates (r=0.53).

\subsection{How prevalent are consent management platforms?}
CMPs are third-party software that website owners can download or buy and then install on their page.
At least 67\% of all consent interfaces are consent management platforms (CMPs), or 44.5\% of all websites analysed (see column \textit{CMP} in Table \ref{tab:PopupStats}).
11.7\% of CMPs were detected using the TCF API, which means they are part of the IAB Transparency \& Consent Framework.
The top three most used CMPs are Usercentrics (headquartered in Germany), CookieYes (in the U.K.), and OneTrust (in the U.S.). 
Together, these three services make up 37.57\% of all CMPs.
Despite this concentration, there is still a long tail of other providers: the total number of unique CMPs that we found was 115 (see Appendix \ref{AllCMPs}).
% TODO: add how many of these are detected based on TCF

\subsubsection{CMPs across countries}
CMPs are quite evenly spread across countries: Denmark has the largest share of banners that are CMPs with 84\%, and Slovenia the smallest with 54\%.
The national CMP markets all exhibit concentration, with just a handful of services being used by most of the websites.
The lowest market share by the top three CMP providers is 36\% in Romania, but the highest is 64\% in Denmark.
In a number of countries, the top three include CMPs that are locally developed. 
For example, ibuenda in Italy, tarteaucitron and Didomi in France, Shoper in Poland, Shoptet in the Czech Republic, Usercentrics in Germany, CookieHub in Iceland, Mozello CookieBar in Latvia, and CookieInformation and Usercentrics (which bought the Danish Cookiebot in 2021) in Denmark.
Interestingly, apart from Usercentrics, CookieInformation, and Shoptet, none of these national CMPs make it in the top three of another country, which means some countries have a national web culture to use locally developed CMPs (and this means those CMPs are within the jurisdiction of national supervisory authorities).

\begingroup
\setlength{\extrarowheight}{2pt}
\begin{table*}[t]
\caption{Prevalence of consent interfaces and CMPs. The countries are ranked based on how many websites have consent interfaces. The colour of the Popularity Rank cells reflects the proportion of interfaces in that popularity bin: darker means more. The CMP percentages are relative to the number of interfaces, not the number of websites. All CMPs without a coloured background are only present once in the top three across all countries. The * indicates that the country had fewer than 10.000 domains available for scraping.}

	\centering
	\resizebox{\textwidth}{!}{%
		\footnotesize
		\begin{tabular}{@{}llllccccccccclllllllllllll@{}}
			\toprule
			& \textbf{Sites} & \multicolumn{2}{c}{\textbf{Interfaces}} & \multicolumn{10}{c}{\textbf{Interfaces \% in Popularity Rank}} & \multicolumn{2}{c}{\textbf{CMP}} & \multicolumn{9}{c}{\textbf{Top 3 CMPs}} \\ 
			\cmidrule(lr){3-4} \cmidrule(lr){5-14} \cmidrule(lr){15-16} \cmidrule(lr){17-25}
			                            &      & \textit{n} & \textit{\%} & \textit{0-1K}                              & \textit{1-2K}                             & \textit{2-3K}                             & \textit{3-4K}                             & \textit{4-5K}                             & \textit{5-6K}                             & \textit{6-7K}                             & \textit{7-8K}                             & \textit{8-9K}                             & \textit{9-10K}                            & \textit{n}                                & \textit{\%}                               & \textit{name}                                        & \textit{n}                                & \textit{\%}                               & \textit{name}                                        & \textit{n}                  & \textit{\%} & \textit{name}                                 & \textit{n} & \textit{\%}                               \\
			\midrule
\ \ \textbf{Germany} & 9386 & 7362 & 78 & {\cellcolor[rgb]{0.796, 0.522, 0.510}} 80 & {\cellcolor[rgb]{0.820, 0.576, 0.569}} 76 & {\cellcolor[rgb]{0.784, 0.490, 0.482}} 82 & {\cellcolor[rgb]{0.776, 0.475, 0.463}} 83 & {\cellcolor[rgb]{0.769, 0.455, 0.447}} 84 & {\cellcolor[rgb]{0.796, 0.522, 0.510}} 80 & {\cellcolor[rgb]{0.851, 0.655, 0.647}} 71 & {\cellcolor[rgb]{0.863, 0.686, 0.675}} 69 & {\cellcolor[rgb]{0.800, 0.533, 0.522}} 79 & {\cellcolor[rgb]{0.788, 0.502, 0.494}} 81 & 4979 & 68 & {\cellcolor[rgb]{0.651, 0.808, 0.890}} Usercentrics & 1591 & 32 & {\cellcolor[rgb]{0.416, 0.239, 0.604}} consentmanager.net & 650 & 13 & {\cellcolor[rgb]{0.698, 0.875, 0.541}} OneTrust & 431 & 9 \\ \cdashline{1-25}[0.5pt/2pt]
\ \ \textbf{Spain} & 8901 & 6869 & 77 & {\cellcolor[rgb]{0.788, 0.502, 0.494}} 81 & {\cellcolor[rgb]{0.776, 0.475, 0.463}} 83 & {\cellcolor[rgb]{0.800, 0.533, 0.522}} 79 & {\cellcolor[rgb]{0.812, 0.561, 0.549}} 77 & {\cellcolor[rgb]{0.800, 0.533, 0.522}} 79 & {\cellcolor[rgb]{0.820, 0.576, 0.569}} 76 & {\cellcolor[rgb]{0.824, 0.588, 0.580}} 75 & {\cellcolor[rgb]{0.824, 0.588, 0.580}} 75 & {\cellcolor[rgb]{0.851, 0.655, 0.647}} 71 & {\cellcolor[rgb]{0.831, 0.608, 0.596}} 74 & 4803 & 70 & {\cellcolor[rgb]{0.698, 0.875, 0.541}} OneTrust & 666 & 14 & {\cellcolor[rgb]{0.122, 0.471, 0.706}} CookieYes & 616 & 13 & {\cellcolor[rgb]{0.651, 0.808, 0.890}} Usercentrics & 586 & 12 \\ \cdashline{1-25}[0.5pt/2pt]
\ \ \textbf{Slovakia} & 9559 & 7388 & 77 & {\cellcolor[rgb]{0.800, 0.533, 0.522}} 79 & {\cellcolor[rgb]{0.784, 0.490, 0.482}} 82 & {\cellcolor[rgb]{0.812, 0.561, 0.549}} 77 & {\cellcolor[rgb]{0.820, 0.576, 0.569}} 76 & {\cellcolor[rgb]{0.808, 0.549, 0.537}} 78 & {\cellcolor[rgb]{0.812, 0.561, 0.549}} 77 & {\cellcolor[rgb]{0.820, 0.576, 0.569}} 76 & {\cellcolor[rgb]{0.824, 0.588, 0.580}} 75 & {\cellcolor[rgb]{0.820, 0.576, 0.569}} 76 & {\cellcolor[rgb]{0.820, 0.576, 0.569}} 76 & 4531 & 61 & {\cellcolor[rgb]{1.000, 1.000, 0.600}} Shoptet & 791 & 17 & {\cellcolor[rgb]{0.792, 0.698, 0.839}} TermsFeed & 686 & 15 & {\cellcolor[rgb]{0.651, 0.808, 0.890}} Usercentrics & 579 & 13 \\ \cdashline{1-25}[0.5pt/2pt]
\ \ \textbf{Austria} & 9467 & 7180 & 76 & {\cellcolor[rgb]{0.800, 0.533, 0.522}} 79 & {\cellcolor[rgb]{0.788, 0.502, 0.494}} 81 & {\cellcolor[rgb]{0.808, 0.549, 0.537}} 78 & {\cellcolor[rgb]{0.808, 0.549, 0.537}} 78 & {\cellcolor[rgb]{0.851, 0.655, 0.647}} 71 & {\cellcolor[rgb]{0.831, 0.608, 0.596}} 74 & {\cellcolor[rgb]{0.831, 0.608, 0.596}} 74 & {\cellcolor[rgb]{0.824, 0.588, 0.580}} 75 & {\cellcolor[rgb]{0.820, 0.576, 0.569}} 76 & {\cellcolor[rgb]{0.843, 0.639, 0.627}} 72 & 4721 & 66 & {\cellcolor[rgb]{0.651, 0.808, 0.890}} Usercentrics & 879 & 19 & Borlabs & 681 & 14 & {\cellcolor[rgb]{0.200, 0.627, 0.173}} Osano & 576 & 12 \\ \cdashline{1-25}[0.5pt/2pt]
\ \ \textbf{Italy} & 8666 & 6562 & 76 & {\cellcolor[rgb]{0.753, 0.416, 0.408}} 87 & {\cellcolor[rgb]{0.796, 0.522, 0.510}} 80 & {\cellcolor[rgb]{0.796, 0.522, 0.510}} 80 & {\cellcolor[rgb]{0.812, 0.561, 0.549}} 77 & {\cellcolor[rgb]{0.863, 0.686, 0.675}} 69 & {\cellcolor[rgb]{0.863, 0.686, 0.675}} 69 & {\cellcolor[rgb]{0.812, 0.561, 0.549}} 77 & {\cellcolor[rgb]{0.831, 0.608, 0.596}} 74 & {\cellcolor[rgb]{0.839, 0.627, 0.616}} 73 & {\cellcolor[rgb]{0.851, 0.655, 0.647}} 71 & 5278 & 80 & iubenda & 1551 & 29 & {\cellcolor[rgb]{0.651, 0.808, 0.890}} Usercentrics & 800 & 15 & {\cellcolor[rgb]{0.698, 0.875, 0.541}} OneTrust & 507 & 10 \\ \cdashline{1-25}[0.5pt/2pt]
\ \ \textbf{Denmark} & 9533 & 7149 & 75 & {\cellcolor[rgb]{0.765, 0.443, 0.435}} 85 & {\cellcolor[rgb]{0.788, 0.502, 0.494}} 81 & {\cellcolor[rgb]{0.788, 0.502, 0.494}} 81 & {\cellcolor[rgb]{0.800, 0.533, 0.522}} 79 & {\cellcolor[rgb]{0.839, 0.627, 0.616}} 73 & {\cellcolor[rgb]{0.851, 0.655, 0.647}} 71 & {\cellcolor[rgb]{0.859, 0.667, 0.659}} 70 & {\cellcolor[rgb]{0.843, 0.639, 0.627}} 72 & {\cellcolor[rgb]{0.859, 0.667, 0.659}} 70 & {\cellcolor[rgb]{0.863, 0.686, 0.675}} 69 & 6000 & 84 & {\cellcolor[rgb]{0.651, 0.808, 0.890}} Usercentrics & 1831 & 31 & {\cellcolor[rgb]{1.000, 0.498, 0.000}} Cookie Information & 1579 & 26 & {\cellcolor[rgb]{0.122, 0.471, 0.706}} CookieYes & 411 & 7 \\ \cdashline{1-25}[0.5pt/2pt]
\ \ \textbf{United Kingdom} & 9163 & 6782 & 74 & {\cellcolor[rgb]{0.796, 0.522, 0.510}} 80 & {\cellcolor[rgb]{0.765, 0.443, 0.435}} 85 & {\cellcolor[rgb]{0.788, 0.502, 0.494}} 81 & {\cellcolor[rgb]{0.820, 0.576, 0.569}} 76 & {\cellcolor[rgb]{0.902, 0.776, 0.769}} 63 & {\cellcolor[rgb]{0.839, 0.627, 0.616}} 73 & {\cellcolor[rgb]{0.812, 0.561, 0.549}} 77 & {\cellcolor[rgb]{0.843, 0.639, 0.627}} 72 & {\cellcolor[rgb]{0.843, 0.639, 0.627}} 72 & {\cellcolor[rgb]{0.910, 0.796, 0.788}} 62 & 4814 & 71 & {\cellcolor[rgb]{0.698, 0.875, 0.541}} OneTrust & 1147 & 24 & {\cellcolor[rgb]{0.651, 0.808, 0.890}} Usercentrics & 610 & 13 & {\cellcolor[rgb]{0.200, 0.627, 0.173}} Osano & 455 & 9 \\ \cdashline{1-25}[0.5pt/2pt]
\ \ \textbf{France} & 9057 & 6626 & 73 & {\cellcolor[rgb]{0.796, 0.522, 0.510}} 80 & {\cellcolor[rgb]{0.812, 0.561, 0.549}} 77 & {\cellcolor[rgb]{0.808, 0.549, 0.537}} 78 & {\cellcolor[rgb]{0.843, 0.639, 0.627}} 72 & {\cellcolor[rgb]{0.878, 0.714, 0.706}} 67 & {\cellcolor[rgb]{0.820, 0.576, 0.569}} 76 & {\cellcolor[rgb]{0.820, 0.576, 0.569}} 76 & {\cellcolor[rgb]{0.839, 0.627, 0.616}} 73 & {\cellcolor[rgb]{0.871, 0.698, 0.686}} 68 & {\cellcolor[rgb]{0.890, 0.745, 0.737}} 65 & 4651 & 70 & tarteaucitron & 868 & 19 & Didomi & 564 & 12 & {\cellcolor[rgb]{0.698, 0.875, 0.541}} OneTrust & 543 & 12 \\ \cdashline{1-25}[0.5pt/2pt]
\ \ \textbf{Romania} & 9392 & 6811 & 73 & {\cellcolor[rgb]{0.808, 0.549, 0.537}} 78 & {\cellcolor[rgb]{0.820, 0.576, 0.569}} 76 & {\cellcolor[rgb]{0.820, 0.576, 0.569}} 76 & {\cellcolor[rgb]{0.820, 0.576, 0.569}} 76 & {\cellcolor[rgb]{0.820, 0.576, 0.569}} 76 & {\cellcolor[rgb]{0.878, 0.714, 0.706}} 67 & {\cellcolor[rgb]{0.859, 0.667, 0.659}} 70 & {\cellcolor[rgb]{0.871, 0.698, 0.686}} 68 & {\cellcolor[rgb]{0.851, 0.655, 0.647}} 71 & {\cellcolor[rgb]{0.871, 0.698, 0.686}} 68 & 4698 & 69 & {\cellcolor[rgb]{0.122, 0.471, 0.706}} CookieYes & 657 & 14 & {\cellcolor[rgb]{0.651, 0.808, 0.890}} Usercentrics & 575 & 12 & {\cellcolor[rgb]{0.200, 0.627, 0.173}} Osano & 477 & 10 \\ \cdashline{1-25}[0.5pt/2pt]
\ \ \textbf{Slovenia} & 9299 & 6676 & 72 & {\cellcolor[rgb]{0.808, 0.549, 0.537}} 78 & {\cellcolor[rgb]{0.824, 0.588, 0.580}} 75 & {\cellcolor[rgb]{0.808, 0.549, 0.537}} 78 & {\cellcolor[rgb]{0.831, 0.608, 0.596}} 74 & {\cellcolor[rgb]{0.831, 0.608, 0.596}} 74 & {\cellcolor[rgb]{0.824, 0.588, 0.580}} 75 & {\cellcolor[rgb]{0.886, 0.733, 0.725}} 66 & {\cellcolor[rgb]{0.890, 0.745, 0.737}} 65 & {\cellcolor[rgb]{0.863, 0.686, 0.675}} 69 & {\cellcolor[rgb]{0.902, 0.776, 0.769}} 63 & 3586 & 54 & {\cellcolor[rgb]{0.122, 0.471, 0.706}} CookieYes & 958 & 27 & {\cellcolor[rgb]{0.984, 0.604, 0.600}} Cookie Notice & 477 & 13 & {\cellcolor[rgb]{0.992, 0.749, 0.435}} Moove & 409 & 11 \\ \cdashline{1-25}[0.5pt/2pt]
\ \ \textbf{Netherlands} & 9529 & 6802 & 71 & {\cellcolor[rgb]{0.812, 0.561, 0.549}} 77 & {\cellcolor[rgb]{0.788, 0.502, 0.494}} 81 & {\cellcolor[rgb]{0.831, 0.608, 0.596}} 74 & {\cellcolor[rgb]{0.910, 0.796, 0.788}} 62 & {\cellcolor[rgb]{0.812, 0.561, 0.549}} 77 & {\cellcolor[rgb]{0.831, 0.608, 0.596}} 74 & {\cellcolor[rgb]{0.886, 0.733, 0.725}} 66 & {\cellcolor[rgb]{0.886, 0.733, 0.725}} 66 & {\cellcolor[rgb]{0.859, 0.667, 0.659}} 70 & {\cellcolor[rgb]{0.878, 0.714, 0.706}} 67 & 4425 & 65 & {\cellcolor[rgb]{0.651, 0.808, 0.890}} Usercentrics & 2032 & 46 & {\cellcolor[rgb]{0.698, 0.875, 0.541}} OneTrust & 371 & 8 & {\cellcolor[rgb]{0.122, 0.471, 0.706}} CookieYes & 241 & 5 \\ \cdashline{1-25}[0.5pt/2pt]
\ \ \textbf{Poland} & 9587 & 6790 & 71 & {\cellcolor[rgb]{0.859, 0.667, 0.659}} 70 & {\cellcolor[rgb]{0.831, 0.608, 0.596}} 74 & {\cellcolor[rgb]{0.851, 0.655, 0.647}} 71 & {\cellcolor[rgb]{0.843, 0.639, 0.627}} 72 & {\cellcolor[rgb]{0.871, 0.698, 0.686}} 68 & {\cellcolor[rgb]{0.859, 0.667, 0.659}} 70 & {\cellcolor[rgb]{0.843, 0.639, 0.627}} 72 & {\cellcolor[rgb]{0.839, 0.627, 0.616}} 73 & {\cellcolor[rgb]{0.863, 0.686, 0.675}} 69 & {\cellcolor[rgb]{0.871, 0.698, 0.686}} 68 & 4192 & 62 & {\cellcolor[rgb]{0.651, 0.808, 0.890}} Usercentrics & 962 & 23 & Shoper & 471 & 11 & {\cellcolor[rgb]{0.122, 0.471, 0.706}} CookieYes & 363 & 9 \\ \cdashline{1-25}[0.5pt/2pt]
\ \ \textbf{Belgium} & 9502 & 6414 & 68 & {\cellcolor[rgb]{0.776, 0.475, 0.463}} 83 & {\cellcolor[rgb]{0.820, 0.576, 0.569}} 76 & {\cellcolor[rgb]{0.863, 0.686, 0.675}} 69 & {\cellcolor[rgb]{0.859, 0.667, 0.659}} 70 & {\cellcolor[rgb]{0.871, 0.698, 0.686}} 68 & {\cellcolor[rgb]{0.918, 0.808, 0.800}} 61 & {\cellcolor[rgb]{0.925, 0.827, 0.820}} 60 & {\cellcolor[rgb]{0.910, 0.796, 0.788}} 62 & {\cellcolor[rgb]{0.918, 0.808, 0.800}} 61 & {\cellcolor[rgb]{0.886, 0.733, 0.725}} 66 & 3870 & 60 & {\cellcolor[rgb]{0.651, 0.808, 0.890}} Usercentrics & 662 & 17 & {\cellcolor[rgb]{0.698, 0.875, 0.541}} OneTrust & 513 & 13 & {\cellcolor[rgb]{0.122, 0.471, 0.706}} CookieYes & 370 & 10 \\ \cdashline{1-25}[0.5pt/2pt]
\ \ \textbf{Croatia} & 9471 & 6336 & 67 & {\cellcolor[rgb]{0.820, 0.576, 0.569}} 76 & {\cellcolor[rgb]{0.843, 0.639, 0.627}} 72 & {\cellcolor[rgb]{0.863, 0.686, 0.675}} 69 & {\cellcolor[rgb]{0.851, 0.655, 0.647}} 71 & {\cellcolor[rgb]{0.886, 0.733, 0.725}} 66 & {\cellcolor[rgb]{0.878, 0.714, 0.706}} 67 & {\cellcolor[rgb]{0.859, 0.667, 0.659}} 70 & {\cellcolor[rgb]{0.910, 0.796, 0.788}} 62 & {\cellcolor[rgb]{0.949, 0.878, 0.871}} 57 & {\cellcolor[rgb]{0.929, 0.839, 0.831}} 59 & 3932 & 62 & {\cellcolor[rgb]{0.122, 0.471, 0.706}} CookieYes & 908 & 23 & {\cellcolor[rgb]{0.890, 0.102, 0.110}} Complianz & 447 & 11 & {\cellcolor[rgb]{0.992, 0.749, 0.435}} Moove & 426 & 11 \\ \cdashline{1-25}[0.5pt/2pt]
\ \ \textbf{Sweden} & 9552 & 6342 & 66 & {\cellcolor[rgb]{0.800, 0.533, 0.522}} 79 & {\cellcolor[rgb]{0.843, 0.639, 0.627}} 72 & {\cellcolor[rgb]{0.851, 0.655, 0.647}} 71 & {\cellcolor[rgb]{0.859, 0.667, 0.659}} 70 & {\cellcolor[rgb]{0.878, 0.714, 0.706}} 67 & {\cellcolor[rgb]{0.918, 0.808, 0.800}} 61 & {\cellcolor[rgb]{0.910, 0.796, 0.788}} 62 & {\cellcolor[rgb]{0.918, 0.808, 0.800}} 61 & {\cellcolor[rgb]{0.925, 0.827, 0.820}} 60 & {\cellcolor[rgb]{0.929, 0.839, 0.831}} 59 & 4413 & 70 & {\cellcolor[rgb]{0.651, 0.808, 0.890}} Usercentrics & 1185 & 27 & {\cellcolor[rgb]{0.122, 0.471, 0.706}} CookieYes & 532 & 12 & {\cellcolor[rgb]{0.698, 0.875, 0.541}} OneTrust & 445 & 10 \\ \cdashline{1-25}[0.5pt/2pt]
\ \ \textbf{Hungary} & 9484 & 6216 & 66 & {\cellcolor[rgb]{0.812, 0.561, 0.549}} 77 & {\cellcolor[rgb]{0.871, 0.698, 0.686}} 68 & {\cellcolor[rgb]{0.878, 0.714, 0.706}} 67 & {\cellcolor[rgb]{0.886, 0.733, 0.725}} 66 & {\cellcolor[rgb]{0.910, 0.796, 0.788}} 62 & {\cellcolor[rgb]{0.871, 0.698, 0.686}} 68 & {\cellcolor[rgb]{0.910, 0.796, 0.788}} 62 & {\cellcolor[rgb]{0.925, 0.827, 0.820}} 60 & {\cellcolor[rgb]{0.902, 0.776, 0.769}} 63 & {\cellcolor[rgb]{0.910, 0.796, 0.788}} 62 & 3555 & 57 & {\cellcolor[rgb]{0.122, 0.471, 0.706}} CookieYes & 585 & 16 & {\cellcolor[rgb]{0.651, 0.808, 0.890}} Usercentrics & 523 & 15 & {\cellcolor[rgb]{0.200, 0.627, 0.173}} Osano & 383 & 11 \\ \cdashline{1-25}[0.5pt/2pt]
\ \ \textbf{Greece} & 9413 & 6105 & 65 & {\cellcolor[rgb]{0.820, 0.576, 0.569}} 76 & {\cellcolor[rgb]{0.859, 0.667, 0.659}} 70 & {\cellcolor[rgb]{0.851, 0.655, 0.647}} 71 & {\cellcolor[rgb]{0.886, 0.733, 0.725}} 66 & {\cellcolor[rgb]{0.890, 0.745, 0.737}} 65 & {\cellcolor[rgb]{0.941, 0.859, 0.851}} 58 & {\cellcolor[rgb]{0.941, 0.859, 0.851}} 58 & {\cellcolor[rgb]{0.918, 0.808, 0.800}} 61 & {\cellcolor[rgb]{0.929, 0.839, 0.831}} 59 & {\cellcolor[rgb]{0.878, 0.714, 0.706}} 67 & 4415 & 72 & {\cellcolor[rgb]{0.122, 0.471, 0.706}} CookieYes & 773 & 18 & InMobi & 629 & 14 & {\cellcolor[rgb]{0.200, 0.627, 0.173}} Osano & 428 & 10 \\ \cdashline{1-25}[0.5pt/2pt]
\ \ \textbf{Finland} & 9510 & 6116 & 64 & {\cellcolor[rgb]{0.796, 0.522, 0.510}} 80 & {\cellcolor[rgb]{0.820, 0.576, 0.569}} 76 & {\cellcolor[rgb]{0.820, 0.576, 0.569}} 76 & {\cellcolor[rgb]{0.910, 0.796, 0.788}} 62 & {\cellcolor[rgb]{0.929, 0.839, 0.831}} 59 & {\cellcolor[rgb]{0.929, 0.839, 0.831}} 59 & {\cellcolor[rgb]{0.929, 0.839, 0.831}} 59 & {\cellcolor[rgb]{0.949, 0.878, 0.871}} 57 & {\cellcolor[rgb]{0.949, 0.878, 0.871}} 57 & {\cellcolor[rgb]{0.929, 0.839, 0.831}} 59 & 4666 & 76 & {\cellcolor[rgb]{0.651, 0.808, 0.890}} Usercentrics & 1474 & 32 & {\cellcolor[rgb]{0.122, 0.471, 0.706}} CookieYes & 524 & 11 & {\cellcolor[rgb]{0.890, 0.102, 0.110}} Complianz & 358 & 8 \\ \cdashline{1-25}[0.5pt/2pt]
\ \ \textbf{Ireland} & 9463 & 6075 & 64 & {\cellcolor[rgb]{0.784, 0.490, 0.482}} 82 & {\cellcolor[rgb]{0.808, 0.549, 0.537}} 78 & {\cellcolor[rgb]{0.890, 0.745, 0.737}} 65 & {\cellcolor[rgb]{0.929, 0.839, 0.831}} 59 & {\cellcolor[rgb]{0.910, 0.796, 0.788}} 62 & {\cellcolor[rgb]{0.925, 0.827, 0.820}} 60 & {\cellcolor[rgb]{0.925, 0.827, 0.820}} 60 & {\cellcolor[rgb]{0.902, 0.776, 0.769}} 63 & {\cellcolor[rgb]{0.929, 0.839, 0.831}} 59 & {\cellcolor[rgb]{0.969, 0.922, 0.914}} 54 & 4360 & 72 & {\cellcolor[rgb]{0.122, 0.471, 0.706}} CookieYes & 917 & 21 & {\cellcolor[rgb]{0.698, 0.875, 0.541}} OneTrust & 719 & 16 & {\cellcolor[rgb]{0.651, 0.808, 0.890}} Usercentrics & 575 & 13 \\ \cdashline{1-25}[0.5pt/2pt]
\ \ \textbf{Portugal} & 9029 & 5759 & 64 & {\cellcolor[rgb]{0.824, 0.588, 0.580}} 75 & {\cellcolor[rgb]{0.831, 0.608, 0.596}} 74 & {\cellcolor[rgb]{0.890, 0.745, 0.737}} 65 & {\cellcolor[rgb]{0.941, 0.933, 0.945}} 46 & {\cellcolor[rgb]{0.910, 0.796, 0.788}} 62 & {\cellcolor[rgb]{0.902, 0.776, 0.769}} 63 & {\cellcolor[rgb]{0.941, 0.859, 0.851}} 58 & {\cellcolor[rgb]{0.890, 0.745, 0.737}} 65 & {\cellcolor[rgb]{0.886, 0.733, 0.725}} 66 & {\cellcolor[rgb]{0.886, 0.733, 0.725}} 66 & 3474 & 60 & {\cellcolor[rgb]{0.122, 0.471, 0.706}} CookieYes & 622 & 18 & {\cellcolor[rgb]{0.698, 0.875, 0.541}} OneTrust & 388 & 11 & {\cellcolor[rgb]{0.651, 0.808, 0.890}} Usercentrics & 323 & 9 \\ \cdashline{1-25}[0.5pt/2pt]
\ \ \textbf{Luxembourg*} & 2243 & 1436 & 64 & {\cellcolor[rgb]{0.863, 0.686, 0.675}} 69 & {\cellcolor[rgb]{0.910, 0.796, 0.788}} 62 & {\cellcolor[rgb]{0.949, 0.878, 0.871}} 57 & - & - & - & - & - & - & - & 992 & 69 & {\cellcolor[rgb]{0.651, 0.808, 0.890}} Usercentrics & 152 & 15 & {\cellcolor[rgb]{0.122, 0.471, 0.706}} CookieYes & 129 & 13 & {\cellcolor[rgb]{0.890, 0.102, 0.110}} Complianz & 110 & 11 \\ \cdashline{1-25}[0.5pt/2pt]
\ \ \textbf{Czech Republic} & 9583 & 6054 & 63 & {\cellcolor[rgb]{0.843, 0.639, 0.627}} 72 & {\cellcolor[rgb]{0.863, 0.686, 0.675}} 69 & {\cellcolor[rgb]{0.902, 0.776, 0.769}} 63 & {\cellcolor[rgb]{0.863, 0.686, 0.675}} 69 & {\cellcolor[rgb]{0.902, 0.776, 0.769}} 63 & {\cellcolor[rgb]{0.918, 0.808, 0.800}} 61 & {\cellcolor[rgb]{0.918, 0.808, 0.800}} 61 & {\cellcolor[rgb]{0.925, 0.827, 0.820}} 60 & {\cellcolor[rgb]{0.941, 0.859, 0.851}} 58 & {\cellcolor[rgb]{0.953, 0.890, 0.882}} 56 & 3480 & 57 & {\cellcolor[rgb]{0.792, 0.698, 0.839}} TermsFeed & 902 & 26 & {\cellcolor[rgb]{1.000, 1.000, 0.600}} Shoptet & 482 & 14 & {\cellcolor[rgb]{0.651, 0.808, 0.890}} Usercentrics & 384 & 11 \\ \cdashline{1-25}[0.5pt/2pt]
\ \ \textbf{Norway} & 9653 & 5884 & 61 & {\cellcolor[rgb]{0.824, 0.588, 0.580}} 75 & {\cellcolor[rgb]{0.871, 0.698, 0.686}} 68 & {\cellcolor[rgb]{0.898, 0.765, 0.757}} 64 & {\cellcolor[rgb]{0.941, 0.859, 0.851}} 58 & {\cellcolor[rgb]{0.941, 0.859, 0.851}} 58 & {\cellcolor[rgb]{0.929, 0.839, 0.831}} 59 & {\cellcolor[rgb]{0.949, 0.878, 0.871}} 57 & {\cellcolor[rgb]{0.929, 0.839, 0.831}} 59 & {\cellcolor[rgb]{0.953, 0.890, 0.882}} 56 & {\cellcolor[rgb]{0.953, 0.890, 0.882}} 56 & 4399 & 75 & {\cellcolor[rgb]{0.651, 0.808, 0.890}} Usercentrics & 1072 & 24 & {\cellcolor[rgb]{1.000, 0.498, 0.000}} Cookie Information & 860 & 20 & {\cellcolor[rgb]{0.122, 0.471, 0.706}} CookieYes & 426 & 10 \\ \cdashline{1-25}[0.5pt/2pt]
\ \ \textbf{Bulgaria} & 9526 & 5812 & 61 & {\cellcolor[rgb]{0.890, 0.745, 0.737}} 65 & {\cellcolor[rgb]{0.878, 0.714, 0.706}} 67 & {\cellcolor[rgb]{0.898, 0.765, 0.757}} 64 & {\cellcolor[rgb]{0.910, 0.796, 0.788}} 62 & {\cellcolor[rgb]{0.902, 0.776, 0.769}} 63 & {\cellcolor[rgb]{0.902, 0.776, 0.769}} 63 & {\cellcolor[rgb]{0.898, 0.765, 0.757}} 64 & {\cellcolor[rgb]{0.949, 0.878, 0.871}} 57 & {\cellcolor[rgb]{0.976, 0.937, 0.933}} 53 & {\cellcolor[rgb]{0.980, 0.945, 0.941}} 52 & 3215 & 55 & {\cellcolor[rgb]{0.200, 0.627, 0.173}} Osano & 562 & 17 & {\cellcolor[rgb]{0.122, 0.471, 0.706}} CookieYes & 481 & 15 & {\cellcolor[rgb]{0.984, 0.604, 0.600}} Cookie Notice & 292 & 9 \\ \cdashline{1-25}[0.5pt/2pt]
\ \ \textbf{Lithuania} & 9425 & 5750 & 61 & {\cellcolor[rgb]{0.851, 0.655, 0.647}} 71 & {\cellcolor[rgb]{0.902, 0.776, 0.769}} 63 & {\cellcolor[rgb]{0.918, 0.808, 0.800}} 61 & {\cellcolor[rgb]{0.902, 0.776, 0.769}} 63 & {\cellcolor[rgb]{0.918, 0.808, 0.800}} 61 & {\cellcolor[rgb]{0.925, 0.827, 0.820}} 60 & {\cellcolor[rgb]{0.910, 0.796, 0.788}} 62 & {\cellcolor[rgb]{0.918, 0.808, 0.800}} 61 & {\cellcolor[rgb]{0.976, 0.937, 0.933}} 53 & {\cellcolor[rgb]{0.965, 0.910, 0.902}} 55 & 3829 & 67 & {\cellcolor[rgb]{0.122, 0.471, 0.706}} CookieYes & 690 & 18 & {\cellcolor[rgb]{0.200, 0.627, 0.173}} Osano & 501 & 13 & {\cellcolor[rgb]{0.651, 0.808, 0.890}} Usercentrics & 471 & 12 \\ \cdashline{1-25}[0.5pt/2pt]
\ \ \textbf{Latvia} & 9487 & 5297 & 56 & {\cellcolor[rgb]{0.839, 0.627, 0.616}} 73 & {\cellcolor[rgb]{0.902, 0.776, 0.769}} 63 & {\cellcolor[rgb]{0.890, 0.745, 0.737}} 65 & {\cellcolor[rgb]{0.925, 0.827, 0.820}} 60 & {\cellcolor[rgb]{0.965, 0.910, 0.902}} 55 & {\cellcolor[rgb]{0.980, 0.945, 0.941}} 52 & {\cellcolor[rgb]{0.976, 0.961, 0.957}} 49 & {\cellcolor[rgb]{0.965, 0.953, 0.957}} 48 & {\cellcolor[rgb]{0.957, 0.949, 0.953}} 47 & {\cellcolor[rgb]{0.941, 0.933, 0.945}} 46 & 2899 & 55 & {\cellcolor[rgb]{0.122, 0.471, 0.706}} CookieYes & 579 & 20 & {\cellcolor[rgb]{0.651, 0.808, 0.890}} Usercentrics & 369 & 13 & Mozello CookieBar & 351 & 12 \\ \cdashline{1-25}[0.5pt/2pt]
\ \ \textbf{Cyprus*} & 1724 & 946 & 55 & {\cellcolor[rgb]{0.910, 0.796, 0.788}} 62 & {\cellcolor[rgb]{0.965, 0.953, 0.957}} 48 & - & - & - & - & - & - & - & - & 656 & 69 & {\cellcolor[rgb]{0.122, 0.471, 0.706}} CookieYes & 200 & 30 & {\cellcolor[rgb]{0.200, 0.627, 0.173}} Osano & 63 & 10 & {\cellcolor[rgb]{0.651, 0.808, 0.890}} Usercentrics & 50 & 8 \\ \cdashline{1-25}[0.5pt/2pt]
\ \ \textbf{Malta*} & 1222 & 600 & 49 & {\cellcolor[rgb]{0.980, 0.945, 0.941}} 52 & {\cellcolor[rgb]{0.847, 0.859, 0.894}} 40 & - & - & - & - & - & - & - & - & 409 & 68 & {\cellcolor[rgb]{0.122, 0.471, 0.706}} CookieYes & 162 & 40 & {\cellcolor[rgb]{0.984, 0.604, 0.600}} Cookie Notice & 47 & 11 & {\cellcolor[rgb]{0.651, 0.808, 0.890}} Usercentrics & 28 & 7 \\ \cdashline{1-25}[0.5pt/2pt]
\ \ \textbf{Estonia} & 9665 & 3946 & 41 & {\cellcolor[rgb]{0.941, 0.859, 0.851}} 58 & {\cellcolor[rgb]{0.976, 0.937, 0.933}} 53 & {\cellcolor[rgb]{0.957, 0.949, 0.953}} 47 & {\cellcolor[rgb]{0.929, 0.925, 0.937}} 45 & {\cellcolor[rgb]{0.859, 0.871, 0.902}} 41 & {\cellcolor[rgb]{0.796, 0.820, 0.871}} 37 & {\cellcolor[rgb]{0.702, 0.749, 0.827}} 31 & {\cellcolor[rgb]{0.753, 0.784, 0.847}} 34 & {\cellcolor[rgb]{0.753, 0.784, 0.847}} 34 & {\cellcolor[rgb]{0.651, 0.710, 0.808}} 28 & 2869 & 73 & {\cellcolor[rgb]{0.122, 0.471, 0.706}} CookieYes & 916 & 32 & {\cellcolor[rgb]{0.651, 0.808, 0.890}} Usercentrics & 332 & 12 & {\cellcolor[rgb]{0.984, 0.604, 0.600}} Cookie Notice & 286 & 10 \\ \cdashline{1-25}[0.5pt/2pt]
\ \ \textbf{Iceland*} & 4535 & 1775 & 39 & {\cellcolor[rgb]{0.969, 0.922, 0.914}} 54 & {\cellcolor[rgb]{0.847, 0.859, 0.894}} 40 & {\cellcolor[rgb]{0.796, 0.820, 0.871}} 37 & {\cellcolor[rgb]{0.753, 0.784, 0.847}} 34 & {\cellcolor[rgb]{0.639, 0.702, 0.804}} 27 & - & - & - & - & - & 1140 & 64 & CookieHub & 389 & 34 & {\cellcolor[rgb]{0.122, 0.471, 0.706}} CookieYes & 185 & 16 & {\cellcolor[rgb]{0.200, 0.627, 0.173}} Osano & 107 & 9 \\ \cdashline{1-25}[0.5pt/2pt]
\ \ \textbf{Liechtenstein*} & 122 & 41 & 34 & {\cellcolor[rgb]{0.753, 0.784, 0.847}} 34 & - & - & - & - & - & - & - & - & - & 23 & 56 & {\cellcolor[rgb]{0.651, 0.808, 0.890}} Usercentrics & 12 & 52 & {\cellcolor[rgb]{0.416, 0.239, 0.604}} consentmanager.net & 3 & 13 & Cookie-Script & 1 & 4 \\ \midrule
\ \ \textbf{Total} & 254148 & 169901 & 67 & {\cellcolor[rgb]{0.831, 0.608, 0.596}} 74 & {\cellcolor[rgb]{0.851, 0.655, 0.647}} 71 & {\cellcolor[rgb]{0.859, 0.667, 0.659}} 70 & {\cellcolor[rgb]{0.878, 0.714, 0.706}} 67 & {\cellcolor[rgb]{0.890, 0.745, 0.737}} 65 & {\cellcolor[rgb]{0.886, 0.733, 0.725}} 66 & {\cellcolor[rgb]{0.890, 0.745, 0.737}} 65 & {\cellcolor[rgb]{0.898, 0.765, 0.757}} 64 & {\cellcolor[rgb]{0.902, 0.776, 0.769}} 63 & {\cellcolor[rgb]{0.910, 0.796, 0.788}} 62 & 113274 & 67 & {\cellcolor[rgb]{0.651, 0.808, 0.890}} Usercentrics & 19549 & 17 & {\cellcolor[rgb]{0.122, 0.471, 0.706}} CookieYes & 14102 & 12 & {\cellcolor[rgb]{0.698, 0.875, 0.541}} OneTrust & 9101 & 8 \\ \bottomrule
		\end{tabular}
	}
     \label{tab:PopupStats}
     \Description[Prevalence of consent interfaces and CMPs]{Overview of the scores for each country in terms of the number of interfaces, how those are distributed across popularity ranks, the CMP prevalence, and which three CMPs are most prevalent. Shows 67 percent of websites have consent interfaces, 67 percent of those have a CMP, and usercentrics, cookieyes, and onetrust are the most popular ones.}
\end{table*}
\endgroup

\vspace{2cm}

%TC:ignore
\begin{figure*}[]
    \centering
    \includegraphics[width=\textwidth]{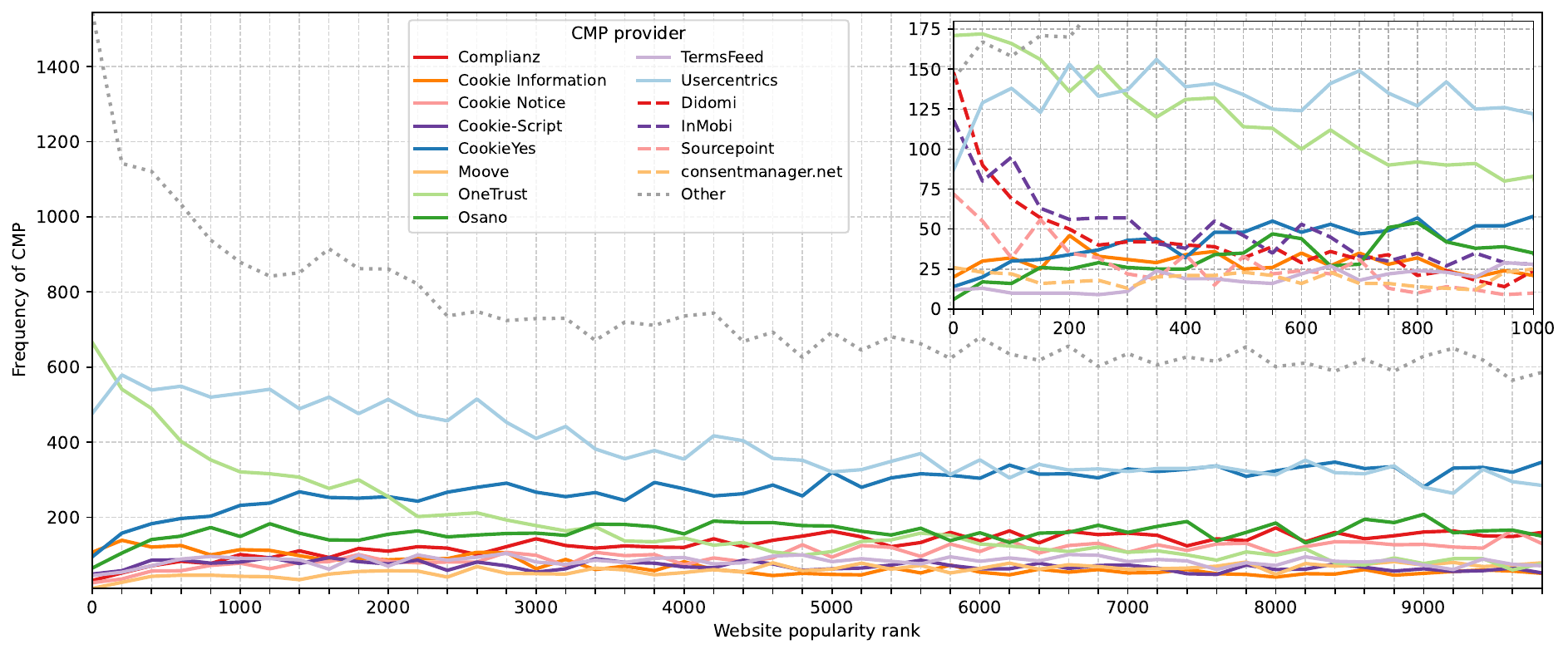}  % For PDF
    \caption{Distribution of top 10 CMP providers by website popularity rank.
    The main plot groups websites into popularity rank bins of 200 (e.g., 1–200, 201–400).
    The inset shows the top 1,000 sites, using bins of 50.
    Dashed lines indicate CMPs that appear in the top 10 only within this smaller subset.
    The dotted line represents the combined count of all other CMPs outside the top 10.
    Grid lines mark bin boundaries.}
    \label{fig:CMPpopularity}
    \Description[CMP prevalence across website popularity]{Shows the top 10 CMPs across popularity ranks.}
\end{figure*}
%TC:endignore

\subsubsection{CMPs across popularity ranks}
% TODO: should there be a table as well with the total number per bin, and a longer table in the appendix where we report the total for the top 20 per bin?
CMPs are used by websites across all popularity ranks (see Figure \ref{fig:CMPpopularity}), but predominately by the most popular sites.
Usercentrics and CookieYes are, with some distance, clearly the two most used CMPs.
However, Usercentrics is used more by the most popular websites, which might be because Usercentrics' cheapest option starts at €50 per month, while CookieYes offers a free option.
% When zooming in on the most popular sites we can see that, curiously, InMobi (formerly Quantcast Choice) is the  second most used CMP by the top fifty websites, but quickly drops off after and remains the least prevalent of the top ten.
OneTrust is another CMP that is considerably more popular in the top 3,000, but is less prevalent in the segments after that compared to other CMPs.
Didomi, InMobi, and Sourcepoint are CMPs that are only popular with the top 200 ranked websites.
% Overall, CMP providers do not appear to have particular website ranks as their main customers, but are evenly spread out over all segments.
Above all, however, we can see that there is a long tail of other CMP providers and their cumulative count is consistently higher than each individual provider in the top ten.
This is particularly evident in the top 1,000, where the share of the long tail is double and even triple that of Usercentrics.

\subsection{What do consent interfaces look like?}
Most consent interfaces have an accept option (88\%), while less than half have a reject option (45\%) (see Table \ref{interfacedesign}).
In fact, more interfaces provide some way to access settings (54\%) -- a button or link that often takes the user to another view with more information and granular controls -- than a way to refuse consent.
Likely, the reject options are hidden on this second screen, a common design practice that is generally considered non-compliant and well-documented to significantly decrease users' opt-out rates~\cite{Nouwens_Liccardi_Veale_Karger_Kagal_2020, bielova2024effect}.
Reject options also consistently score lower on visual prominence compared to accept options.
Pay buttons that ask users to subscribe if they do not want to consent are still vanishingly rare (0.24\%).
The European Data Protection Board recently opined that this ``should not be the way forward''~\cite{EDPBPay} in response to attempts from Meta to implement this across their platforms, but websites are increasingly developing first-party data collection methods such as login walls as alternatives to third-party tracking.
Purpose controls -- checkboxes or toggles -- are not very common on the initial screen (only 16\% of interfaces have them), and surprisingly there are even fewer user options to save whatever the user toggles (8\%).
Roughly 17\% of these purpose controls are pre-checked by default, another design choice that is firmly established as non-compliant~\cite{planet49ag}.
A notable proportion of interfaces (9\%) lack any options altogether, which means these are likely merely notification banners that inform users about the use of cookies.

There is considerable homogeneity in the labels that are used for the user options (see Table \ref{interfacedesign}). 
For each category, only four or five phrases represent nearly all labels that are used across \textasciitilde 170,000 websites.
The most commonly used labels clearly indicate what clicking the option will do (``Accept all'', ``Change settings''), but there are some notable exceptions. 
For example, more than a quarter of all reject options (27\%) use the rather ambiguous ``Accept necessary''  text (or comparable variations).
Similarly, for settings labels, phrases like ``See details'' or ``Learn more'' do not clearly convey that clicking will lead to more controls.

\setlength{\extrarowheight}{2pt}
\begin{table*}[]
\caption{Prevalence of user interface elements in consent interfaces. The table is ordered based on the compliance score. Compliance is defined as an interface with both an accept and reject option that are visually equally as prominent (if not identical), and has either a settings option or purpose controls without anything pre-checked. The `score' is the median of all the elements' visual prominence. The \ding{51} is the number of checkboxes with at least one optional purpose already pre-checked. Countries marked with * have ePD and/or GDPR regulators which have issued guidance~\cite{bielova2024two, noybCookies}, and \textsuperscript{\dag} have imposed fines~\cite{noyb}.}
\centering
\resizebox{\textwidth}{!}{%
\setlength{\tabcolsep}{4pt}
\begin{tabular}{@{}llllllllllllllllllllllllll@{}}
\toprule
 & \textbf{Interfaces} & \multicolumn{2}{c}{\textbf{Compliant}} & \multicolumn{2}{c}{\textbf{No options}} & \multicolumn{3}{c}{\textbf{Accept}} & \multicolumn{3}{c}{\textbf{Reject}} & \multicolumn{3}{c}{\textbf{Settings}} & \multicolumn{3}{c}{\textbf{Save}} & \multicolumn{3}{c}{\textbf{Pay}} & \multicolumn{4}{c}{\textbf{Purpose controls}} \\
 
\cmidrule(lr){3-4} \cmidrule(lr){5-6} \cmidrule(lr){7-9} \cmidrule(lr){10-12}  \cmidrule(lr){13-15} \cmidrule(lr){16-18} \cmidrule(lr){19-21} \cmidrule(lr){22-25}

&      & \textit{n} & \textit{\%} & \textit{n} & \textit{\%} & \textit{n} & \textit{\%} & \textit{score} & \textit{n} & \textit{\%} & \textit{score} & \textit{n} & \textit{\%} & \textit{score} & \textit{n} & \textit{\%} & \textit{score} & \textit{n} & \textit{\%} & \textit{score} & \textit{n} & \textit{\%} & \textit{\ding{51}} & \textit{\%}\\
\midrule
\ \ \textbf{Spain*\textsuperscript{\dag}} & 6869 & 1894 & 28 & 401 & 6 & 6312 & 92 & 1.55 & 3983 & 58 & 1.13 & 4462 & 65 & 0.56 & 325 & 5 & 1.67 & 86 & 1.25 & 0.77 & 579 & 8 & 119 & 21 \\ \cdashline{1-25}[0.5pt/2pt] 
\ \ \textbf{Slovakia} & 7388 & 1815 & 25 & 333 & 5 & 6828 & 92 & 1.59 & 4075 & 55 & 0.95 & 5355 & 72 & 0.52 & 397 & 5 & 1.63 & - & - & - & 1199 & 16 & 133 & 11 \\ \cdashline{1-25}[0.5pt/2pt] 
\ \ \textbf{Italy*\textsuperscript{\dag}} & 6562 & 1536 & 23 & 466 & 7 & 5935 & 90 & 1.74 & 3637 & 55 & 1.10 & 4337 & 66 & 0.77 & 457 & 7 & 1.78 & 58 & 0.88 & -0.10 & 990 & 15 & 122 & 12 \\ \cdashline{1-25}[0.5pt/2pt] 
\ \ \textbf{France*\textsuperscript{\dag}} & 6626 & 1539 & 23 & 752 & 11 & 5602 & 85 & 1.56 & 4196 & 63 & 0.72 & 4464 & 67 & 0.58 & 179 & 3 & 1.44 & 29 & 0.44 & 1.18 & 303 & 5 & 63 & 21 \\ \cdashline{1-25}[0.5pt/2pt] 
\ \ \textbf{Austria*} & 7180 & 1485 & 21 & 269 & 4 & 6452 & 90 & 1.80 & 3781 & 53 & 1.24 & 4388 & 61 & 0.67 & 1724 & 24 & 1.48 & 10 & 0.14 & 1.20 & 1761 & 25 & 170 & 10 \\ \cdashline{1-25}[0.5pt/2pt] 
\ \ \textbf{United Kingdom*\textsuperscript{\dag}} & 6782 & 1416 & 21 & 545 & 8 & 6028 & 89 & 1.55 & 2872 & 42 & 1.10 & 4448 & 66 & 0.55 & 382 & 6 & 1.64 & 2 & 0.03 & 3.26 & 762 & 11 & 188 & 25 \\ \cdashline{1-25}[0.5pt/2pt] 
\ \ \textbf{Ireland*} & 6075 & 1217 & 20 & 522 & 9 & 5430 & 89 & 1.47 & 2926 & 48 & 0.99 & 3862 & 64 & 0.47 & 507 & 8 & 1.74 & 1 & 0.02 & 3.26 & 738 & 12 & 102 & 14 \\ \cdashline{1-25}[0.5pt/2pt] 
\ \ \textbf{Germany*} & 7362 & 1430 & 19 & 302 & 4 & 6616 & 90 & 1.95 & 3955 & 54 & 1.19 & 4809 & 65 & 0.60 & 1328 & 18 & 1.31 & 168 & 2.28 & 1.56 & 1457 & 20 & 131 & 9 \\ \cdashline{1-25}[0.5pt/2pt] 
\ \ \textbf{Luxembourg*} & 1436 & 265 & 18 & 110 & 8 & 1285 & 89 & 1.47 & 793 & 55 & 0.86 & 810 & 56 & 0.52 & 117 & 8 & 1.37 & - & - & - & 154 & 11 & 27 & 18 \\ \cdashline{1-25}[0.5pt/2pt] 
\ \ \textbf{Finland*} & 6116 & 1087 & 18 & 229 & 4 & 5777 & 94 & 1.96 & 3815 & 62 & 1.33 & 3429 & 56 & 0.69 & 921 & 15 & 1.80 & - & - & - & 1411 & 23 & 161 & 11 \\ \cdashline{1-25}[0.5pt/2pt] 
\ \ \textbf{Cyprus} & 946 & 166 & 18 & 96 & 10 & 828 & 88 & 1.33 & 449 & 47 & 0.76 & 517 & 55 & 0.26 & 76 & 8 & 1.31 & - & - & - & 122 & 13 & 30 & 25 \\ \cdashline{1-25}[0.5pt/2pt] 
\ \ \textbf{Denmark*\textsuperscript{\dag}} & 7149 & 1210 & 17 & 279 & 4 & 6710 & 94 & 2.05 & 5119 & 72 & 1.25 & 1750 & 24 & 0.43 & 1699 & 24 & 1.66 & 48 & 0.67 & 0.25 & 3766 & 53 & 311 & 8 \\ \cdashline{1-25}[0.5pt/2pt] 
\ \ \textbf{Czech Republic*} & 6054 & 1046 & 17 & 243 & 4 & 5554 & 92 & 1.66 & 2732 & 45 & 0.82 & 4416 & 73 & 0.45 & 265 & 4 & 1.34 & - & - & - & 1189 & 20 & 113 & 10 \\ \cdashline{1-25}[0.5pt/2pt] 
\ \ \textbf{Iceland} & 1775 & 280 & 16 & 190 & 11 & 1544 & 87 & 1.34 & 507 & 29 & 1.25 & 718 & 40 & 0.59 & 25 & 1 & 1.78 & - & - & - & 44 & 2 & 9 & 20 \\ \cdashline{1-25}[0.5pt/2pt] 
\ \ \textbf{Belgium*\textsuperscript{\dag}} & 6414 & 1057 & 16 & 594 & 9 & 5516 & 86 & 1.62 & 2977 & 46 & 0.91 & 3551 & 55 & 0.55 & 515 & 8 & 1.43 & 1 & 0.02 & 3.31 & 863 & 13 & 169 & 20 \\ \cdashline{1-25}[0.5pt/2pt] 
\ \ \textbf{Sweden*} & 6342 & 881 & 14 & 323 & 5 & 5874 & 93 & 1.83 & 3119 & 49 & 1.29 & 3847 & 61 & 0.63 & 852 & 13 & 1.91 & - & - & - & 1662 & 26 & 259 & 16 \\ \cdashline{1-25}[0.5pt/2pt] 
\ \ \textbf{Latvia*} & 5297 & 688 & 13 & 725 & 14 & 4373 & 83 & 1.44 & 2000 & 38 & 1.11 & 2465 & 47 & 0.32 & 236 & 4 & 1.79 & - & - & - & 603 & 11 & 134 & 22 \\ \cdashline{1-25}[0.5pt/2pt] 
\ \ \textbf{Greece*} & 6105 & 782 & 13 & 424 & 7 & 5592 & 92 & 1.25 & 2534 & 42 & 0.75 & 3144 & 51 & 0.35 & 406 & 7 & 1.46 & - & - & - & 709 & 12 & 169 & 24 \\ \cdashline{1-25}[0.5pt/2pt] 
\ \ \textbf{Portugal*} & 5759 & 711 & 12 & 795 & 14 & 4820 & 84 & 1.33 & 2059 & 36 & 0.86 & 2462 & 43 & 0.46 & 267 & 5 & 1.13 & 1 & 0.02 & 3.27 & 457 & 8 & 80 & 18 \\ \cdashline{1-25}[0.5pt/2pt] 
\ \ \textbf{Liechtenstein} & 41 & 5 & 12 & 4 & 10 & 37 & 90 & 1.47 & 25 & 61 & 1.50 & 16 & 39 & 0.34 & 7 & 17 & 1.59 & - & - & - & 7 & 17 & 1 & 14 \\ \cdashline{1-25}[0.5pt/2pt] 
\ \ \textbf{Malta} & 600 & 71 & 12 & 76 & 13 & 516 & 86 & 1.20 & 215 & 36 & 0.73 & 302 & 50 & 0.24 & 20 & 3 & 1.44 & - & - & - & 33 & 6 & 7 & 21 \\ \cdashline{1-25}[0.5pt/2pt] 
\ \ \textbf{Romania\textsuperscript{\dag}} & 6811 & 807 & 12 & 712 & 10 & 5872 & 86 & 1.53 & 2477 & 36 & 0.99 & 3339 & 49 & 0.36 & 325 & 5 & 2.05 & 1 & 0.01 & -0.18 & 817 & 12 & 269 & 33 \\ \cdashline{1-25}[0.5pt/2pt] 
\ \ \textbf{Norway} & 5884 & 624 & 11 & 625 & 11 & 5033 & 86 & 1.83 & 3005 & 51 & 1.07 & 2382 & 40 & 0.58 & 709 & 12 & 1.69 & - & - & - & 1578 & 27 & 237 & 15 \\ \cdashline{1-25}[0.5pt/2pt] 
\ \ \textbf{Lithuania} & 5750 & 604 & 11 & 423 & 7 & 5165 & 90 & 1.25 & 1651 & 29 & 0.72 & 2088 & 36 & 0.37 & 290 & 5 & 1.87 & - & - & - & 584 & 10 & 136 & 23 \\ \cdashline{1-25}[0.5pt/2pt] 
\ \ \textbf{Poland} & 6790 & 658 & 10 & 1062 & 16 & 5437 & 80 & 1.75 & 1980 & 29 & 1.04 & 3763 & 55 & 0.58 & 351 & 5 & 2.08 & - & - & - & 1177 & 17 & 317 & 27 \\ \cdashline{1-25}[0.5pt/2pt] 
\ \ \textbf{Croatia\textsuperscript{\dag}} & 6336 & 658 & 10 & 731 & 12 & 5330 & 84 & 1.39 & 2137 & 34 & 0.90 & 3234 & 51 & 0.36 & 312 & 5 & 1.51 & - & - & - & 660 & 10 & 85 & 13 \\ \cdashline{1-25}[0.5pt/2pt] 
\ \ \textbf{Netherlands*\textsuperscript{\dag}} & 6802 & 620 & 9 & 475 & 7 & 5944 & 87 & 1.97 & 2904 & 43 & 0.97 & 3848 & 57 & 0.60 & 804 & 12 & 1.65 & 1 & 0.01 & 3.31 & 1150 & 17 & 462 & 40 \\ \cdashline{1-25}[0.5pt/2pt] 
\ \ \textbf{Estonia} & 3946 & 318 & 8 & 349 & 9 & 3485 & 88 & 1.39 & 1436 & 36 & 0.86 & 1757 & 45 & 0.28 & 185 & 5 & 1.79 & - & - & - & 394 & 10 & 76 & 19 \\ \cdashline{1-25}[0.5pt/2pt] 
\ \ \textbf{Hungary\textsuperscript{\dag}} & 6216 & 466 & 7 & 502 & 8 & 5440 & 88 & 1.52 & 2240 & 36 & 0.78 & 3335 & 54 & 0.43 & 278 & 4 & 1.87 & 2 & 0.03 & -0.18 & 681 & 11 & 203 & 30 \\ \cdashline{1-25}[0.5pt/2pt] 
\ \ \textbf{Bulgaria} & 5812 & 255 & 4 & 734 & 13 & 4912 & 85 & 1.36 & 1134 & 20 & 0.89 & 2194 & 38 & 0.41 & 237 & 4 & 1.17 & 1 & 0.02 & -0.87 & 424 & 7 & 127 & 30 \\ \cdashline{1-25}[0.5pt/2pt] 
\ \ \textbf{Slovenia} & 6676 & 198 & 3 & 1478 & 22 & 4897 & 73 & 1.20 & 1313 & 20 & 0.73 & 1484 & 22 & 0.40 & 43 & 1 & 1.32 & - & - & - & 595 & 9 & 92 & 15 \\ \midrule
\ \ \textbf{Total} & 169901 & 25789 & 15 & 14769 & 9 & 149144 & 88 & 1.61 & 76046 & 45 & 1.02 & 90976 & 54 & 0.52 & 14239 & 8 & 1.62 & 409 & 0.24 & 0.98 & 26869 & 16 & 4502 & 17 \\
\bottomrule
\end{tabular}\label{interfacedesign}
\Description[Prevalence of user interface elements in consent interfaces across countries]{Shows only 15 percent is compliant, 9 percent have no options, 88 percent have accept options, 45 percent have reject options, 54 percent have settings, 8 percent have save, 0 percent have pay, 16 percent have checkboxes and 17 percent of those are pre-checked.}
}
\end{table*}

%TC:ignore
\setlength{\extrarowheight}{2pt}
\begin{table*}
\caption{The most common labels for different user options. The labels were translated using deep-translator~\cite{deep-translator} with the Google Translator engine, fuzzy matched using RapidFuzz~\cite{rapidfuzz}, and then manually clustered.}
\centering

\begin{minipage}[t]{0.50\textwidth}
\vspace{0pt}
\centering
\begin{tabular}{@{}p{7cm}p{0.3cm}@{}}
\toprule
\ \ \textbf{Accept label} & \textbf{\%} \\
\midrule
\ \ Accept/allow/enable/approve/agree (all) & 65 \\ \cdashline{1-2}[0.5pt/2pt]
\ \ Accept cookies & 20 \\\cdashline{1-2}[0.5pt/2pt]
\ \ Ok & 5 \\\cdashline{1-2}[0.5pt/2pt]
\ \ I see/understand & 4 \\\cdashline{1-2}[0.5pt/2pt]
\ \ Consent & 2\\ \cdashline{1-2}[0.5pt/2pt]
\ \ Other & 4 \\
\midrule
\ \ \textbf{Reject label} & \textbf{\%} \\
\midrule
\ \ Reject/refuse/deny/decline/disagree (all) & 65 \\\cdashline{1-2}[0.5pt/2pt]
\ \ Accept (only) necessary/essential/mandatory (cookies) & 27\\\cdashline{1-2}[0.5pt/2pt]
\ \ I don't accept/do not allow & 4 \\\cdashline{1-2}[0.5pt/2pt]
\ \ Continue without accepting & 2 \\\cdashline{1-2}[0.5pt/2pt]
\ \ Other & 2 \\
\midrule
\ \ \textbf{Save label} & \textbf{\%} \\
\midrule
\ \ Allow/accept selection/choice & 78 \\\cdashline{1-2}[0.5pt/2pt]
\ \ Save settings/preferences & 19 \\\cdashline{1-2}[0.5pt/2pt]
\ \ Other & 3 \\
\midrule
% \midrule
\end{tabular}
\end{minipage}
\hfill
\begin{minipage}[t]{0.45\textwidth}
\vspace{0pt}
\centering
\begin{tabular}{@{}p{6.5cm}p{0.3cm}@{}}
\midrule
\ \ \textbf{Settings label} & \textbf{\%} \\
\midrule
\ \ Change settings/preferences & 44\\\cdashline{1-2}[0.5pt/2pt]
\ \ See details/learn more/more info & 32 \\\cdashline{1-2}[0.5pt/2pt]
\ \ Customize/adjust/personalize/configure/manage & 13 \\\cdashline{1-2}[0.5pt/2pt]
\ \ More options & 5 \\\cdashline{1-2}[0.5pt/2pt]
\ \ Set/configure cookies & 2 \\ \cdashline{1-2}[0.5pt/2pt]
\ \ Other & 4 \\
\midrule
\ \ \textbf{Pay label} & \textbf{\%} \\
\midrule
\ \ Ad-free for  \EUR{}x/month  & 42\\\cdashline{1-2}[0.5pt/2pt]
\ \ (Refuse and) subscribe & 25 \\\cdashline{1-2}[0.5pt/2pt]
\ \ Log in with contentpass & 13 \\\cdashline{1-2}[0.5pt/2pt]
\ \ Buy access & 12 \\\cdashline{1-2}[0.5pt/2pt]
\ \ Reject and pay & 8\\
\bottomrule
\end{tabular}\label{tab:buttonLabels}
\end{minipage}
\Description[Labels used in user options]{Accept all, Reject all, Accept selection, Change Settings, and Ad free for x euro per month are the most popular labels across user options.}
\end{table*}
%TC:endignore

\vspace{10pt}

%TC:ignore
\begin{figure*}
    \centering
    \includegraphics[width=1\textwidth]{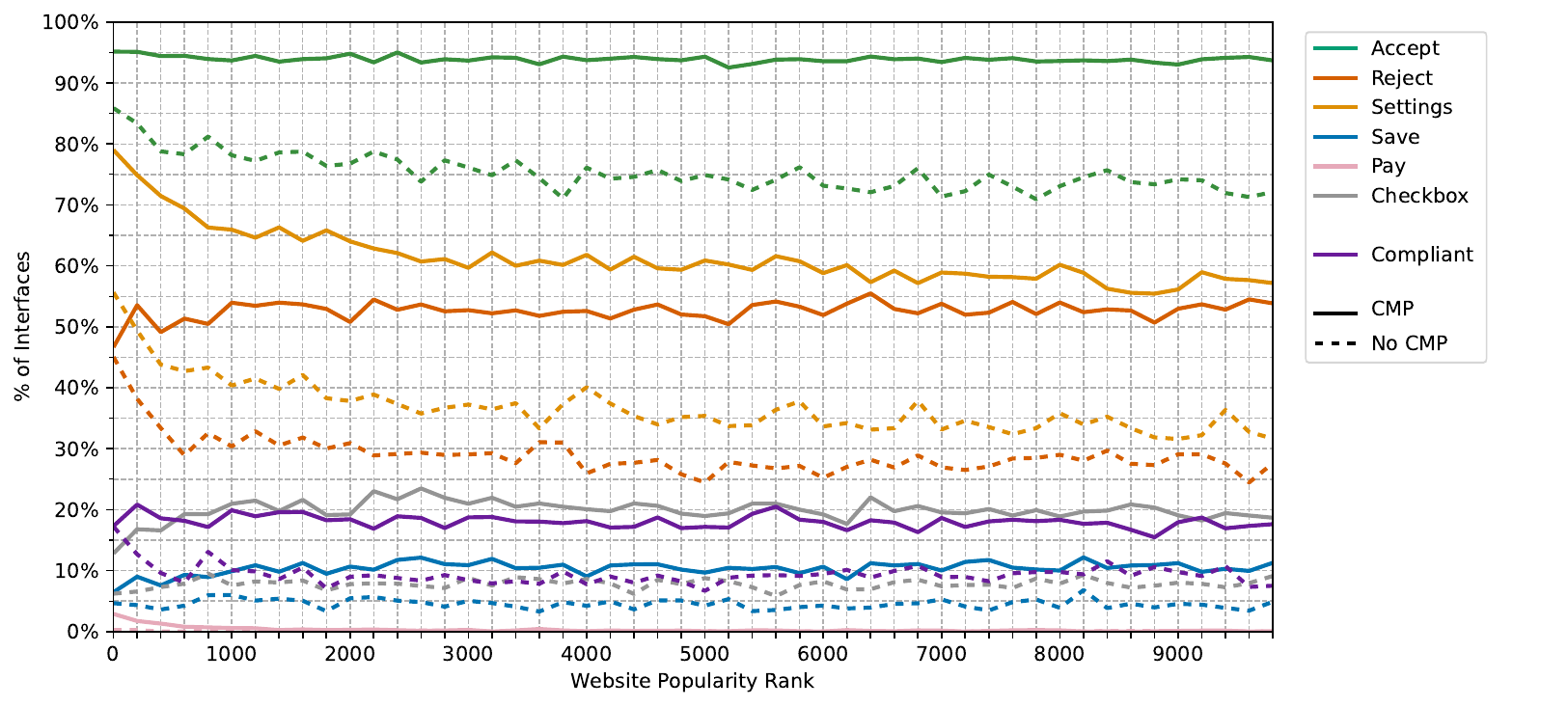}  % For PDF
    \caption{Frequency of user options and compliance across website popularity ranks and CMPs. Solid lines are options from interfaces where a CMP was detected. Dashed lines are either not a CMP or undetected. The plot uses bins of size 200. The grid lines show the edges of the bins.}
    \label{fig:ButtonPopularity}
    \Description[Frequency of user options across website popularity]{Shows that there are no large difference in user options across popularity ranks, but that there are differences between CMPs and not-CMPs}
\end{figure*}
%TC:endignore

\subsubsection{User options across countries}
% TODO: Is there some way of charictarising the entropy/noise/coherence of the labels? Like.. some labels are clearly used across all countries, there's a lot of overlap.. I now did this table with percentages, but I feel like there should be some info that is useful if one were to do a detection script with minimal amount of label matching, how much would you cover with just a small amount? I guess it's a measure of diversity? It's not standard deviation because that's for numerical, this is categorical
% From chatGPT:
% Uniformity vs. Diversity: Entropy, Gini-Simpson Index, and Shannon Diversity are better for comparing how evenly distributed categories are.
% Dominance: Use Dominance Index if you're concerned about whether one category overwhelms the dataset.
The proportion of accept options is relatively consistent across all countries, ranging from 94\% (Finland and Denmark) to 73\% (Slovenia), with a standard deviation of 4.3 percentage points. 
In contrast, there are substantial variations in other user options. 
For reject options, the difference is much larger, with Denmark leading at 72\% while Bulgaria and Slovenia lag behind at just 20\%, and a standard deviation of 12.6 points. 
This is similar for settings options (Czech Republic: 73\%, Slovenia: 22\%, $\sigma = 12.6$).
The option to pay for website access is clearly a country-specific practice, only meaningfully present in six out of thirty-one countries.
Even in these cases, their prevalence is usually below 1\%, with the exception of Germany, where they are present on 2.28\% of interfaces. 
Purpose controls are also relatively scarce on the initial screen, with the exception of Denmark where more than half of all interfaces (53\%) include toggles or checkboxes. 
The countries with a higher proportion of purpose controls also have a noticeably higher share of save options and, in the case of Denmark, also a significantly lower number of settings options, suggesting there is a dominant design pattern.
Pre-selected purposes are present in all countries, but the Netherlands stands out as an outlier with 40\% of purpose controls pre-selected, significantly higher than second-ranked countries Hungary and Bulgaria with 30\%.

\subsubsection{User options across popularity ranks and CMPs}
The popularity of a website does not seem to have a very large impact on what kind of user options they have in their interface--percentages are mostly stable--with the exception of setting options which are more prevalent in the top 3,000 and checkboxes which are slightly less prevalent on the top 1,000 (see Figure \ref{fig:ButtonPopularity}).
The main differences can be seen between interfaces that use a CMP or those without a CMP (that we could detect): non-CMP consent interfaces consistently have fewer user options.
Non-CMP interfaces also show more variation across popularity ranks: there are more accept and reject options in the top 1,000 most popular sites before dropping off and leveling out.

\subsection{How compliant are consent interfaces?}
\subsubsection{Defining compliance}
The compliance of consent interfaces with the GDPR and ePD depends on more things than what we measure here.
Some requirements are not included in our analysis--such as what is present on the second layer of the interface, the ability to withdraw consent at a later stage, or whether any processing happens before a user has given their answer--while other requirements cannot be automated because they rely on qualitative interpretation--such as the quality of the information provided to the website visitor.
Instead, we take a generous approach to compliance and only consider measurable design choices that serve as a minimum threshold. These conditions for minimal compliance are:
\begin{itemize}
    \item \textit{Accepting is as easy as rejecting:} the interface has both an accept and reject option on the first layer and their visual prominence is similar;
    \item \textit{No pre-checked purposes}: the interface does not have any optional purpose controls which are already toggled on;
    \item \textit{Granular controls}: the interface gives the option to consent to specific purposes or vendors through checkbox controls on the first layer or a settings option that we assume gives access to granular controls.
\end{itemize}

% This is a generous definition: we allow for some design differences between the accept and reject button that might affect their prominence (a strict definition would reduce the  compliance rate by 3 points), and we consider that having a settings button means there is some way for the user to access granular options that control specific data processing purposes (to fulfill the "specific" requirement for valid consent under the GDPR).
% This also does not take other requirements into account, such as the quality of the information provided to the website visitor, the ability to withdraw consent at a later stage, whether any processing happens before a user has given their answer, etc.
Meeting these conditions gives insight in the \textit{maximum} level of compliance, but the real compliance rate is likely lower when other obligations are taken into account.

\subsubsection{Compliance across countries, CMPs, and website popularity}
The average compliance rate of consent interfaces is only 15\% (see the column \textit{Compliant} in Table \ref{interfacedesign}).
Compliance rates do not appear to be correlated with the popularity rank of the website (see Figure \ref{fig:ButtonPopularity}).
Compliance rates do vary considerably between countries: Spain has the highest compliance rate with 28\% of consent interfaces, while only 3\% of interfaces are compliant in Slovenia ($\sigma=5.9$).
The main reason for non-compliance is the lack of a reject option (56\%, see Figure \ref{fig:ComplianceConditions}), but many interfaces are also missing granular consent controls (30\%) or have reject options that are harder to see than accept options (24\%). 
Compliance rates vary even more across CMPs ($\sigma=17.5$, see Table \ref{tab:CMPcompliance}).
In the top twenty most compliant CMPs, the compliance rate ranges from 65\% for Shopify (an e-commerce platform that offers an integrated consent interface for their business clients), to merely 4\% for Osano (a platform that offers automated cookie classification and integrates with many other services).
Usercentrics and CookieYes -- the most and second-most used services responsible for nearly a third of all CMPs -- have a compliance rate of merely 17\% and 8\%.
% CookieYes, the second most popular CMP, is only compliant 13\% of the time.}

%TC:ignore
\begin{figure*}
    \centering
    \includegraphics[width=0.80\textwidth]{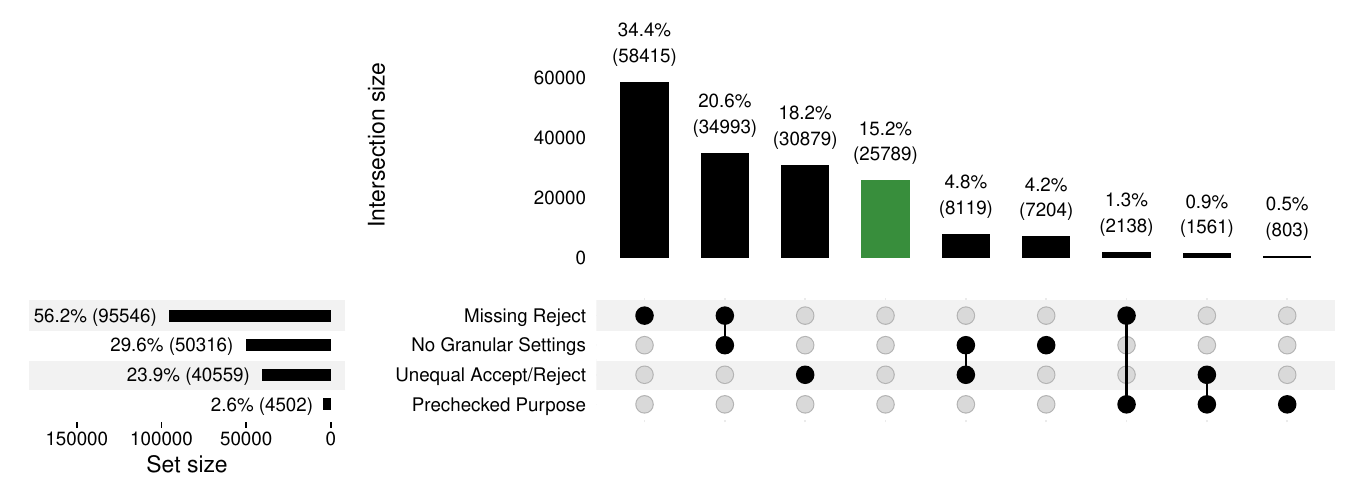}  % For PDF
    \caption{Distribution of the reasons a consent interface is considered non-compliant. The set sizes represent the total count for each individual condition, while the intersections display the exclusive combinations of conditions. The green bar represents interfaces that do not meet any condition and are considered compliant~\cite{lex2014upset,krassowski}.}
    \label{fig:ComplianceConditions}
    \Description[Reasons for non-compliance]{Shows an UpSet diagram that breaks down the reason for non-compliance, with 56 percent because of missing reject buttons, 30 percent because no granular settings, 24 percent because of unequal accept and reject, and 3 percent because of pre-checked purposes.}
\end{figure*}
%TC:endignore

\setlength{\extrarowheight}{0.5pt}
\begin{table*}[]
\caption{The compliance rate of identified consent management platforms. This table shows only the top twenty most compliant CMPs (accounting for 78\% of all CMPs detected), ranked from most to least compliant.}
\centering
\begin{minipage}[t]{0.49\textwidth}
\vspace{0pt}
\centering
\begin{tabular}{@{}p{3cm}p{0.5cm}p{0.5cm}p{0.5cm}p{0.5cm}@{}}
\toprule
\ \  & \multicolumn{2}{c}{\textbf{Compliance}} & \multicolumn{2}{c}{\textbf{Prevalence}} \\
\ \ \textbf{CMP} & \textit{n} & \textit{\%} & \textit{n} & \textit{\%} \\
\midrule
\ \ \textbf{Shopify} & 927 & 65 & 1431 & 1.26 \\ \cdashline{1-5}[0.5pt/2pt]
\ \ \textbf{tarteaucitron} & 623 & 54 & 1153 & 1.02 \\ \cdashline{1-5}[0.5pt/2pt]
\ \ \textbf{iubenda} & 1036 & 48 & 2143 & 1.89 \\ \cdashline{1-5}[0.5pt/2pt]
\ \ \textbf{CookieHub} & 501 & 47 & 1068 & 0.94 \\ \cdashline{1-5}[0.5pt/2pt]
\ \ \textbf{CookieConsent} & 702 & 45 & 1573 & 1.39 \\ \cdashline{1-5}[0.5pt/2pt]
\ \ \textbf{CIVIC} & 219 & 44 & 499 & 0.44 \\ \cdashline{1-5}[0.5pt/2pt]
\ \ \textbf{OneTrust} & 3729 & 41 & 9101 & 8.03 \\ \cdashline{1-5}[0.5pt/2pt]
\ \ \textbf{consentmanager.net} & 528 & 29 & 1830 & 1.62 \\ \cdashline{1-5}[0.5pt/2pt]
\ \ \textbf{Borlabs} & 304 & 28 & 1069 & 0.94 \\ 
\midrule
\end{tabular}
\end{minipage}
\hfill
\begin{minipage}[t]{0.49\textwidth}
\vspace{0pt}
\centering
\begin{tabular}{@{}p{3cm}p{0.5cm}p{0.5cm}p{0.5cm}p{0.8cm}@{}}
\midrule
\ \ \textbf{Shoptet} & 364 & 27 & 1372 & 1.21 \\ \cdashline{1-5}[0.5pt/2pt]
\ \ \textbf{Moove} & 767 & 26 & 2990 & 2.64 \\ \cdashline{1-5}[0.5pt/2pt]
\ \ \textbf{TermsFeed} & 1072 & 26 & 4164 & 3.68 \\ \cdashline{1-5}[0.5pt/2pt]
\ \ \textbf{Cookie-Script} & 812 & 23 & 3502 & 3.09 \\ \cdashline{1-5}[0.5pt/2pt]
\ \ \textbf{Usercentrics} & 3404 & 17 & 19549 & 17.26 \\ \cdashline{1-5}[0.5pt/2pt]
\ \ \textbf{Didomi} & 373 & 17 & 2208 & 1.95 \\ \cdashline{1-5}[0.5pt/2pt]
\ \ \textbf{Cookie Information} & 361 & 10 & 3556 & 3.14 \\ \cdashline{1-5}[0.5pt/2pt]
\ \ \textbf{Complianz} & 591 & 9 & 6444 & 5.69 \\ \cdashline{1-5}[0.5pt/2pt]
\ \ \textbf{CookieYes} & 1126 & 8 & 14102 & 12.45 \\ \cdashline{1-5}[0.5pt/2pt]
\ \ \textbf{InMobi} & 208 & 8 & 2678 & 2.38 \\ \cdashline{1-5}[0.5pt/2pt]
\ \ \textbf{Osano} & 359 & 4 & 8037 & 7.10 \\
\bottomrule
\end{tabular}
\end{minipage}
\label{tab:CMPcompliance}
\Description[Compliance across CMPs]{Shows the top 20 most compliant CMPs ranked based on their compliance rates. Shows Shopify has a compliance of 65 percent, and the most popular CMPs UserCentrics, CookieYes, and Onetrust have average compliance rates.}
\end{table*}

\subsubsection{Explaining variations in compliance}
The compliance of a consent interface is significantly predicted by the CMP that a website uses.
% It's not possible to pinpoint the effect of country overall because there is not enough variation in country after taking into account guidance and fines. For that reason, country was omitted as an independent variable.
We conducted an ANOVA to examine the impact on compliance of 1) the CMP, 2) the existence of regulatory guidance, 3) a history of fines, and 4) the website's popularity rank (categorised into buckets of 500, e.g., 1–500, 501–1,000, and so on).\footnote{We chose these bin sizes because the popularity ranking is based on a combination of four top lists that use different ranking methodologies and bucket ranges (see Section \ref{listgeneration}), so smaller bins would misrepresent the precision of the ranking while larger bins would hide variance. We excluded country as an independent variable due to its high multicollinearity with guidance and fines. This overlap made it difficult to isolate the unique effect of country, as the variation in compliance explained by country was largely accounted for by guidance and fines.}
%OLD: It showed there was a significant effect for the CMP $(F(96,160504)=279.07,p<0.001)$, guidance $(F(1,160476)=1274.71,p<0.001)$, and fines $(F(1,160476)=110.17,p<0.001)$, but no effect for website popularity $(F(19,160476)=1.34,p<0.115)$.
It showed there was a significant effect of the CMP $(F(115,270052)=489.9,p<0.001)$, website popularity $(F(19,270052)=2.2,p=0.002)$, and guidance $(F(1,270052)=681.8,p<0.001)$, but no effect from fines $(F(1,270052)=0.8,p=0.357)$.\footnote{The reader should be aware that there is no official complete dataset of regulator guidance and fines, nor are fines always made public, which affects the quality of this independent variable. Other factors that likely affect whether regulatory activity would have a visible impact are the time since guidance or fines were given out and this data was collected, the height of the fine, the media attention it received, etc. These results should be interpreted with caution and not as proof that fines are ineffective.}
To estimate the contribution of each variable to the compliance rate, we performed separate linear regressions.
This showed that CMPs explain 17,8\% of the variance in compliance ($F(115,270253)=509.9,p<0.001, R^2 = 0.178$, adjusted $R^2 = 0.178$), guidance accounts for 0.8\% of the variance ($F(1,270367)=2052.0,p<0.001, R^2 = 0.008$, adjusted $R^2 = 0.008$), and website popularity explains 0.1\% of the variance ($F(19,270349)=14.1,p<0.001, R^2 = 0.001$, adjusted $R^2 = 0.001$).

% and fines explain 0.7\% of the variance ($F(1,160620)=1213.0,p<0.001, R^2 = 0.007$, adjusted $R^2 = 0.007$).
% There are many possible explanations for this, such as how recently guidance or fines were given, whether the guidance addressed all the components we take into account for our compliance score, how high the fine was, who was fined, whether the fine or guidance increased websites' perceived risk, whether the guidance or fines were publicised widely, etc.

We conducted a multiple linear regression to examine the direction and strength of the effects of individual CMPs, website popularity, and guidance on the compliance rate, while including fines to account for its potential confounding effects.
% We also conducted a multiple linear regression to predict compliance based on the country, CMP used, website ranking, regulatory guidance, and regulator fines as possible explanatory variables.
We used non-CMP consent interfaces as the reference category for CMPs (i.e., consent interfaces likely created by the website owner themselves), because we assume that a CMP developed by an expert third-party should considerably increase its compliance.
We used an absence of guidance and fines, and the first popularity ranking bin (top 500) as the reference category for the other variables.

The model is statistically significant ($F(136,270232)=437.7, p=0.00$) and explains 18.1\% of the variance in compliance ($R^2 = 0.181$, adjusted $R^2 = 0.180$).
%TODO: Note that the intercept is 0.117, this means that for all sites that are not a CMP, in top 500, for a country with no guidance, with no fines, has a compliance rate of 11,7\%.
The intercept of 0.0255 indicates that, for websites that do not use a third-party CMP, are ranked in the top 500, and are located in countries without guidance or fines, the compliance rate is 2.6\%.
% Of all ninety-six CMPs, fifty-four (56\%) demonstrated statistically significant differences in compliance rates compared to the reference category.

The model shows that the compliance rate increases by 2.8 percentage points if the regulator has released guidance ($\beta=0.0284, p<.001$).
It also shows that websites that fall outside the top 500 most popular sites are less compliant, but not increasingly so.
The compliance rate is between -1.4 points (rank 2000-2500, $\beta=0.0135, p<.001$) and -0.7 points (rank 9000-9500, $\beta=0.007, p=0.033$) lower than the most popular websites.
% Similarly, a history of fines is associated with an, albeit smaller, increase in compliance ($\beta=0.0228, p<.001$), suggesting that compliance is 2.3 percentage points higher if fines have been imposed by the national regulator. 
Additionally, the model shows substantial differences in how specific CMPs impact compliance.
The effect sizes range widely, from strongly positive (up to +97 percentage points) to slightly negative (-5 percentage points).
Among the twenty most commonly used CMPs—covering 82\% of all identified CMPs (see Table \ref{tab:CMPcompliance})—Shopify stands out with the largest positive association ($\beta=0.611, p<.001$), followed by CookieHub ($\beta=0.438, p<.001$).
In contrast, Osano ($\beta=0.011, p<.001$) and InMobi ($\beta=0.037, p<.001$) had the lowest coefficients, with a mere 1-3 points improvement in compliance compared to not using a CMP.
The second most popular CMP CookieYes, representing 13\% of all consent interfaces, was similarly associated with only a minor improvement ($\beta=0.048, p<.001$).
Usercentrics, the most used CMP, increased compliance by 13.6 percentages ($\beta=0.136, p<.001$) points and OneTrust, the third most, by 36.9 points ($\beta=0.369, p<.001$).

% The most widely used CMP, Usercentrics, demonstrated a modestly positive effect on compliance rates, increasing compliance by 13.6 percentage points  ($\beta=0.136, p<.001$). 
% In contrast, the second most popular CMP, CookieYes, was associated with a modest decrease in compliance, reducing it by 3.4 percentage points ($\beta=0.048, p<.001$).
% The third most popular CMP, OneTrust, also had a positive impact, raising compliance by 27.7 percentage points ($\beta=0.369, p<.001$).}

These results indicate that websites in countries where national authorities have provided guidance tend to have slightly higher compliance rate, but that the choice of third-party CMP can have wildly varying effects, with some services increasing compliance by as much as 97 percentage points, while others reduce it by up to 5 percentage points, relative to not using a CMP. 
Notably, the top three most commonly used CMPs—accounting for 40\% of those analysed—were decidedly mediocre at improving compliance.

\section{Discussion}

\subsection{How have consent interfaces changed over time?}
\subsubsection{Consent interfaces}
Our findings reveal 67\% of websites have consent interfaces, a higher prevalence compared to all previous studies (see Table \ref{litresultsoverview}), although it is unclear whether this reflects differences in detection methods or an actual increase in their adoption. 
For instance, \citet{rasaii2023exploring} reported that 47\% of websites had pop-ups in 2023, using a detection method which they report had a 99\% accuracy. 
The 20-point difference with our results may stem from their inclusion of non-EU countries, making it difficult to conclude whether the increase is due to temporal trends or geographic differences.
On the other end, \citet{Degeling_Utz_Lentzsch_Hosseini_Schaub_Holz_2019} reported a similar prevalence of 63\% already back in 2019, based on a manual analysis of the top 500 across all EU countries.
It is unclear whether the similarity with our results means that little has changed in the intervening years, or whether consent interfaces were mostly prevalent on the most popular sites 
% so shortly after the GDPR went into effect 
and the other 9,500 sites have now caught up.
% TODO: do comparisons with specific countries here: Ireland, Greece/UK, Degeling has all EU countries

\subsubsection{CMPs}
We find significantly more consent management platforms compared to previous work: 67\% of all consent interfaces, or 45\% of all websites.
% TODO: degeleing = of total sites, zhang = also US sites
The previously highest detected proportion was 15\% by \citet{Degeling_Utz_Lentzsch_Hosseini_Schaub_Holz_2019} in 2019, while the most recent study in 2024 by \citet{Zhang_Meng_Zhou_Ren_2024} found only 8\%. 
Earlier research generally focused on fewer CMPs~\cite{Nouwens_Liccardi_Veale_Karger_Kagal_2020} or specific frameworks like the IAB TCF~\cite{Matte_Bielova_Santos_2020}, so our broader detection method is likely the reason for this considerable increase.
% TODO:
% However, if we focus specifically on the CMPs that were previously measured, such as OneTrust, Quantcast, [.. we can see a difference?]
% This might indicate that CMPs have indeed increased significantly in the intervening five years.
Previous work by \citet{Hils_Woods_Bohme_2020} found that CMPs were primarily used by websites in the 100-10,000 segment of popularity.
This sounded reasonable: one would expect the top 100 popular websites, likely belonging to organisations with significant resources, to implement their own compliance mechanisms rather than outsource such important designs.
Instead, our data shows that CMPs are actually most prevalent in the top 100 and their popularity decreases steadily across popularity ranks.
This could be explained by the fact that \citet{Hils_Woods_Bohme_2020} only looked at six CMP companies, while we considered more, except that three of the companies they considered were also included in our study -- OneTrust, InMobi (previously QuantCast), and Cookiebot (now absorbed by Usercentrics) -- and these are actually some of the most popular in the 100. 
This suggests that, between 2020 and 2024, popular websites have now also started to use consent interfaces developed by external companies.

\subsubsection{Accept options}
Our results show that 88\% of interfaces include accept options, a level consistent with findings from prior studies examining consent interfaces.
Keeping all the methodological differences in mind, it is noteworthy to see that the average proportion of accept options has remained relatively similar across multiple studies since 2020~\cite{Mehrnezhad_2020}.

\subsubsection{Reject options}
Reject options show an upward trend, from less than 10\% across most studies in 2020~\cite{Mehrnezhad_2020, Bornschein_Schmidt_Maier_2020, Soe_Nordberg_Guribye_Slavkovik_2020} to somewhere between 22-61\% in recent work~\cite{Khandelwal_Nayak_Harkous_Fawaz_2023, Gundelach_Herrmann_2023, Bouhoula_Kubicek_Zac_Cotrini_Basin_2024}.
The average of those numbers seems to align with our results, where we find 45\% of consent interfaces have reject options.
While a positive development, this is still only about half of the number of accept options, and our visual analysis shows they are consistently designed to be less prominent.

\subsubsection{Purpose controls}
The proportion of pre-checked purpose controls, although relatively understudied, has shown a downward trend across previous work, starting at roughly \textasciitilde 50\%~\cite{Matte_Bielova_Santos_2020, Nouwens_Liccardi_Veale_Karger_Kagal_2020}.
This design pattern was one of the early battlegrounds after the GDPR went into effect, even though the GDPR specifically mentioned that "pre-ticked boxes [...] should not therefore constitute consent"~\cite{gdpr}, a point reinforced by case law~\cite{planet49ag}. 
Our results now show an average of 17\% of pre-checked purposes controls, suggesting some improvement.

\subsection{How do consent interface designs impact users?}
This paper provides the most comprehensive and robust analysis of consent interfaces to date on 250,000 websites across thirty-one countries, drawing from and innovating on efforts and methodological insights from a large body of academic studies.
But what do these statistics mean in practice?
How do the design choices of consent interfaces influence user behaviour, and what are the implications for people's data protection and privacy?
% What is the link between interface design and people's privacy/data protection/online tracking? How important are those in the grand scheme of other factors, and which bits?

\subsubsection{Interfaces vs. other factors}
Controlled and natural experiments consistently show that the design of consent interfaces significantly impacts user decisions, more so than other factors studied~\cite{Bouma-Sims_Li_Lin_Sakura-Lemessy_Nisenoff_Young_Birrell_Cranor_Habib_2023}, which means that quantitatively measuring the prevalence of these interfaces is essential to assess the scale of user manipulation and the extent to which individuals' data protection rights are compromised.
Beyond the interface, studies show device type also plays a role, with higher consent rates on mobile devices~\cite{Bouma-Sims_Li_Lin_Sakura-Lemessy_Nisenoff_Young_Birrell_Cranor_Habib_2023, Utz_Degeling_Fahl_Schaub_Holz_2019, Bauer_Bergstrøm_Foss-Madsen_2021, bermejo2021website}, likely because consent interfaces block more content on smaller screens. 
A longitudinal natural experiment across hundreds of websites further indicates that EU and UK users have lower consent rates than other geographic areas, and that Android users are less likely to consent than iOS users~\cite{Jha_Trevisan_Mellia_Irarrazaval_Fernandez_2023}. 
Other factors, such as privacy attitudes~\cite{Machuletz_Böhme_2020, bermejo2021website} and risk appetite~\cite{coventry2016personality} show some effects, while self-reported technology and security expertise~\cite{Bouma-Sims_Li_Lin_Sakura-Lemessy_Nisenoff_Young_Birrell_Cranor_Habib_2023, bermejo2021website} show mixed results.
There is no evidence that age or gender is correlated with consent choices~\cite{Bouma-Sims_Li_Lin_Sakura-Lemessy_Nisenoff_Young_Birrell_Cranor_Habib_2023, Habib_Li_Young_Cranor_2022}.

\subsubsection{Presence of user options}
The initial screen of the consent interface, which we focus on in our study, is crucial to understand user behavior, as multiple studies indicate that users rarely engage with secondary pages that require additional clicks~\cite{Nouwens_Liccardi_Veale_Karger_Kagal_2020, Bouma-Sims_Li_Lin_Sakura-Lemessy_Nisenoff_Young_Birrell_Cranor_Habib_2023, Jha_Trevisan_Mellia_Irarrazaval_Fernandez_2023}.
This implies that any choice options that are not on the first page might as well not exist.
% , which implies that any choice options that are moved to a secondary page might as well not exist.
For example, our study reveals that roughly 60\% of consent interfaces do not have a reject option on the initial screen, and multiple user studies have clearly demonstrated that this significantly inflates consent rates and increases unwanted processing of people's personal data.
\citet{Nouwens_Liccardi_Veale_Karger_Kagal_2020} found that removing the reject button from the first page increased acceptance rate by 22 percentage points (from 55\% to 77\%), %22 > 40%
\citet{bielova2024effect} found it increased acceptance by 13 points (from 83\% to 96\%), %13 > 16%
\citet{Habib_Li_Young_Cranor_2022} by 30 points (from \textasciitilde 60\% to 90\%), %30 > 50%
and \citet{Jha_Trevisan_Mellia_Irarrazaval_Fernandez_2023} by 19 points (from 79\% to \textasciitilde 98\%). %19 > 24%
While they all report different effect sizes, they consistently demonstrate that removing the reject option from the first page substantially increases consent.

\subsubsection{Design of user options}
In addition to requiring more clicks, another way in which interfaces are designed that makes options more or less prominent is through their visual design, such as contrast, colour, or clickability.
Our study shows that, on average, accept options are designed to be 1.5 times more prominent than reject options and 3 times more prominent than settings options.
Results from user studies strongly suggest this will nudge people to allow more data processing than if options were equally as prominent.
% User studies show that these differences in prominence can have significant impacts on consent rates.
The most impactful is when one option is designed as a button and the other as a link in the text: \citet{Bouma-Sims_Li_Lin_Sakura-Lemessy_Nisenoff_Young_Birrell_Cranor_Habib_2023} find participants are roughly six times more likely to edit their settings if this option is presented as a button rather than a link, and \citet{berens2022cookie} find that participants are between five and twelve times more likely to reject consent if it is a button. 
There is conflicting evidence whether making the background colour of one button more prominent nudges people compared to two buttons with the same background: ~\citet{Utz_Degeling_Fahl_Schaub_Holz_2019} showed that highlighting the accept button increased consent on mobile devices (and non-response on desktop), \citet{bielova2024effect} found no effect of highlighting the accept button but instead found a significant increase in rejections when the reject button was more prominent, while \citet{berens2022cookie} found no effect either way.
The choice of colour might also matter: \citet{bielova2024effect} showed that using red-orange-green to signal semantic meaning (where green equals reject) made participants 2.3 times more likely to reject consent (using these specific "traffic light" colours has been deemed non-compliant by the Danish authority~\cite{datatilsynetJP}).
When combined, these unequal designs might also exacerbate each other: \citet{Bauer_Bergstrøm_Foss-Madsen_2021} find a 17 percentage point difference in acceptance rate when comparing a completely equal design to a design where accept is a green button and reject is a link in the text.
\subsubsection{Labels of user options}
The text labels on the choice options is another design element that seems to affect users' consent behaviour.
Our study shows that 65\% of reject options use clearly descriptive labels such as "reject", "decline", or "disagree", while 27\% instead label the option as accepting some "necessary", "essential", or "mandatory" cookies.
Perhaps counter-intuitively, \citet{berens2022cookie} found participants were less likely to use clearly labelled options (e.g., "reject" or "no additional cookies"), while ambiguous labels like "only necessary cookies" increased rejection rates, possibly due to fears of losing access or functionality if they chose to reject cookies.
This fear-driven behaviour aligns with \citet{Ma_Birrell_2022}, who showed that labels explicitly stating how a choice would "degrade your experience" reduced rejection rates more than "deny cookies." 
Conversely, labels highlighting the negative consequences for the participant's privacy were able to nudge users toward privacy-preserving decisions~\cite{Ma_Birrell_2022, bielova2024effect}. 
For instance, \citet{bielova2024effect} found that labels like "accept being tracked" or "continue without being tracked" increased rejection rates by as much as 30 points.
% Our study shows that 27\% of reject buttons have a label which, rather than using clearly descriptive words like "reject", "decline", or "disagree", instead present the choice to users as accepting some "necessary", "essential", or "mandatory" cookies.
% Perhaps counterintuitively, \citet{berens2022cookie} shows that using such ambiguous language actually makes people more likely to use it compared to more direct language, which their qualitative data suggests is because users are scared they will be denied access or limit the website's functionality if they "reject". 
Ambiguous language appears less impactful for accept options; \citet{Habib_Li_Young_Cranor_2022} found no difference in consent rates between the labels "Okay" (which we found on 5\% of accept options) and "Allow all cookies" (found on 65\%).

% button labels 
% - habib: no significant difference between Submit+Okay vs Allow selected cookies+Allow all cookies
% - ma: We fnd that for both possible slants, a negative framing is signifcantly more efective at nudging user decisions, and we fnd that the combination of slant and framing impact cookie opt-out rates by a factor of three.
% - berens: We detected smaller effects with regards to the “Label” variable. As such, labeling the “reject” option “Only necessary cookies” made it 2.5 times less likely to accept cookies compared labeling this option as “Reject” (OR = 0.404[0.183, 0.893]). Similarly, changing the label from “No additional cookies” to “Only necessary cookies” makes it more than twice as likely for the users to accept cookies (OR = 2.25[1.049, 4.828]).
% - bielova: Label that explains consequence (‘accept to be tracked/continue without being tracked") increases reject * 3 
% - inal: "they did not read the text in the presented cookie consent interfaces; most skipped over the text regarding the cookie notice, and the rest just skimmed it"

\subsubsection{Presence of granular choices}
The presence of granular options on the initial screen of the interface, such as the purposes for which data is collected, has multiple positive effects on user behaviour and experience.
However, our results show that such options are actually quite rare and only present on 16\% of interfaces.
Having granular options is associated with greater user engagement in decision-making~\cite{Habib_Li_Young_Cranor_2022} and lower consent rates~\cite{Nouwens_Liccardi_Veale_Karger_Kagal_2020, Habib_Li_Young_Cranor_2022}.
Paradoxically, these effects exist even though very few participants actually interact with the options (merely 1.3\%~\cite{Nouwens_Liccardi_Veale_Karger_Kagal_2020}, 2.7\%~\cite{Habib_Li_Young_Cranor_2022}, and \textasciitilde 3\%~\cite{Bouhoula_Kubicek_Zac_Cotrini_Basin_2024}).
% people don't understand functional and performance cookies (bouma, habib)
% Impact of having granular options
% - habib: "the absence of inline options impact people's decision, comprehension, and sentiment towards UI"
% - habib: "The most common interactions were changing in-line options in the initial consent interface"
% "We found that the absence of in-line options within the initial screen of the interface impacted participants' consent decision, comprehension of available cookie options, as well as sentiment toward the consent interface"
% "This suggests that the absence of in-line options within the initial screen of the consent interface may have reduced participants’ investment in their consent decision."
% "However, we did fnd a negative impact of providing in-line options on participants’ comprehension of choices when explicitly instructed to revisit the consent interfaces" 
% - multiple options increase response time (inal, machuletz)
% - Nouwens shows people are less likely to consent when granular options are present
% - Nouwens: only 17 answers out of 1280 (1.3\%) represent a participant consenting to a specific selection of purposes or vendors
% habib: only 2.7\% actually make specific choices
Importantly, these positive effects are negated if options are pre-selected by default, which our study shows is the case in 17\% of interfaces.
Controlled experiments show that preselected purposes or vendors dramatically increase consent rates, from less than 0.1\% to 10\% on desktops~\cite{Habib_Li_Young_Cranor_2022}, and from less than 0.1\% to around 30\% on mobile devices~\cite{Utz_Degeling_Fahl_Schaub_Holz_2019}.
The biasing effect of such default states, in interfaces and more broadly, is well documented~\cite{thaler2004save, johnson2002defaults, lai2006internet, anaraky2020exacerbating}.
% Impact of having prechecked granular options (and more generally, defaults?)
% - Habib?: Preselecting purposes/vendors increased consent rate from <0.1\% to 10\% (desktop)     
% - Utz: The pre-selected versions led around 30\% of mobile users and 10\% of desktop users to accept all third parties. In contrast, only a small fraction (< 0.1\%) allowed all third parties when given the opt-in choice and 1 to 4\% allowed one or more third parties (“other” in Figure 4), indicating that some users still engaged with the offered choices.\cite{Utz_Degeling_Fahl_Schaub_Holz_2019}
% - Mazumdar: Only 0.45\% opt-out when default is opt-in.In the default opt-out, 12,2\% opted in (so that should be their preference)
% - machuletz: "Permissive default options lead to more types of cookies being accepted, with users being less sure of what options they had selected and less content with their choice when informed about what they chose" (cited in Bouma)

\subsubsection{Affective experience}
% AS A RESULT OF THE DESIGNS: PEOPLE FEEL DISEMPOWERED
In general, what user studies indicate is that consent interfaces are experienced as disempowering and do not help people control their personal data in line with their preferences.
% It is difficult to say to what degree people are manipulated away from their preferred answer, because this requires some measurement of a person's ideal level of control, which has proven to be conceptually and methodologically challenging to capture~\cite{preibusch2013guide}. 
% Some user studies have tried to capture the distance between the consent given and consent desired through survey questions (with the caveat that the accuracy of such self-report measures rely to a large extent on the conceptual understanding of online tracking of the participant).
For example, \citet{Habib_Li_Young_Cranor_2022} report that less than half of participants (45\%) who participated in their experiment felt the answer they gave was actually their ideal answer, \citet{Nouwens_Liccardi_Veale_Karger_Kagal_2020} reported that 63\% were not satisfied with their answer, \citet{bielova2024effect} show how the level of participants' satisfaction decreased the more the interface nudged towards more data sharing, and \citet{Machuletz_Böhme_2020} described that participants exposed to more deceptive interfaces had higher levels of regret. 
% Inal: no one reads the text
In line with these self-reported affective experiences, multiple studies demonstrate that when given the chance users overwhelmingly ignore consent interfaces altogether when they do not block too much of the screen~\cite{Inal_Volden_Carlsen_Hjelmtveit_2024, Habib_Li_Young_Cranor_2022, Utz_Degeling_Fahl_Schaub_Holz_2019, Ma_Birrell_2022, Nouwens_Liccardi_Veale_Karger_Kagal_2020, Kulyk_Hilt_Gerber_Volkamer_2018, Bouma-Sims_Li_Lin_Sakura-Lemessy_Nisenoff_Young_Birrell_Cranor_Habib_2023} and that it is difficult to encourage intentional interactions~\cite{bavelpriego, Gerber_Stöver_Peschke_Zimmermann_2024}, likely because the repetitive nature of these interfaces has habituated users into simply repeating the same interaction pattern (referred to as "consent fatigue")~\cite{bravo2014harder}, which early evidence suggests might already have developed after just two exposures to an interface~\cite{bielova2024effect}.

\vspace{5pt}

Ultimately, the statistical results of this study combined with the empirical insights from user studies demonstrate that some of the most common design choices of consent interfaces on the European web are oriented towards maximising the consent rate of people using deceptive designs.
% Whether this is because there is no reject button, or that reject button is designed to be visually less prominent than the accept option, or its text label scares users away from using it. 
% Or because there are no granular options to choose which data processing purposes a user might want to allow, or because they are already pre-selected by default. 
% As a result, the status quo of consent interfaces is currently still more likely than not to lead to unintended, unwanted, and illegal processing of people's personal data.

Further user studies on the effect of dominant designs are important, for example to compare how users from different countries experience them, or to go beyond individual features and instead study interfaces holistically (perhaps from specific CMPs).
To establish what designs are most prevalent, quantitative overviews such as this study provide the necessary empirical grounding. 
However, building scrapers and detection algorithms requires a high level of technical expertise and resources.
% These groups might lack the skills, time, or desire to develop their own detection tools, even if such information might be very relevant for their work (such as assessing the impact of enforcement actions).
To address this barrier, we developed \href{https://consent-observatory.eu/}{consent-observatory.eu}, an online scraping service powered by our open source detection code. 
This platform allows users to provide a list of URLs and choose which elements they want to analyse, making the study of consent interfaces more accessible to a broader audience.
\subsection{Untangling the influence of CMPs}
% are they controller or processor : santos et al + disconnect.me
% then>privacy/dp by default > force them to make the default design compliant (toth)
% processor liability: if they know an instruction is manifestly non-compliant, they need to tell the controller, and if they don't, dpa can go after them. But won't have massive effect
% processors that know for sure something is noncompliant: do they have .. by design obligations? technical and organisational measures? not clear in the GDPR, would have to take them to court > art 32?
% standard contractual clauses by the DPA? ofc can't enforce
% perhaps move the description of the different types of CMPs here and how they are more or less likely to be controllers
% can DPA request the contracts?
% IAB case means unlikely CMPs can claim not to process personal data (unless just a JS stub?)

The online tracking ecosystem is marked by a complex web of intermediaries between website owners who want to sell advertising space and advertisers who seek to promote products~\cite{vealeAdtechRealTimeBidding2022}. It includes tracking services, publishers, supply-side platforms, auction houses, data brokers, demand-side platforms, standard-setting bodies, and other ad-tech vendors.
% (see Figure \ref{fig:CMPposition}).
% Supply-side platforms help website owners sell their advertising space while demand-side platforms aggregate advertisers and their ads, with auction houses in the middle that try to dynamically and in real-time match these parties. 
% Tracking services help website owners collect information about their visitors' behaviour and generate unique identifiers, which are then passed along with the website's requests for ads to entice specific advertisers to bid on their space, and advertisers take this information and enrich it by contacting data brokers who have compiled complex profiles connected to those identifiers.
% Entangled throughout all of this are dominant infrastructural players who set standards, provide legal frameworks, and develop technical infrastructure to orchestrate the flows of data and commodify consent~\cite{Woods_Böhme_2022} (e.g., Google, Meta, Interactive Advertising Bureau, Digital Advertising Alliance).
Consent interfaces, often provided by consent management platforms, are supposed to be the main gateway through which the collection and processing of all the (personal) data is controlled by the user and brought into compliance.
Generally, CMPs can be divided into four types of organisations: 1) comprehensive CMPs that offer a variety of services (such as analytics and website scanning) across different devices and integrated with various platforms for a monthly subscription fee (e.g., Usercentrics, OneTrust); 2) plug-in CMPs with basic features that integrate with popular platforms such as WordPress (e.g. CookieYes, Complianz); 3) Integrated solutions developed by the platforms themselves (often eCommerce, e.g., Shoper, Shoptet), and; 4) open source CMPs that are basic scripts developed by entrepreneurial individuals (e.g., tarteaucitron, CookieConsent).
Our analysis shows that all these types of CMPs are present in the top three across countries.

Because there are so many actors involved in collecting and processing the (personal) data for online tracking, it becomes difficult to understand who is actually in control and how liability is distributed.
% It makes the tracking industry opaque, not only for outsiders (users, scholars, regulators) but also for the very participants within it. 
% Key knowledge such as who has what data about which user or the actual effectiveness of personal data for advertising conversion disappears in the cracks between the various players.
This has allowed many actors to plausibly dodge responsibility by arguing that, actually, someone else is in control.
% Website owners say they are trapped by the lack of competitive alternatives to this data-centric monetisation structure, advertisers claim they simply purchase a distribution service from adtech vendors, adtech vendors wilfully operate under the assumption that the frameworks they use but are established by others are legal, and 
The Internet Advertising Bureau (IAB), for example, a standard-setting body that provides technical infrastructure and legal frameworks for (among others) CMPs, has long denied they are a controller.
However, in European law, the concept of `joint controllership'~\cite{mahieuResponsibilityDataProtection2019} captures situations where more than one actor contributes to determining the how and why of a data processing practice, and therefore shares responsibility for it in some configuration. 
Case law has shown courts take an expansive view on the concept of joint controllership: it also includes situations where a controller never has access to the data but has influence on what and how other actors collect and process data downstream~\cite{jehovahs_witnesses_2018}.
The CJEU recently clarified that IAB was indeed not just a mere bystander, but a joint-controller because of the influence it exerted through its development of the Transparency and Consent Framework (TCF), which establishes guidelines and technical standards for obtaining, recording, and managing consent, effectively shaping how personal data was processed across participating websites~\cite{IABvsBelgium}.

Similarly, CMPs have traditionally held they simply provide a technological service and the website owner is in full control of the configuration and thus "determines the means and purposes" of the data collection.
Our results provide evidence that consent management platforms are in fact influential actors and have an impact on the compliance rate of consent interfaces.
Firstly, our analysis shows that CMPs are central to the online tracking industry: at least 67\% of all consent interfaces are provided by CMPs rather than built by website owners themselves.
The proportion of CMPs is particularly high on the most popular websites where web browsing is most heavily concentrated, meaning they intermediate a significant portion of all web traffic (the number one ranked website in a country typically represents 20\% of all page loads~\cite{Ruth_Fass_Azose_Pearson_Thomas_Sadowski_Durumeric_2022}).
Secondly, 37\% of the CMP use is concentrated in the hands of just three organisations: Usercentrics, CookieYes and OneTrust (i.e., 25\% of all consent interfaces, or 17\% of all websites).
This concentration is also present at a national scale, where in extreme cases three CMP intermediaries hold 64\% of the market. 
% TODO: check this claim with Luke?
Thirdly, there is a significant correlation between consent management platforms and compliance of the interface -- more than guidance or fines -- and specific CMPs have a sizeable positive or negative effect on compliance.

This evidence strengthens the case made by other studies that consent management platforms are not merely neutral facilitators.
\citet{Toth_Bielova_Roca_2022} show how CMPs often nudge their clients toward less compliant, data-maximizing configurations: the default designs when first configuring a CMP have no or less prominent reject buttons, they often only scan a website once a month for any new trackers for which consent is required, they make marketing claims about maximizing consent, and they provide A/B testing services to compare the consent rates of different interface designs.
% By their default configurations, CMPs exert direct influence on the designs of consent interfaces, the subsequent consent rates, and the other actors the data flows to.
\citet{santos2021consent} provide a techno-legal analysis and discuss how CMPs act as controllers when they add additional processing activities to their tools (for example, when they add their own tracking pixels), when they automatically sort installed trackers into specific purpose categories, and when they automatically include pre-populated lists of third-party vendors in the interface.
These dynamics raise important questions about the rule-taking nature of website owners, who may simply follow industry defaults rather than actively shaping their own compliance strategies, and the position of CMPs as joint-controllers rather than mere service providers.
% TODO: add disconnect.me report when it is public

Researchers and regulators must take an infrastructural perspective on the problem of online tracking. 
Going after a website might seem tempting, and can give a detailed understanding of what compliance looks like and what technologies are used, but such a whack-a-mole approach has limited transformative potential, and brings huge enforcement costs through investigation time and duplicated legal proceedings. 
Consent management platforms should be considered points of leverage, and viable targets for regulation in different ways.
The exact legal classification and strategy may change depending on the features and integration level of the CMP, as described at the start of this section.
As `joint controllers' co-deciding on data processing, CMPs can be targeted directly by regulators in the same jurisdiction.
As data processors (entities that only do what they are told), they have fewer legal obligations, but must still inform controllers of potential illegal instructions, which could be creatively interpreted and used.
Even as just openly published code (such as in free website templates), regulators could signal that they will use certain bad code as a flag for potential website or app investigation, disincentivising their use.
This study helps all these aims by highlighting which CMPs take a central role in what jurisdiction, identifying significant leverage points for regulatory action.
In particular, this is the case where there is either an internationally widespread CMP in a specific jurisdiction, or a highly prevalent national CMP.

\subsection{Beyond better interfaces}
More than two-thirds of European websites now have consent interfaces. 
A lot of effort has been spent, both by researchers and regulators, to try and improve consent interfaces by measuring their designs, demonstrating their impacts, evaluating their compliance, or proposing new and more empowering designs.
% New studies and guidelines are provided about yet another specific design element (the placement of the interface, the labels on the buttons, the meaning of a close button), and new legal bases proposed by adtech companies are evaluated and rejected~\cite{}.
Slowly, haltingly, over the course of 15 years, consent interface designs have evolved from purely informational notifications, to cookie walls without reject buttons, to banners with, and then without, pre-checked purposes~\cite{Nouwens_Liccardi_Veale_Karger_Kagal_2020}, to pop-ups with boldly coloured accept buttons besides ambiguous reject buttons, to interfaces where, even when consent is rejected, the same processing continues under the basis of legitimate interest~\cite{Kyi_2023}, and, most recently, the invention of consent-or-pay designs, where rejecting costs money~\cite{Morel_Santos_Lintao_Human_2022}.

While the focus has been on improving this most visible aspect of online tracking, studies show that
% while rejecting consent interfaces does reduce tracking~\cite{Jha_Trevisan_Vassio_Mellia_2022}, 
data is already collected and shared before users interact with the interface~\cite{papadogiannakis2021user, Sanchez-Rola_Dell’Amico_Kotzias_Balzarotti_Bilge_Vervier_Santos_2019}, that once given consent answers are sometimes ignored \cite{Bouhoula_Kubicek_Zac_Cotrini_Basin_2024}, and that attempts to revoke it later never propagate through the networks~\cite{vealeImpossibleAsksCan2022}.
It begs the question whether any improvements to the interface ultimately result in improvements to people's privacy and data protection, or whether this tug-of-war over the interface simply creates performative, surface-level consent theatre while the difficult to observe but structurally more consequential practices remain under-examined and unquestioned.

Caught up in the narrative that better interfaces are the answer,
we risk losing sight of the fact that disempowerment is not a design flaw, but an inherent feature.
The law does not require it--consent in EU law is a last-resort when no other legal basis can be established, not an important prerequisite.
Because online tracking cannot be legalised in any other way in Europe, the industry response to a \emph{de facto} legal prohibition has been to use design to stretch consent to and beyond its conceptual limits.
The ineffectiveness of enforcement compounds the problem. 
Guidance and fines have so far failed to foster substantial compliance, and the persistent frustration erodes public trust in the regulation and regulators.
A narrative has emerged in which the GDPR is blamed for both the rise of consent interfaces and the failure to eliminate them.
The public frustration has fuelled political calls for stronger measures, including proposals from the EDPB~\cite{EDPBBan} and members of the European Parliament~\cite{MEPban} to ban behavioural advertising entirely.
% and preliminary proposals by the European Commission to break up Google's various advertising services.

What such efforts need are not better interfaces, but counter narratives and alternative visions.
For example, recent work has evaluated the usability and proposed interfaces for browser-level consent interfaces and HTTP header signals~\cite{zimmeck2023usability, zimmeck2024generalizable}.
Future work could consider the performance and financial viability of contextual advertising (as demonstrated in at least the case of the Dutch public broadcaster~\cite{STER}), or critically examine the actual effectiveness of online tracking based targeting and personalisation, for which little independent evidence exists.
% TODO: add reference to DMA/DSA perhaps, since it will force transparency?
The goals of these studies should be to provide the necessary knowledge to immunise policy makers and citizens against unsubstantiated claims about the demise of free services, shift the window of discussion, and break out of the current myopic interpretation of what the Web can be.

\section{Conclusion}
Online tracking remains a controversial practice on the Web that, despite years of effort, legislators and regulators struggle to bring into legal compliance or moral alignment with their populations.
Consent interfaces are the most visible component of this industry and, as such, have received the most attention from researchers, journalists, politicians, and regulators. 
Over the years, automated methods to study these interfaces have improved.
Quantitative data on the prevalence of consent interfaces and whether they meet some metric of compliance have helped focus attention on specific designs and informed regulatory interventions.

In this paper we present robust methods to detect the presence, design, and suppliers of consent interfaces. We provide \href{https://consent-observatory.eu/}{consent-observatory.eu}: a web service that makes these methods accessible for a wider audience to further evaluate consent interfaces.
Using these methods, we scraped the top 10,000 most popular websites for 31 countries across Europe where the ePrivacy Directive and General Data Protection Regulation are in effect.
We detected the presence of consent interfaces, identified the use of consent management platforms (CMPs), and analysed the proportion and visual prominence of user options such as accept, reject, settings, save, and pay buttons and purpose-level controls such as checkboxes and their checked status.

We find that 67\% all websites now have consent interfaces, and that two-thirds of those are provided by CMPs.
In particular the most visited websites are likely to use CMPs, which means that the vast majority of web traffic flows through their products.
There is significant concentration in the use of CMPs: only three organisations supply 37\% of all CMPs used on the web.
Compliance rates of these interfaces continue to lag behind, and there is little effect of guidance and fines on compliance.
Only 15\% of interfaces meet our very limited definition of compliance, which is as a banner where accepting is as easy as rejecting, there are granular consent options, and there are no pre-checked purposes. 
Most non-compliance stems from not having a reject option on the first layer, and most of the variance in compliance is explained by the use of particular CMPs.

We suggest that future studies on consent interfaces that might use \href{https://consent-observatory.eu/}{consent-observatory.eu} should be careful not to think of non-compliance and disempowerment as an unintentional design problem that better interfaces can resolve.
Instead, studies should map the different actors in this space and their relative power, with an eye to provide insight into the leverage points for impactful action.

% \section{Acknowledgments}
% \begin{acks}
% \end{acks}

\bibliographystyle{ACM-Reference-Format}
\bibliography{sample-base}

\clearpage  
\appendix
%TC:ignore
\section*{Appendices}
\section{Top lists}\label{toplists}
The list of top 10,000 domains scraped for each country can be found at the following permanent links:
\begin{table}[h]
\setlength{\extrarowheight}{1pt}
\begin{tabular}{ll}
\toprule
& \textbf{Permalink to domain list} \\
\midrule
\ \ \textbf{Austria} & \url{https://tranco-list.eu/list/N3YKW/10000} \\ \cdashline{1-2}[0.5pt/2pt]
\ \ \textbf{Belgium} & \url{https://tranco-list.eu/list/7X7VX/10000} \\ \cdashline{1-2}[0.5pt/2pt]
\ \ \textbf{Bulgaria} & \url{https://tranco-list.eu/list/LJ8Q4/10000} \\ \cdashline{1-2}[0.5pt/2pt]
\ \ \textbf{Croatia} & \url{https://tranco-list.eu/list/KJ3QW/10000} \\ \cdashline{1-2}[0.5pt/2pt]
\ \ \textbf{Cyprus} & \url{https://tranco-list.eu/list/83K4V/10000} \\ \cdashline{1-2}[0.5pt/2pt]
\ \ \textbf{Czech Republic} & \url{https://tranco-list.eu/list/G6KJK/10000} \\ \cdashline{1-2}[0.5pt/2pt]
\ \ \textbf{Denmark} & \url{https://tranco-list.eu/list/XJL6N/10000} \\ \cdashline{1-2}[0.5pt/2pt]
\ \ \textbf{Estonia} & \url{https://tranco-list.eu/list/24PZ9/10000} \\ \cdashline{1-2}[0.5pt/2pt]
\ \ \textbf{Finland} & \url{https://tranco-list.eu/list/4QPWX/10000} \\ \cdashline{1-2}[0.5pt/2pt]
\ \ \textbf{France} & \url{https://tranco-list.eu/list/58L7N/10000} \\ \cdashline{1-2}[0.5pt/2pt]
\ \ \textbf{Greece} & \url{https://tranco-list.eu/list/3N5LL/10000} \\ \cdashline{1-2}[0.5pt/2pt]
\ \ \textbf{Germany} & \url{https://tranco-list.eu/list/PNX9J/10000} \\ \cdashline{1-2}[0.5pt/2pt]
\ \ \textbf{Hungary} & \url{https://tranco-list.eu/list/66WKX/10000} \\ \cdashline{1-2}[0.5pt/2pt]
\ \ \textbf{Iceland} & \url{https://tranco-list.eu/list/83KGV/10000} \\ \cdashline{1-2}[0.5pt/2pt]
\ \ \textbf{Ireland} & \url{https://tranco-list.eu/list/PNXQJ/10000} \\ \cdashline{1-2}[0.5pt/2pt]
\ \ \textbf{Italy} & \url{https://tranco-list.eu/list/YXWNG/10000} \\ \cdashline{1-2}[0.5pt/2pt]
\ \ \textbf{Latvia} & \url{https://tranco-list.eu/list/93VX2/10000} \\ \cdashline{1-2}[0.5pt/2pt]
\ \ \textbf{Liechtenstein} & \url{https://tranco-list.eu/list/N3YQW/10000} \\ \cdashline{1-2}[0.5pt/2pt]
\ \ \textbf{Lithuania} & \url{https://tranco-list.eu/list/QGJQ4/10000} \\ \cdashline{1-2}[0.5pt/2pt]
\ \ \textbf{Luxembourg} & \url{https://tranco-list.eu/list/J93QY/10000} \\ \cdashline{1-2}[0.5pt/2pt]
\ \ \textbf{Malta} & \url{https://tranco-list.eu/list/LJ8P4/10000} \\ \cdashline{1-2}[0.5pt/2pt]
\ \ \textbf{Netherlands} & \url{https://tranco-list.eu/list/4QPJX/10000} \\ \cdashline{1-2}[0.5pt/2pt]
\ \ \textbf{Norway} & \url{https://tranco-list.eu/list/W8VY9/10000} \\ \cdashline{1-2}[0.5pt/2pt]
\ \ \textbf{Poland} & \url{https://tranco-list.eu/list/7X72X/10000} \\ \cdashline{1-2}[0.5pt/2pt]
\ \ \textbf{Portugal} & \url{https://tranco-list.eu/list/58LZN/10000} \\ \cdashline{1-2}[0.5pt/2pt]
\ \ \textbf{Romania} & \url{https://tranco-list.eu/list/KJ3PW/10000} \\ \cdashline{1-2}[0.5pt/2pt]
\ \ \textbf{Slovakia} & \url{https://tranco-list.eu/list/V9KYN/10000} \\ \cdashline{1-2}[0.5pt/2pt]
\ \ \textbf{Slovenia} & \url{https://tranco-list.eu/list/G6KNK/10000} \\ \cdashline{1-2}[0.5pt/2pt]
\ \ \textbf{Spain} & \url{https://tranco-list.eu/list/XJLQN/10000} \\ \cdashline{1-2}[0.5pt/2pt]
\ \ \textbf{Sweden} & \url{https://tranco-list.eu/list/Z378G/10000} \\ \cdashline{1-2}[0.5pt/2pt]
\ \ \textbf{United Kingdom} & \url{https://tranco-list.eu/list/3N54L/10000} \\
\bottomrule
\end{tabular}\label{permalinks}
\end{table}

\clearpage  

\onecolumn
\section{Consent Interface Detection Corpus}\label{popupcorpus}
The words used to check potential consent interfaces (based on heuristics) and determine whether it is indeed a consent interface.
\setlength{\extrarowheight}{1pt}
\begin{longtable}{p{3cm}p{11cm}}
% \begin{tabular}{p{3cm}p{11cm}}
\toprule
& \textbf{Trigger words} \\
\midrule
\endfirsthead

\toprule
& \textbf{Trigger words} \\
\midrule
 \endhead

\ \ International & cookie, cookies, gdpr \\ \cdashline{1-2}[0.5pt/2pt]
\ \ Austria & alle akzeptieren, einstellungen verwalten, zwecke anzeigen, ablehnen, datenschutzerklärung \\ \cdashline{1-2}[0.5pt/2pt]
\ \ Belgium & accepter, en savoir plus, akkoord, meer informatie, alle cookies aanvaarden, paramètres, accepteren, d'accord \\ \cdashline{1-2}[0.5pt/2pt]
\ \ Bulgaria &\foreignlanguage{russian}{политика за поверителност, приемане, затваряне, настройки, отхвърли всички, приеми всички, научете повече, приемане и затваряне, приемам, към сайта, опции за управление, подробни настройки, продължи, бисквитки, бисквитките, приемете, политика за защита на личните данни, политика за бисквитките, съгласие, научете повече, политика за използване на бисквитки, други възможности, приемате, декларацията за поверителност, съгласявате, персонализираме съдържанието}\\ \cdashline{1-2}[0.5pt/2pt]
\ \ Croatia & prihvati i zatvori, prihvaćam, saznaj više, saznajte više, prihvati sve kolačiće, prihvaćam sve, postavke, postavke kolačića, slažem se, pogledajte naše partnere, upravljanje opcijama, ne prihvaćam, više informacija, politika privatnosti, pravila privatnosti, odbaci sve, prihvati i zatvori, prihvati, na stranicu, opcije za upravljanje, detaljne postavke, nastavi, kolačići, kolačići, prihvati, pravila o kolačićima, pristanak, pravila o kolačićima, druge opcije, prihvaćam, izjava o privatnosti, slažem se, prilagodite sadržaj \\ \cdashline{1-2}[0.5pt/2pt]
\ \ Cyprus & \foreignlanguage{greek}{αποδοχη ολων, διαδοχη ολων, απορριψη ολων, συμφωνω, ρυθμίσεις cookies, αποδοχή όλων, διαφωνω, προτιμησεις, πολιτική απορρήτου} \\ \cdashline{1-2}[0.5pt/2pt]
\ \ Czech Republic & podrobné nastavení, povolit vše, souhlasím, odmítnout, rozumím, povolit nezbytné, další volby, přijmout vše, upravit mé předvolby, nastavení, zásady ochrany osobních údajů \\ \cdashline{1-2}[0.5pt/2pt]
\ \ Germany & datenschutz, akzeptieren, stimme zu, zustimmen, berechtigtes interesse, privatsphäre \\ \cdashline{1-2}[0.5pt/2pt]
\ \ Denmark & privatliv, samtykke, acceptér, tillad, legitim interesse \\ \cdashline{1-2}[0.5pt/2pt]
\ \ Estonia & nõustun, keeldu, luba kõik, kohanda, küpsiste seaded, küpsiste sätted, küpsised, nõustu, halda, privaatsus, küpsiseid, küpsistega, küpsistest, privaatsuspoliitika, sulge, seaded, rohkem teavet, keeldun, kuva eesmärgid, muudan küpsiste seadistusi, küpsiste seadetega, sain aru, loen veel, privaatsuspõhimõtete, nõustun kõigi küpsistega, selge, lisainfo, isikupärastamiseks, isikupärastatud, isikupärasem, seaded, tingimused, tingimustega, seadistusi \\ \cdashline{1-2}[0.5pt/2pt]
\ \ England & privacy, consent, accept, agree, legitimate interest \\ \cdashline{1-2}[0.5pt/2pt]
\ \ Spain & privacidad, acept, acceptar, acordar, interés legítimo \\ \cdashline{1-2}[0.5pt/2pt]
\ \ Finland & evästeitä, evästeiden, tietosuoja, hyväksy, hylkää, asetukset, suostumustasi, suostumuksesi \\ \cdashline{1-2}[0.5pt/2pt]
\ \ France & confidentialité, accepter, accord, intérêt légitime \\ \cdashline{1-2}[0.5pt/2pt]
\ \ Greece & \foreignlanguage{greek}{περισσοτερες επιλογες, συμφωνω, διαφωνω, αποδοχη, απορριψη, περισσότερα, απορρητην, πολιτική απορρήτου}\\ \cdashline{1-2}[0.5pt/2pt]
\ \ Hungary & cookie-kat, elfogadom, további opciók, nem elfogadom, további információ, elfogadás és bezárás, beállítások, beállítások kezelése, hozzájárulás, összes engedélyezése, mindent elfogadok,  adatvédelmi szabályzat, elfogadás, adatvédelmi szabályzat, sütik, az ön adatainak védelme fontos számunkra, tartalom testreszabása, lehetőségek, további lehetőségek, részletek, cookie-k, információ, cookie-szabályzat, kapcsolódó sütikkel kapcsolatos információk \\ \cdashline{1-2}[0.5pt/2pt]
\ \ Iceland & vefkökur, kökur, vafrakökur, samþykkja, hafna, vefköku stillingar, leyfa, vista val, fótspor \\ \cdashline{1-2}[0.5pt/2pt]
\ \ Ireland & fianáin, cuacha, lean ar aghaidh, cosanta sonraí, socruithe fianán, glac le gach fianán, diúltú neamhriachtanach, bainistigh fianáin \\ \cdashline{1-2}[0.5pt/2pt]
\ \ Italy & politica, consenso, accetta, concordare, interesse legittimo \\ \cdashline{1-2}[0.5pt/2pt]
\ \ Latvia & 
\begingroup
\fontencoding{T2A}\selectfont
piekrītu, pielāgot, papildu opcijas, uzzināt vairāk, atļaut visas sīkdatnes, apstiprināt, pārvaldības iespējas, apstiprināt, pārvaldības iespējas, согласен, nepiekriţu,\foreignlanguage{russian}{дополнительные параметры}, privātuma politika, piekrist, aizvērt, iestatījumi, noraidīt visu, pieņemt visu, uzzināt vairāk, pieņemt un aizvērt, piekrist, opcijas pārvaldība, detalizēti iestatījumi, turpināt, sīkfaili, pieņemt, piekrišana, uzzināt vairāk, sīkfailu politika, cits opcijas, es piekrītu, paziņojums par konfidencialitāti, es piekrītu, pielāgot saturu 
\endgroup
\\ \cdashline{1-2}[0.5pt/2pt]
\ \ Lithuania & sutinku, tvarkyti parinktis, leisti visus slapukus, daugiau pasirinkimų, atsisakyti visų, supratau, slapukų nustatymai, sutikimas, rodyti informaciją, patvirtinti, privatumo politika, rinktis, slapuku politikoje, nesutinku, tinkinti, priimti, slapukai, slapukų politika, privatumo pareiškimas, nustatymai, rodyti paskirtis, privatumas, slapukuose, tvarkyti parinktis, slapuku politikoje, nuostatos, rinkodara, slapukus \\ \cdashline{1-2}[0.5pt/2pt]
\ \ Luxembourg & j'accepte, je refuse, gérer les cookies, paramètres des cookies, accepter tout, afficher toutes les finalités, privatsphär \\ \cdashline{1-2}[0.5pt/2pt]
\ \ Malta & il-privatezza, il-cookies, tal-cookies, naqbel, naċċetta, aktar dwar il cookies, aċċetta, irrifjuta \\ \cdashline{1-2}[0.5pt/2pt]
\ \ Netherlands & accepteren, afwijzen, akkoord, instellen, toestemming, privacy-instellingen, instellingen, cookiebeleid, privacyverklaring \\ \cdashline{1-2}[0.5pt/2pt]
\ \ Norway & informasjonskapsler, personvern, godta, avvis \\ \cdashline{1-2}[0.5pt/2pt]
\ \ Poland & plików, plikach, akceptuję, odrzucenie wszystkich, zaakceptuj, ordzuć, prwatność \\ \cdashline{1-2}[0.5pt/2pt]
\ \ Portugal & privacidade, consentimento, aceitar, concordo, interesse legítimo \\ \cdashline{1-2}[0.5pt/2pt]
\ \ Romania & cookie-uri, accept toate, vreau sa modific setarile individual, modific setările, mai multe opțiuni, respinge toate, gestionajți opțiunile, consimțământ, setari cookie-uri, setări cookies, politica de confidențialitate \\ \cdashline{1-2}[0.5pt/2pt]
\ \ Slovakia & pokračovať s nevyhnutnými cookies, nastavenia, súhlasím, prijať všetko, akceptovať, zamietnuť, nastavenie cookies, nastavenia cookies, ďalšie informácie, bližšie informácie, zásady ochrany osobných údajov \\ \cdashline{1-2}[0.5pt/2pt]
\ \ Slovenia & strinjam se, več možnosti, nastavitve, sprejmi, sprejmem, ne strinjam se, nastavitve piškotov, sprejmem vse, dovoli vse in zapri, prilagodi, politika zasebnosti, zavrni vse, namesti vse, po meri, vi redu, razumem, piškotkov, piškotke, piškotki, piškotkih \\ \cdashline{1-2}[0.5pt/2pt]
\ \ Sweden & acceptera, godkänn, kakor \\
\bottomrule
% \end{tabular}\label{popupcorpus}
\end{longtable}

\clearpage 

\section{CMP CSS selectors}\label{selectors}
% \begin{table}
\setlength{\extrarowheight}{1pt}
\begin{longtable}{ll}

\toprule
\textbf{Regex pattern} & \textbf{CMP provider} \\
\midrule
\endfirsthead

\toprule
\textbf{Regex pattern} & \textbf{CMP provider} \\
\midrule
\endhead

\ \ \code{eightworks-cookie-consent} & 8works \\ \cdashline{1-2}[0.5pt/2pt]
\ \ \code{acris-cookie-settings} & Acris \\ \cdashline{1-2}[0.5pt/2pt]
\ \ \code{amgdprcookie-button} & Amasty (Magento plugin) \\ \cdashline{1-2}[0.5pt/2pt]
\ \ \code{\^{}axeptio} & Axeptio \\ \cdashline{1-2}[0.5pt/2pt]
\ \ \code{\^{}borlabs} & Borlabs \\ \cdashline{1-2}[0.5pt/2pt]
\ \ \code{ccm-modal-inner} & CCM19 \\ \cdashline{1-2}[0.5pt/2pt]
\ \ \code{ccc-overlay} & CIVIC \\ \cdashline{1-2}[0.5pt/2pt]
\ \ \code{cmplz} & Complianz \\ \cdashline{1-2}[0.5pt/2pt]
\ \ \code{cs-privacy-content-text} & Consent Magic \\ \cdashline{1-2}[0.5pt/2pt]
\ \ \code{cb-enable} & Cookie Bar (generic script adapted by many) \\ \cdashline{1-2}[0.5pt/2pt]
\ \ \code{cookieinfo-close} & Cookie Info Script \\ \cdashline{1-2}[0.5pt/2pt]
\ \ \code{\^{}coi} & Cookie Information \\ \cdashline{1-2}[0.5pt/2pt]
\ \ \code{cn-notice-text} & Cookie Notice \\ \cdashline{1-2}[0.5pt/2pt]
\ \ \code{\^{}cookiescript\_} & Cookie-Script \\ \cdashline{1-2}[0.5pt/2pt]
\ \ \code{cm\_\_desc} & CookieConsent \\ \cdashline{1-2}[0.5pt/2pt]
\ \ \code{cf3E9g} & CookieFirst \\ \cdashline{1-2}[0.5pt/2pt]
\ \ \code{cookiecontent} & CookieHint \\ \cdashline{1-2}[0.5pt/2pt]
\ \ \code{\^{}ch2-} & CookieHub \\ \cdashline{1-2}[0.5pt/2pt]
\ \ \code{-cli-} & CookieYes \\ \cdashline{1-2}[0.5pt/2pt]
\ \ \code{cookie-law-info-bar} & CookieYes \\ \cdashline{1-2}[0.5pt/2pt]
\ \ \code{cookie\_action\_close\_header} & CookieYes \\ \cdashline{1-2}[0.5pt/2pt]
\ \ \code{\^{}cky} & CookieYes \\ \cdashline{1-2}[0.5pt/2pt]
\ \ \code{didomi} & Didomi \\ \cdashline{1-2}[0.5pt/2pt]
\ \ \code{\^{}CookieReports} & Digital Control Room \\ \cdashline{1-2}[0.5pt/2pt]
\ \ \code{cc-cookie-accept} & Django Cookie Consent \\ \cdashline{1-2}[0.5pt/2pt]
\ \ \code{eu-cookie-compliance-categories} & Drupal \\ \cdashline{1-2}[0.5pt/2pt]
\ \ \code{pea\_cook\_btn} & FireCask (formerly Peadig) \\ \cdashline{1-2}[0.5pt/2pt]
\ \ \code{\_\_gomagCookiePolicy} & Gomag \\ \cdashline{1-2}[0.5pt/2pt]
\ \ \code{\^{}hs-en-cookie-} & HubSpot \\ \cdashline{1-2}[0.5pt/2pt]
\ \ \code{gdpr-cookie-accept} & I Have Cookies by Ketan Mistry \\ \cdashline{1-2}[0.5pt/2pt]
\ \ \code{iqitcookielaw} & IQIT commerce \\ \cdashline{1-2}[0.5pt/2pt]
\ \ \code{iai\_cookie} & IdoSell \\ \cdashline{1-2}[0.5pt/2pt]
\ \ \code{qc-cmp2-ui} & InMobi \\ \cdashline{1-2}[0.5pt/2pt]
\ \ \code{cookie-settings-necessary} & Jimdo \\ \cdashline{1-2}[0.5pt/2pt]
\ \ \code{id-cookie-notice} & Klaro \\ \cdashline{1-2}[0.5pt/2pt]
\ \ \code{\^{}moove-gdpr} & Moove \\ \cdashline{1-2}[0.5pt/2pt]
\ \ \code{cookie-notification-text} & Mozello CookieBar \\ \cdashline{1-2}[0.5pt/2pt]
\ \ \code{ot-sdk-container} & OneTrust \\ \cdashline{1-2}[0.5pt/2pt]
\ \ \code{onetrust} & OneTrust \\ \cdashline{1-2}[0.5pt/2pt]
\ \ \code{optanon} & OneTrust \\ \cdashline{1-2}[0.5pt/2pt]
\ \ \code{cc-window} & Osano \\ \cdashline{1-2}[0.5pt/2pt]
\ \ \code{cc\_container} & Osano \\ \cdashline{1-2}[0.5pt/2pt]
\ \ \code{cookieconsent:desc} & Osano \\ \cdashline{1-2}[0.5pt/2pt]
\ \ \code{osano} & Osano \\ \cdashline{1-2}[0.5pt/2pt]
\ \ \code{\^{}ppms\_cm} & Piwik \\ \cdashline{1-2}[0.5pt/2pt]
\ \ \code{sf-cookie-settings} & Serviceform \\ \cdashline{1-2}[0.5pt/2pt]
\ \ \code{consents\_\_advanced-buttons} & Shoper \\ \cdashline{1-2}[0.5pt/2pt]
\ \ \code{shopify-pc\_\_banner} & Shopify \\ \cdashline{1-2}[0.5pt/2pt]
\ \ \code{nanobar-buttons} & Shoprenter \\\cdashline{1-2}[0.5pt/2pt] 
\ \ \code{siteCookies} & Shoptet \\ \cdashline{1-2}[0.5pt/2pt]
\ \ \code{page-wrap--cookie-permission} & Shopware \\ \cdashline{1-2}[0.5pt/2pt]
\ \ \code{cookie-permission--container} & Shopware \\ \cdashline{1-2}[0.5pt/2pt]
\ \ \code{cookie-consent--header} & Shopware \\ \cdashline{1-2}[0.5pt/2pt]
\ \ \code{sp\_message\_container} & Sourcepoint \\ \cdashline{1-2}[0.5pt/2pt]
\ \ \code{sqs-cookie-banner-v2-cta} & Squarespace \\ \cdashline{1-2}[0.5pt/2pt]
\ \ \code{termly} & Termly \\ \cdashline{1-2}[0.5pt/2pt]
\ \ \code{cc\_div} & TermsFeed \\ \cdashline{1-2}[0.5pt/2pt]
\ \ \code{cc-nb-text} & TermsFeed \\ \cdashline{1-2}[0.5pt/2pt]
\ \ \code{\^{}truste} & TrustArc \\ \cdashline{1-2}[0.5pt/2pt]
\ \ \code{ct-ultimate-gdpr-} & Unidentified CMP (possibly createIT) \\ \cdashline{1-2}[0.5pt/2pt]
\ \ \code{w-cookie-modal} & Unidentified CMP 001 \\ \cdashline{1-2}[0.5pt/2pt]
\ \ \code{bemCookieOverlay} & Unidentified CMP 002 \\ \cdashline{1-2}[0.5pt/2pt]
\ \ \code{consents\_\_wrapper} & Unidentified CMP 003 \\ \cdashline{1-2}[0.5pt/2pt]
\ \ \code{\^{}cookie-policy-overlay} & Unidentified CMP 004 \\ \cdashline{1-2}[0.5pt/2pt]
\ \ \code{\^{}cookie-policy-details} & Unidentified CMP 005 \\ \cdashline{1-2}[0.5pt/2pt]
\ \ \code{\^{}popup-text\$} & Unidentified CMP 006 \\ \cdashline{1-2}[0.5pt/2pt]
\ \ \code{lgcookieslaw} & Unidentified CMP 007 \\ \cdashline{1-2}[0.5pt/2pt]
\ \ \code{module-notification-137} & Unidentified CMP 008 \\ \cdashline{1-2}[0.5pt/2pt]
\ \ \code{cookieNoticeContent} & Unidentified CMP 009 \\ \cdashline{1-2}[0.5pt/2pt]
\ \ \code{cNkVwm} & Usercentrics \\ \cdashline{1-2}[0.5pt/2pt]
\ \ \code{CybotCookiebot} & Usercentrics \\ \cdashline{1-2}[0.5pt/2pt]
\ \ \code{\^{}usercentrics} & Usercentrics \\ \cdashline{1-2}[0.5pt/2pt]
\ \ \code{\^{}uc-heading-title} & Usercentrics \\ \cdashline{1-2}[0.5pt/2pt]
\ \ \code{ccsu-banner-text-container} & Wix \\ \cdashline{1-2}[0.5pt/2pt]
\ \ \code{consent-banner-root-container} & Wix \\ \cdashline{1-2}[0.5pt/2pt]
\ \ \code{fusion-privacy-bar} & WordPress Theme Avada \\ \cdashline{1-2}[0.5pt/2pt]
\ \ \code{avia-cookie-} & WordPress Theme Enfold \\ \cdashline{1-2}[0.5pt/2pt]
\ \ \code{flatsome-cookies} & WordPress Theme Flatsome \\ \cdashline{1-2}[0.5pt/2pt]
\ \ \code{wd-cookies-inner} & WordPress Theme WoodMart by xtemos \\ \cdashline{1-2}[0.5pt/2pt]
\ \ \code{cmpwelcomebtnsave} & consentmanager.net \\ \cdashline{1-2}[0.5pt/2pt]
\ \ \code{cmpbox} & consentmanager.net \\ \cdashline{1-2}[0.5pt/2pt]
\ \ \code{\^{}cookiesplus} & idnovate \\ \cdashline{1-2}[0.5pt/2pt]
\ \ \code{iubenda} & iubenda \\ \cdashline{1-2}[0.5pt/2pt]
\ \ \code{eupopup-body} & jQuery EU Cookie Law popup by wimagguc \\ \cdashline{1-2}[0.5pt/2pt]
\ \ \code{tarteaucitron} & tarteaucitron \\
\bottomrule
\end{longtable}\label{CSSselectors}
% \end{table}

\clearpage  

\section{Detected consent management platforms}\label{AllCMPs}
\begin{longtable}{@{}llll@{}}
% \begin{tabular}{@{}llll@{}}
\toprule
\textbf{CMP} & \textbf{count} & \textbf{\%} & \textbf{rank} \\ 
 \midrule
\endfirsthead

\toprule
\textbf{CMP} & \textbf{count} & \textbf{\%} & \textbf{rank} \\ 
 \midrule
\endhead
 
\ \ Usercentrics & 19549 & 17.26 & 1 \\ \cdashline{1-4}[0.5pt/2pt]
\ \ CookieYes & 14102 & 12.45 & 2 \\ \cdashline{1-4}[0.5pt/2pt]
\ \ OneTrust & 9101 & 8.03 & 3 \\ \cdashline{1-4}[0.5pt/2pt]
\ \ Osano & 8037 & 7.10 & 4 \\ \cdashline{1-4}[0.5pt/2pt]
\ \ Complianz & 6444 & 5.69 & 5 \\ \cdashline{1-4}[0.5pt/2pt]
\ \ Cookie Notice & 5090 & 4.49 & 6 \\ \cdashline{1-4}[0.5pt/2pt]
\ \ TermsFeed & 4164 & 3.68 & 7 \\ \cdashline{1-4}[0.5pt/2pt]
\ \ Cookie Information & 3556 & 3.14 & 8 \\ \cdashline{1-4}[0.5pt/2pt]
\ \ Cookie-Script & 3502 & 3.09 & 9 \\ \cdashline{1-4}[0.5pt/2pt]
\ \ Moove & 2990 & 2.64 & 10 \\ \cdashline{1-4}[0.5pt/2pt]
\ \ InMobi & 2678 & 2.36 & 11 \\ \cdashline{1-4}[0.5pt/2pt]
\ \ Didomi & 2208 & 1.95 & 12 \\ \cdashline{1-4}[0.5pt/2pt]
\ \ iubenda & 2143 & 1.89 & 13 \\ \cdashline{1-4}[0.5pt/2pt]
\ \ Google & 1992 & 1.76 & 14 \\ \cdashline{1-4}[0.5pt/2pt]
\ \ consentmanager.net & 1830 & 1.62 & 15 \\ \cdashline{1-4}[0.5pt/2pt]
\ \ CookieConsent & 1573 & 1.39 & 16 \\ \cdashline{1-4}[0.5pt/2pt]
\ \ Shopify & 1431 & 1.26 & 17 \\ \cdashline{1-4}[0.5pt/2pt]
\ \ Shoptet & 1372 & 1.21 & 18 \\ \cdashline{1-4}[0.5pt/2pt]
\ \ CookieFirst & 1192 & 1.05 & 19 \\ \cdashline{1-4}[0.5pt/2pt]
\ \ tarteaucitron & 1153 & 1.02 & 20 \\ \cdashline{1-4}[0.5pt/2pt]
\ \ Sourcepoint & 1146 & 1.01 & 21 \\ \cdashline{1-4}[0.5pt/2pt]
\ \ Borlabs & 1069 & 0.94 & 22 \\ \cdashline{1-4}[0.5pt/2pt]
\ \ CookieHub & 1068 & 0.94 & 23 \\ \cdashline{1-4}[0.5pt/2pt]
\ \ TrustArc & 589 & 0.52 & 24 \\ \cdashline{1-4}[0.5pt/2pt]
\ \ Drupal & 576 & 0.51 & 25 \\ \cdashline{1-4}[0.5pt/2pt]
\ \ I Have Cookies by Ketan Mistry & 576 & 0.51 & 26 \\ \cdashline{1-4}[0.5pt/2pt]
\ \ Unidentified CMP 007 & 546 & 0.48 & 27 \\ \cdashline{1-4}[0.5pt/2pt]
\ \ Amasty (Magento plugin) & 534 & 0.47 & 28 \\ \cdashline{1-4}[0.5pt/2pt]
\ \ Cookie Bar (generic script adapted by many) & 501 & 0.44 & 29 \\ \cdashline{1-4}[0.5pt/2pt]
\ \ Axeptio & 499 & 0.44 & 30 \\ \cdashline{1-4}[0.5pt/2pt]
\ \ CIVIC & 499 & 0.44 & 31 \\ \cdashline{1-4}[0.5pt/2pt]
\ \ Wix & 498 & 0.44 & 32 \\ \cdashline{1-4}[0.5pt/2pt]
\ \ Shoper & 493 & 0.44 & 33 \\ \cdashline{1-4}[0.5pt/2pt]
\ \ WordPress Theme WoodMart by xtemos & 447 & 0.39 & 34 \\ \cdashline{1-4}[0.5pt/2pt]
\ \ idnovate & 442 & 0.39 & 35 \\ \cdashline{1-4}[0.5pt/2pt]
\ \ Mozello CookieBar & 389 & 0.34 & 36 \\ \cdashline{1-4}[0.5pt/2pt]
\ \ Unidentified CMP 009 & 377 & 0.33 & 37 \\ \cdashline{1-4}[0.5pt/2pt]
\ \ Django Cookie Consent & 372 & 0.33 & 38 \\ \cdashline{1-4}[0.5pt/2pt]
\ \ IdoSell & 369 & 0.33 & 39 \\ \cdashline{1-4}[0.5pt/2pt]
\ \ SIRDATA & 355 & 0.31 & 40 \\ \cdashline{1-4}[0.5pt/2pt]
\ \ Gomag & 352 & 0.31 & 41 \\ \cdashline{1-4}[0.5pt/2pt]
\ \ CCM19 & 326 & 0.29 & 42 \\ \cdashline{1-4}[0.5pt/2pt]
\ \ Squarespace & 324 & 0.29 & 43 \\ \cdashline{1-4}[0.5pt/2pt]
\ \ WordPress Theme Avada & 306 & 0.27 & 44 \\ \cdashline{1-4}[0.5pt/2pt]
\ \ Objectis & 301 & 0.27 & 45 \\ \cdashline{1-4}[0.5pt/2pt]
\ \ WordPress Theme Flatsome & 295 & 0.26 & 46 \\ \cdashline{1-4}[0.5pt/2pt]
\ \ Clickio & 294 & 0.26 & 47 \\ \cdashline{1-4}[0.5pt/2pt]
\ \ Cookie Info Script & 278 & 0.25 & 48 \\ \cdashline{1-4}[0.5pt/2pt]
\ \ Unidentified CMP 008 & 270 & 0.24 & 49 \\ \cdashline{1-4}[0.5pt/2pt]
\ \ IQIT commerce & 257 & 0.23 & 50 \\ \cdashline{1-4}[0.5pt/2pt]
\ \ Unidentified CMP 004 & 255 & 0.23 & 51 \\ \cdashline{1-4}[0.5pt/2pt]
\ \ CookieHint & 248 & 0.22 & 52 \\ \cdashline{1-4}[0.5pt/2pt]
\ \ jQuery EU Cookie Law popup by wimagguc & 246 & 0.22 & 53 \\ \cdashline{1-4}[0.5pt/2pt]
\ \ WordPress Theme Enfold & 242 & 0.21 & 54 \\ \cdashline{1-4}[0.5pt/2pt]
\ \ Shopware & 232 & 0.20 & 55 \\ \cdashline{1-4}[0.5pt/2pt]
\ \ Piwik & 227 & 0.20 & 56 \\ \cdashline{1-4}[0.5pt/2pt]
\ \ Unidentified CMP 002 & 219 & 0.19 & 57 \\ \cdashline{1-4}[0.5pt/2pt]
\ \ FireCask (formerly Peadig) & 217 & 0.19 & 58 \\ \cdashline{1-4}[0.5pt/2pt]
\ \ Consent Magic & 216 & 0.19 & 59 \\ \cdashline{1-4}[0.5pt/2pt]
\ \ Unidentified CMP 001 & 215 & 0.19 & 60 \\ \cdashline{1-4}[0.5pt/2pt]
\ \ Termly & 214 & 0.19 & 61 \\ \cdashline{1-4}[0.5pt/2pt]
\ \ Unidentified CMP 005 & 210 & 0.19 & 62 \\ \cdashline{1-4}[0.5pt/2pt]
\ \ Ezoic & 176 & 0.16 & 63 \\ \cdashline{1-4}[0.5pt/2pt]
\ \ HubSpot & 166 & 0.15 & 64 \\ \cdashline{1-4}[0.5pt/2pt]
\ \ Gravito & 159 & 0.14 & 65 \\ \cdashline{1-4}[0.5pt/2pt]
\ \ Traffective & 142 & 0.13 & 66 \\ \cdashline{1-4}[0.5pt/2pt]
\ \ Mediavine Inc. & 132 & 0.12 & 67 \\ \cdashline{1-4}[0.5pt/2pt]
\ \ Jimdo & 125 & 0.11 & 68 \\ \cdashline{1-4}[0.5pt/2pt]
\ \ Seznam & 103 & 0.09 & 69 \\ \cdashline{1-4}[0.5pt/2pt]
\ \ AppConsent by SFBX & 97 & 0.09 & 70 \\ \cdashline{1-4}[0.5pt/2pt]
\ \ Serviceform & 85 & 0.08 & 71 \\ \cdashline{1-4}[0.5pt/2pt]
\ \ Klaro & 71 & 0.06 & 72 \\ \cdashline{1-4}[0.5pt/2pt]
\ \ Setupad & 70 & 0.06 & 73 \\ \cdashline{1-4}[0.5pt/2pt]
\ \ Transfon & 68 & 0.06 & 74 \\ \cdashline{1-4}[0.5pt/2pt]
\ \ Digital Control Room & 60 & 0.05 & 75 \\ \cdashline{1-4}[0.5pt/2pt]
\ \ DPG Media & 56 & 0.05 & 76 \\ \cdashline{1-4}[0.5pt/2pt]
\ \ Tri-table Sp. z o.o. & 52 & 0.05 & 77 \\ \cdashline{1-4}[0.5pt/2pt]
\ \ NextRoll & 47 & 0.04 & 78 \\ \cdashline{1-4}[0.5pt/2pt]
\ \ FastCMP & 43 & 0.04 & 79 \\ \cdashline{1-4}[0.5pt/2pt]
\ \ Pubtech & 40 & 0.04 & 80 \\ \cdashline{1-4}[0.5pt/2pt]
\ \ Wirtualna Polska Media S.A. & 33 & 0.03 & 81 \\ \cdashline{1-4}[0.5pt/2pt]
\ \ Snigel Web Services Limited & 29 & 0.03 & 82 \\ \cdashline{1-4}[0.5pt/2pt]
\ \ Gemius SA & 25 & 0.02 & 83 \\ \cdashline{1-4}[0.5pt/2pt]
\ \ Papoo Software \& Media GmbH & 25 & 0.02 & 84 \\ \cdashline{1-4}[0.5pt/2pt]
\ \ devowl.io GmbH & 24 & 0.02 & 85 \\ \cdashline{1-4}[0.5pt/2pt]
\ \ Mozilor Limited & 22 & 0.02 & 86 \\ \cdashline{1-4}[0.5pt/2pt]
\ \ 8works & 17 & 0.02 & 87 \\ \cdashline{1-4}[0.5pt/2pt]
\ \ Seven.One Entertainment Group GmbH & 16 & 0.01 & 88 \\ \cdashline{1-4}[0.5pt/2pt]
\ \ Ensighten/Cheq & 16 & 0.01 & 89 \\ \cdashline{1-4}[0.5pt/2pt]
\ \ ShareThis Inc. & 15 & 0.01 & 90 \\ \cdashline{1-4}[0.5pt/2pt]
\ \ Axel Springer Deutschland GmbH & 14 & 0.01 & 91 \\ \cdashline{1-4}[0.5pt/2pt]
\ \ TRUENDO Technologies GmbH & 14 & 0.01 & 92 \\ \cdashline{1-4}[0.5pt/2pt]
\ \ WebAds B.V & 14 & 0.01 & 93 \\ \cdashline{1-4}[0.5pt/2pt]
\ \ optAd360 & 13 & 0.01 & 94 \\ \cdashline{1-4}[0.5pt/2pt]
\ \ Agilitation & 12 & 0.01 & 95 \\ \cdashline{1-4}[0.5pt/2pt]
\ \ AdOpt & 11 & 0.01 & 96 \\ \cdashline{1-4}[0.5pt/2pt]
\ \ Lawwwing & 11 & 0.01 & 97 \\ \cdashline{1-4}[0.5pt/2pt]
\ \ AVACY & 10 & 0.01 & 98 \\ \cdashline{1-4}[0.5pt/2pt]
\ \ illow & 8 & 0.01 & 99 \\ \cdashline{1-4}[0.5pt/2pt]
\ \ Admiral & 8 & 0.01 & 100 \\ \cdashline{1-4}[0.5pt/2pt]
\ \ Ethyca Inc. & 6 & 0.01 & 101 \\ \cdashline{1-4}[0.5pt/2pt]
\ \ GG Software LLC & 5 & 0.00 & 102 \\ \cdashline{1-4}[0.5pt/2pt]
\ \ Associated Newspapers Ltd & 5 & 0.00 & 103 \\ \cdashline{1-4}[0.5pt/2pt]
\ \ Adnuntius AS & 5 & 0.00 & 104 \\ \cdashline{1-4}[0.5pt/2pt]
\ \ RCS MediaGroup S.p.A. & 4 & 0.00 & 105 \\ \cdashline{1-4}[0.5pt/2pt]
\ \ Cloudflare & 4 & 0.00 & 106 \\ \cdashline{1-4}[0.5pt/2pt]
\ \ Commanders Act & 4 & 0.00 & 107 \\ \cdashline{1-4}[0.5pt/2pt]
\ \ Clym Inc. & 3 & 0.00 & 108 \\ \cdashline{1-4}[0.5pt/2pt]
\ \ Onesecondbefore B.V. & 3 & 0.00 & 109 \\ \cdashline{1-4}[0.5pt/2pt]
\ \ SIBBO & 3 & 0.00 & 110 \\ \cdashline{1-4}[0.5pt/2pt]
\ \ Kleinanzeigen GmbH & 2 & 0.00 & 111 \\ \cdashline{1-4}[0.5pt/2pt]
\ \ mobile.de GmbH & 2 & 0.00 & 112 \\ \cdashline{1-4}[0.5pt/2pt]
\ \ CookieMan & 1 & 0.00 & 113 \\ \cdashline{1-4}[0.5pt/2pt]
\ \ Adlane LTD & 1 & 0.00 & 114 \\ \cdashline{1-4}[0.5pt/2pt]
\ \ Shoprenter & 1 & 0.00 & 115 \\
\bottomrule
% \end{tabular}
\end{longtable}
%TC:endignore

\clearpage  

\end{document}